# Prediction of 5-hydroxytryptamine Transporter Inhibitor based on Machine Learning

Weikaixin Kong[1], Wenyu Wang[2], Jinbing An[3*]

[1] School of Pharmaceutical Sciences, Peking University, Beijing 100191, China

[2] School of Nursing, Peking University, Beijing 100191, China

[3] Department of Natural Sciences for Medicine, Peking University, Beijing 100191, China

Corresponding author: Jinbing An ： Email:kwkxtg@163.com

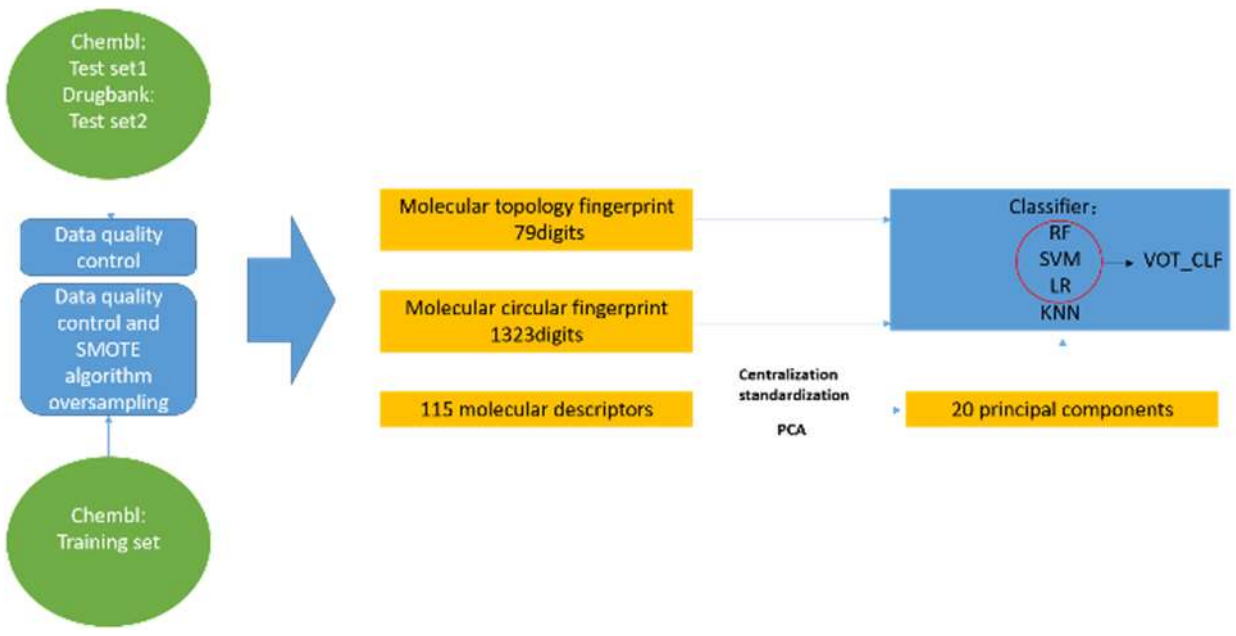

**Abstract:** In patients with depression, the use of 5-HT reuptake inhibitors can improve the condition. Topological fingerprints, ECFP4, and molecular descriptors were used. Some SERT and small molecules combined prediction models were established by using 5 machine learning methods. We selected the higher accuracy models(RF, SVM, LR) in five-fold cross-validation of training set to establish an integrated model (VOL_CLF). The training set is from Chembl database and oversampled by SMOTE algorithm to eliminate data imbalance. The unbalanced data from same sources (Chembl) was used as Test set 1; the unbalanced data with different sources(Drugbank) was used as Test set 2 . The prediction accuracy of SERT inhibitors in Test set 1 was 90.7%~93.3%(VOL_CLF method was the highest); the inhibitory recall rate was 84.6%-90.1%(RF method was the highest); the non-inhibitor prediction accuracy rate was 76.1%~80.2%(RF method is the highest); the non-inhibitor predictive recall rate is 81.2%~87.5% (SVM and VOL_CLF methods were the highest) The RF model in Test Set 2 performed better than the other models. The SERT inhibitor predicted accuracy rate, recall rate, non-inhibitor predicted accuracy rate, recall rate were 42.9%, 85.7%, 95.7%, 73.3%.This study demonstrates that machine learning methods effectively predict inhibitors of serotonin transporters and accelerate drug screening.

## 1 Introduction

Depression is a type of mood disorder characterized by a marked and persistent state of mind. It is one of the most common mental disorders. At present, the etiology and pathogenesis of depression are still unclear. It is generally believed that the onset of depression is closely related to three factors: genetic factors, biochemical factors such as norepinephrine (norepinephrine, NE), serotonin (5-hydroxytryptamine, 5-HT) and dopamine (dopamine, DA), as well as social and environmental factors [1]. Depression not only leads to a series of physical, psychological, social dysfunction, related complications, and potentially high suicide risk, but also increases the burden on patients, families, and society, and seriously reduces the quality of life of patients. According to a World Health Organization (WHO) survey, there are approximately 322 million people with depression worldwide, accounting for 4.4% of the world's population [2]. Studies have also shown that depression has become the world's largest disabling disease [3] and is expected to rise to the top of the world's disease burden by 2030 [4].

Serotonin (5-HT, serotonin) is a highly conserved chemical signal that is widely distributed in vertebrates and invertebrates. It exists in the brain and digestive tract and plays a key role in various regulatory processes. It can stimulate the target organ and participate in a variety of processes, including mood, motivation, thinking, diet and nociception [5]. After the physiological action is applied, 5-HT is inactivated to avoid sustained excitation of the target organ and desensitization of the 5-HT receptor, which is mainly accomplished by the 5-HT transporter. 5-HT can make the brain feel happy, 5-HT transporter (5-HTT / 5-hydroxytryptamine transporter, SERT) is a transmembrane transporter with high affinity for 5-HT, 5-HT is widely present in the intestine chromaffin cell membrane, mast cells and serotonergic neurons presynaptic membrane [6]. SERT is a component of the synapse at the nerve endings of the presynaptic membrane at the nerve endings. It re-enters the 5-HT of presynaptic neurons from the synaptic cleft, directly reducing the concentration of serotonin in the synaptic cleft [7], preventing the occurrence of adverse reactions. SERT removes 5-HT from the synaptic cleft and affects the number and duration of postsynaptic receptor-mediated signaling, which plays a key role in the fine-tuning of overall pulse delivery [8].

If the SERT structure changes during physiological processes, the synthesis, clearance, and function of the 5-HT and 5-HT receptors will be greatly affected. These changes have important clinical implications for SERT inhibitors. The mechanism of action of selective serotonin reuptake inhibitors (SSRIs) in antidepressants is selective inhibition of the reuptake effect of 5-HT in central nervous system presynaptic membranes, increasing the 5-HT concentration in the synaptic cleft, exciting the brain, thereby achieving an antidepressant effect. It is characterized by strong specificity and selectivity for 5-HT, limited effect on other neurotransmitters, good oral absorption and less adverse reactions [9]. In general, studying molecular biology and pharmacological properties of SERT and exploring SSRIs with selective high affinity for SERT have brought bright prospects for the treatment of these diseases.

In this research, the machine learning methods were used to predict the binding ability of the ligand molecule and the SERT protein, based on the molecular descriptors and the molecular fingerprints, and the obtained model can be applied to the drug screening. In the process, the cost of the experiment can be reduced and the drug development cycle is shortened. This work has certain practical application value.

# 2 Materials and methods

## 2.1 Data acquisition and processing

This study selected the training set, validation set and Test set 1 data from the Chembl database. The experimental model of the selected data was Hek293 cells. The experimental method was the isotope [3H] labeling method, and all the molecular smiles provided by the Chembl database were read and written by the RDkit package in Python, and the standard smiles were generated and the duplicate data was removed. Considering that the half-inhibitory concentration IC50 of the drugs that have significant inhibition of SERT protein published in the Drugbank database is less than 500 nmol • $L^{-1}$, and in order to make the established model have better generalization performance, this study will regard the molecules with IC50 smaller than 500 nmol • $L^{-1}$ as inhibitors of SERT protein, and molecules with IC50 greater than 1000 nmol • $L^{-1}$ as non-inhibitors of SERT protein, which is similar to the method has been reported[10].. We screened 812 inhibitors and 400 non-inhibitors using the above method; 162 inhibitors and 80 non-inhibitors were randomly selected as Test set 1. Since many machine learning algorithms are sensitive to unbalanced data, we used the SMOTE algorithm for oversampling the remaining 650 inhibitors and 320 non-inhibitors to eliminate data imbalance [11]. Finally, we obtained 960 samples of inhibitor and 960 samples of non-inhibitor, and then used machine learning methods to establish several classification models, and verified the models by five-fold cross-validation.

The test set 2 is an external data set. The test set 2 was constructed using compounds which were reported by Drugbank and not used in the training set. Seven IC50s with IC50 less than 500 nmol • $L^{-1}$ SERT were found, and 30 drugs without SERT protein binding ability were selected as non-inhibitors of SERT protein. Both Test Set 1 and Test Set 2 are unbalanced data sets that can be used to test the robustness and generalization capabilities of the model. The molecular smiles of the training data set, the test 1 data set, and the test 2 data set are in supplementary materials (Supplementary Table 1-3).

## 2.2 The chemical space

We obtained the ECFP4 fingerprint of the molecule, and then randomly selected 100 molecules in all data sets to find the Tanimoto correlation coefficient between every two molecules [12] [13]. We finally found the average Tanimoto correlation coefficient of 100 molecules. The Tanimoto correlation coefficient can effectively measure the similarity between molecules, which in turn reflects the size of the chemical space occupied by the data. In the training data set, we calculated the relative molecular mass (MW), the lipid-water partition coefficient (ALogP), the hydrogen bond acceptor number (nHBAcc), the hydrogen bond donor number (nHBDon) , and the number of rotatable keys(nRotB), according to the Lipinski rules of five[14] [15] [16]. MW and ALogP are used to plot scatter plots between data with different labels. The Lipinski rules of five are plotted into a radar chart to observe the chemical space distribution of the compound molecules[17]. The calculation of the five properties is done by the "rcdk" package in the

R language (3.5.3)[18].

## 2.3 Extraction of molecular features

In this study, 115 descriptors of molecules were extracted using the RDkit package as a characterization of molecular properties [19], using topological fingerprints and circular fingerprints (ECFP4) [20], [21] as molecular structure characterization. Here we do not use traditional hash coding methods for molecular fingerprints. When the molecular fingerprint is obtained, the molecular fingerprint appearing in all the molecules of the training set is taken as the complete set U, and one-hot encoding is performed, and the fingerprint features related to the complete set U are obtained in the test set 1 and the test set 2, such as the formula(1) (2).

$$a_i = \cup_{k=1}^{m} f_{ik} \quad (1)$$

$$\mathrm{U} = \cup_{i=1}^{n} ai \quad (2)$$

The number of samples in the training set was n. $f_{ik}$ was the fingerprint feature k of the molecule i. Then m was the number of features of the molecule i. And $a_1$、$a_2$、……、 $a_n$ were the sets of all fingerprint features of each molecule in the training set. Finally, 79 topological fingerprints and 1323 circular fingerprints were obtained.

## 2.4 Machine learning methods

In this study, the classification models were built using the support vector machine (SVM), logistic regression (LR), random forest (RF), and k nearest neighbor (KNN). An in-depth description of these four methods can be obtained from some excellent works and research papers. Here, only the main ideas of the four methods are briefly described.

LR uses the idea of maximum likelihood estimation to minimize the loss function of the regression to obtain the unknown parameters [22]. Using the sigmoid function on the final result of the binary classification problem, the predicted value range is mapped to [0, 1], and finally, the classification is realized.

SVM is a machine learning method based on the principle of structural risk minimization [23]. It maps the input variables to higher-dimensional spaces by changing the kernel function, and then uses the so-called "support vector" to find the maximum interval in the new space, and clarifies the optimal classification hyperplane. Finally, the SVM completes the classification of the input data. The SVM model of this study uses a linear kernel function.

RF is a collection of multiple decision trees. The process of RF establishment is to construct multiple decision trees by randomly extracting different features and different samples, that is, multiple weak classifiers. The results of multiple weak classifiers are voted to get the final result. For out-of-bag data that has not been extracted, it can be used to test the generalization performance of the model [24].

KNN is based on the idea of maximum likelihood estimation and finds the distance between the new input sample point and the training sample points in the multidimensional space formed by the features. First, the n training sample points closest to the new sample point are selected, and then the distance between the samples and the number of labels of each category in the n training sample points are comprehensively considered to determine the final classification result [25].

In addition, this study builds the voting classification model VOT_CLF based on the idea of ensemble learning. Considering the class probability of the predictions of multiple models, the result with the largest class probability is the final classification result. The study of Bauer et al[26] shows that when the classification principles of multiple sub-classifiers are different and the performance is good enough, the voting model often has the performance improvement.

## 2.5 Prediction results evaluation

In the classification problem, for the A category molecular, the accuracy of the A category in the final result is: the number of real A in the molecules predicted as A category/the number of the moleculars predicted as A category; the recall rate is: the number of predicted A in the real A moleculars/ the number of the real A moleculars; f1-score is: f1=2×precision×recall÷(precision+recall). The f1 score is the harmonic mean of the accuracy and recall rate. In the drug screening problem, the recall rate of SERT protein inhibitor (SERTI) is usually more concerned. In addition, the overall accuracy rate = the number of correct samples in all the classification results / the total number of samples.

In this study, we first performed five-fold cross-validation on the training set. With the overall accuracy rate, we selected a better classifier for the generation of the voter and predict the results of the test 1 and test 2 data sets.

### 2.6 Structural alerts

We use python's library "bioalerts"[27] to find substructures related to molecular neutralization activity. This method searches and counts substructures in a molecule by setting a certain search radius. The molecular fingerprint used in the search for substructures is ECFP4. The number of occurrences of a molecular substructure can be seen as obeying the hypergeometric distribution[28]. When the number of occurrences of a molecular substructure in inhibitors is significantly higher than the number of occurrences in non-inhibitors, it can be considered that this substructure plays an important role in the binding process with SERT protein. We used the training data set before oversampled to find structural alerts.

## 3 Results and discussion

### 3.1 The distribution of the chemical space

The number of molecules in the data set is shown in Table 1. The heat map of the Tanimoto correlation coefficient of 100 randomly selected molecules is shown in Fig. 1. The average Tanimoto correlation coefficient is 0.192, which indicates that the differences between molecules are large, and the model trained may have strong generalization ability. In the radar chart, the inside is a small value and the outside is a large value. It can be seen from Fig. 2 that the molecular distribution of the training set is broad and does not exhibit preference, whether it is a discrete variable or a continuous variable, which is consistent with the results of the Tanimoto correlation coefficient. From the scatter plot (Fig. 3), we can see that there is no difference in the distribution of inhibitors and non-inhibitors between the two continuous indicators of MW and ALogP. This suggests that it is difficult to successfully distinguish between the two types of molecules by simply relying on these simple chemical properties, and the methods we used in this study can solve this problem.

**Table 1 The statistics of chemicals in the data set**

| Data sets | Training set | Train set after oversampled | Test set 1 | Test set 2 |
|---|---|---|---|---|
| Non-inhibitor | 320 | 960 | 80 | 30 |
| Inhibitors | 650 | 960 | 162 | 7 |
| Total | 970 | 1920 | 242 | 37 |

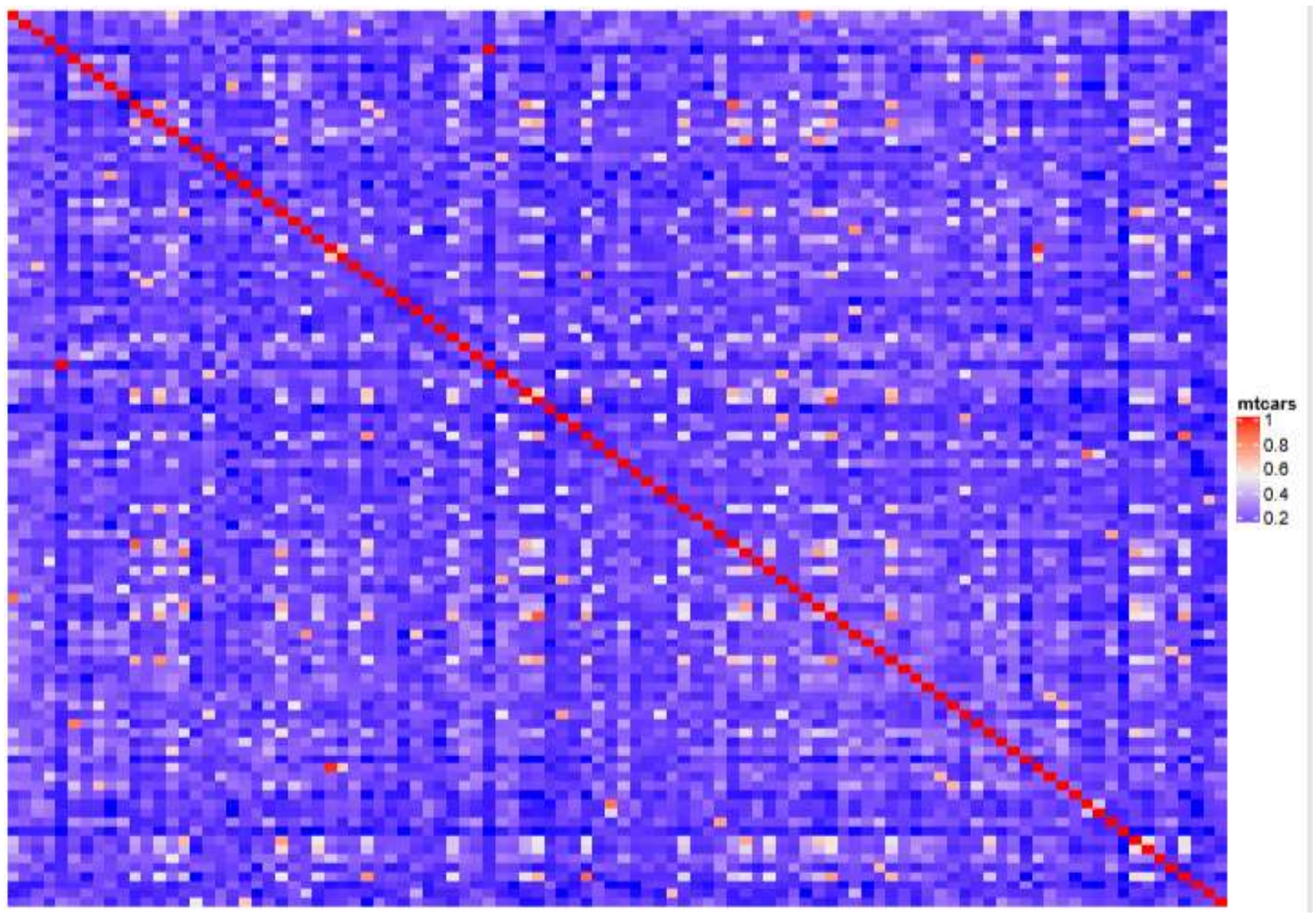

**Fig.1 Heat map of Tanimoto correlation coefficient**

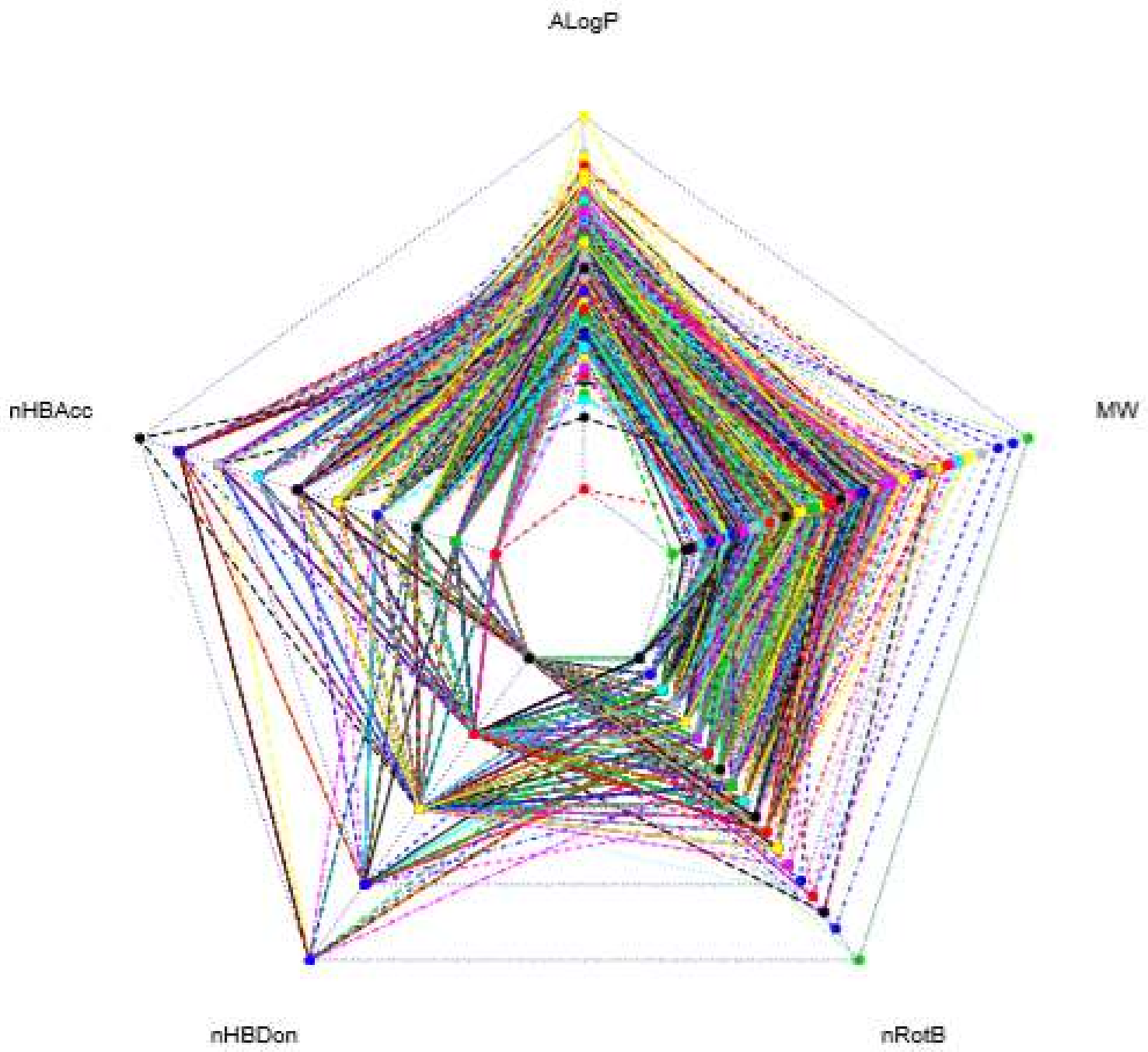

**Fig.2 Radar map of molecular properties of training data sets**

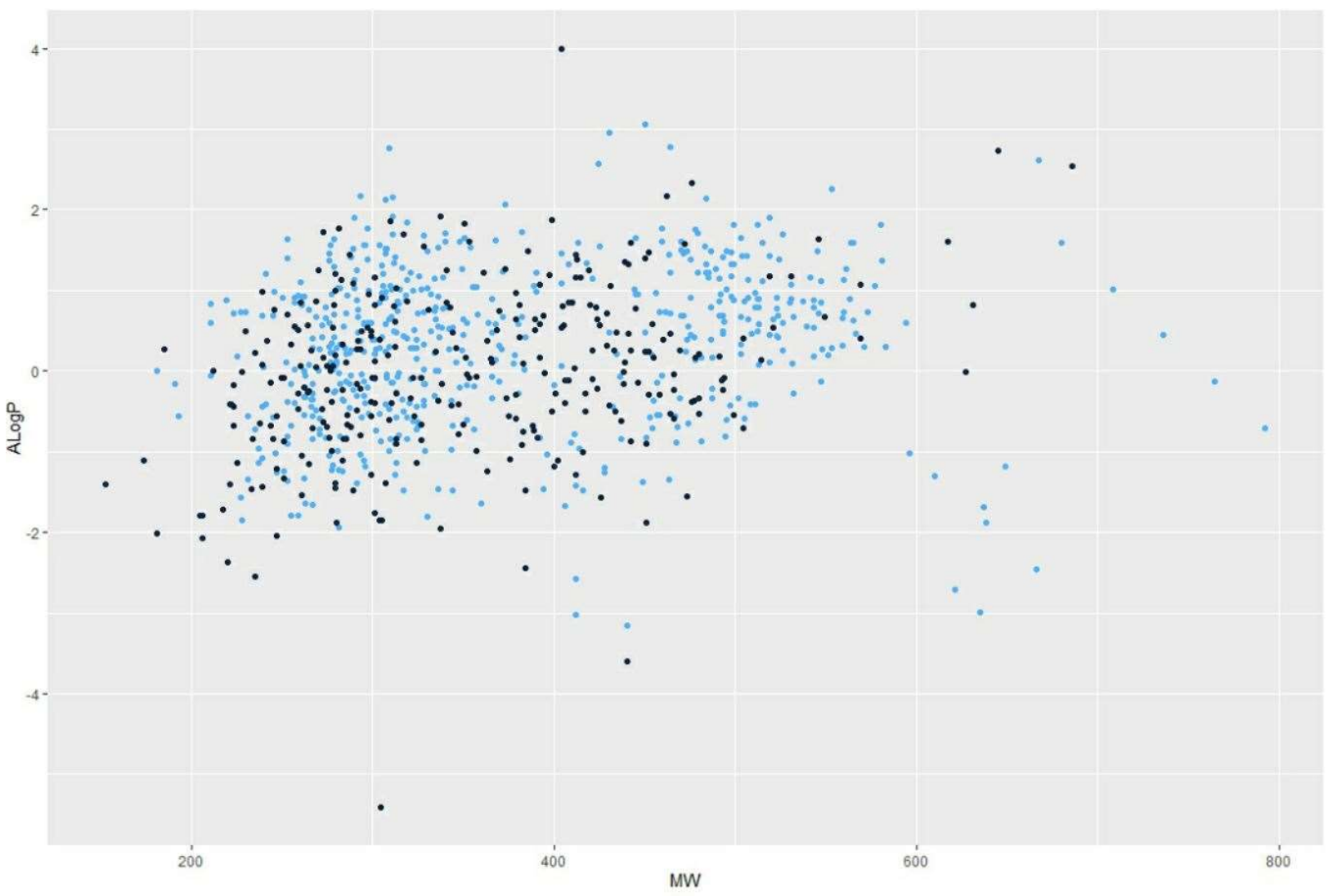

**Fig.3 Scatter plots of molecular properties in different categories (the training set before oversampled, blue for inhibitors, black for non-inhibitors)**

## 3.2 Cross-validation result

In the different machine learning models, the training set was spilt with the ratio 1:4 five times to complete the cross-validation. The results of the cross-validation are shown in Table 2.

Based on the overall accuracy criteria, the RF, SVM, and LR models were selected to construct the ensemble voter VOT_CLF, and the five-fold cross-validation was also performed.

**Table 2 Five-fold cross-validation of different machine learning methods**

| Method | RF | SVM | LR | KNN | VOT_CLF |
|---|---|---|---|---|---|
| accuracy/% | 92.6 | 94.8 | 94.5 | 91.8 | 95.4 |
| std [a] | 0.08 | 0.029 | 0.035 | 0.023 | 0.039 |

[a] Std represents the standard deviation of the five-fold cross-validation

The cross-validation accuracy of all models can reach more than 90%, indicating that the feature extraction method of molecular fingerprints combined with molecular descriptors can fully characterize the molecules to some extent, which can provide more effective information than just using fingerprints or descriptors.

## 3.3 Test set 1 results

After cross-validation, we trained the models with all the data of the training set. The ratio of SERT protein inhibitors to non-inhibitors in the test set 1 was close to 2:1, which is an unbalanced data set. The relevant prediction results are shown in Table 3. The RF model performed best in the SERTI recall rate, and the VOT_CLF model performed best in the SERTI accuracy rate. In non-SERT protein inhibitors (non-SERTI), the SVM and VOT_CLF models have higher recall rates, and the RF model has higher accuracy. In the Receiver Operating Characteristic Curve (ROC) (Fig. 4), except for the SVM model, the other three model curves have a larger area under the curve (AUC). In summary, the VOT_CLF model and the RF model performed better in the test set 1 prediction results.

At the same time, the test set 1 is the unbalanced data set with the majority of positive samples. It can be seen that the VOT_CLF model and the RF model are robust to this imbalance.

**Table3 Unbalanced test set 1 prediction result**

| Method | SERTI | | | | non-SERTI | | | |
|---|---|---|---|---|---|---|---|---|
| | Precision/% | Recall/% | f1-score/% | Support[a] | Precision/% | Recall/% | f1-score/% | Support |
| RF | 90.7 | 90.1 | 90.4 | 162 | 80.2 | 81.2 | 80.7 | 80 |
| SVM | 93.2 | 84.6 | 88.7 | 162 | 73.7 | 87.5 | 80 | 80 |
| LR | 91.4 | 85.2 | 88.2 | 162 | 73.6 | 83.8 | 78.4 | 80 |
| VOT_CLF | 93.3 | 86.4 | 89.7 | 162 | 76.1 | 87.5 | 81.4 | 80 |

### 3.4 Test set 2 results

The ratio of SERT protein inhibitor to non-inhibitor in the test set 2 was 7:30, which was also an unbalanced data set, and the negative sample accounted for the majority. The relevant prediction results are shown in Table 4. The RF model has the best performance in the SERTI recall rate, and the RF model has a high accuracy and recall rate in the non-SERTI, so the RF model may be better used in the drug screening process. The test set 2 and the training set are from different databases, so the generalization performance and robustness of the RF model are better than the other three models.

At present, how to find the possible active drug from the severe imbalance data is a major problem in machine learning to predict the combination of protein and small molecule [29], [30], the researchers tried to solve this problem, and then adopted oneshot-learning method, etc[31]. It can be seen from Test Set 1 and Test Set 2 that the RF model has a higher recall rate than the SVM, LR, and voting models. This indicates that the RF model is more conducive to processing unbalanced data sets characterized by molecular fingerprints and molecular descriptors. This is consistent with some literature results [32], [33].

**Table4 Unbalanced test set 2 prediction result**

| Method | SERTI | | | | non-SERTI | | | |
|---|---|---|---|---|---|---|---|---|
| | Precision/% | Recall/% | f1-score/% | Support | Precision/% | Recall/% | f1-score/% | Support |
| RF | 42.9 | 85.7 | 57.1 | 7 | 95.7 | 73.3 | 83.1 | 30 |
| SVM | 30.8 | 57.1 | 39.9 | 7 | 87.5 | 70.1 | 77.8 | 30 |
| LR | 33.3 | 57.1 | 42.1 | 7 | 88.1 | 73.3 | 80.1 | 30 |
| VOT_CLF | 33.3 | 57.1 | 42.1 | 7 | 88.1 | 73.3 | 80.1 | 30 |

### 3.5 The wrongly predicted drugs

In Test Set 2, the RF model of this study successfully predicted six molecules (Fig.5). in 7 SERTIs and erroneously predicted one molecular (Fig.6). The reasons for Cocaine's wrong prediction may be as follows: The lack of compounds containing nitrogen bridged rings in the training samples, that is, the limited space of the training samples, makes the extracted molecular fingerprints more limited, and it is difficult to characterize the molecular structure of the aza bridges. The lack of descriptors in the molecular descriptors used to accurately characterize the properties of the aza bridges makes the model's predictive ability limited and limited coverage of predicted compounds. Therefore, it is possible to explore new descriptors, enhance data collection, increase training samples, or develop model application domain assessment methods to effectively define the application scope of the model and improve model performance. In addition, we will try other combinations of nonlinear features that are more in line with molecular structural features to further improve the efficiency of machine learning.

### 3.5 Molecular structural alerts for SERT protein

It can be seen from these structures(Table 5) that when the molecular structure tends to shift to piperidine (S2-S5) or tends to form a conjugated structure (S6, S9, S11), it has a higher binding ability to the SERT protein. At this time, the molecule is more basic and more easily forms a hydrogen bond with the amino acid residue on the SERT to function. Many inhibitors of enzymes also contain imidazole structures[34], which is very similar to S12. In summary, these 12 important substructures can provide a reference for the structural design of SERT inhibitors.

**Table5 Molecular Structural Alert for SERT Protein Binding Ability**

| Substructure | Substructure ID | *p* | Compounds with substr. | Comp. with substr. active | Comp. with substr. inactive |
|---|---|---|---|---|---|
| 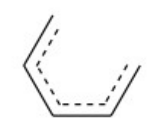 | S1 | <0.001 | 278 | 255 | 23 |
| 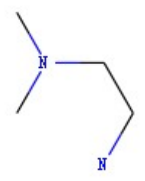 | S2 | 0.0053 | 203 | 179 | 24 |
| 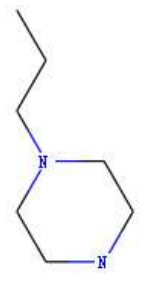 | S3 | 0.0041 | 135 | 122 | 13 |
| 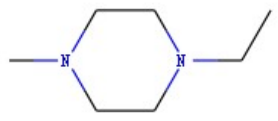 | S4 | <0.001 | 148 | 135 | 13 |
| 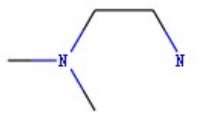 | S5 | <0.001 | 192 | 176 | 16 |
| 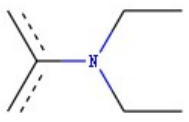 | S6 | <0.001 | 171 | 162 | 9 |
| 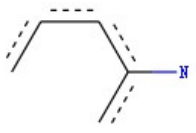 | S7 | <0.001 | 183 | 172 | 11 |
| 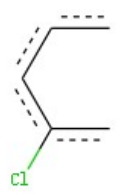 | S8 | <0.001 | 132 | 126 | 6 |
| 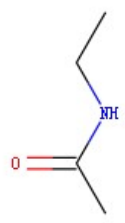 | S9 | <0.001 | 143 | 136 | 7 |
| 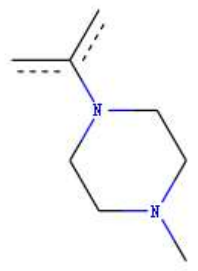 | S10 | <0. 001 | 171 | 162 | 9 |
| 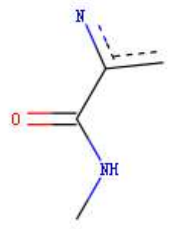 | S11 | <0. 001 | 108 | 107 | 1 |
| 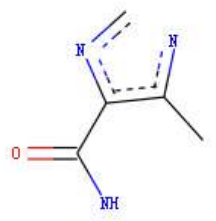 | S12 | <0. 001 | 105 | 105 | 0 |

## 4 Conclusion

Studies have shown that for ligand-based SERTI prediction problems, the use of machine learning methods, especially random forests, has a higher recall rate for new test sets. The RF model is robust to unbalanced data and data sets from different sources and has a high generalization ability. The model obtained by the institute can be applied to the initial screening process of the drug. But so far, the SERTI training data is still very limited, the sample space is small, and the molecular descriptors are not comprehensive. However, we believe that in the near future, with the accumulation of data and the improvement of molecular characterization methods, this shortcoming will gradually be solved. Besides, the improvement of the machine learning method to get higher accuracy and recall rate of SERTI will effectively solve the effective drug-protein binding prediction problem, making the machine learning method a powerful tool in the drug development process.

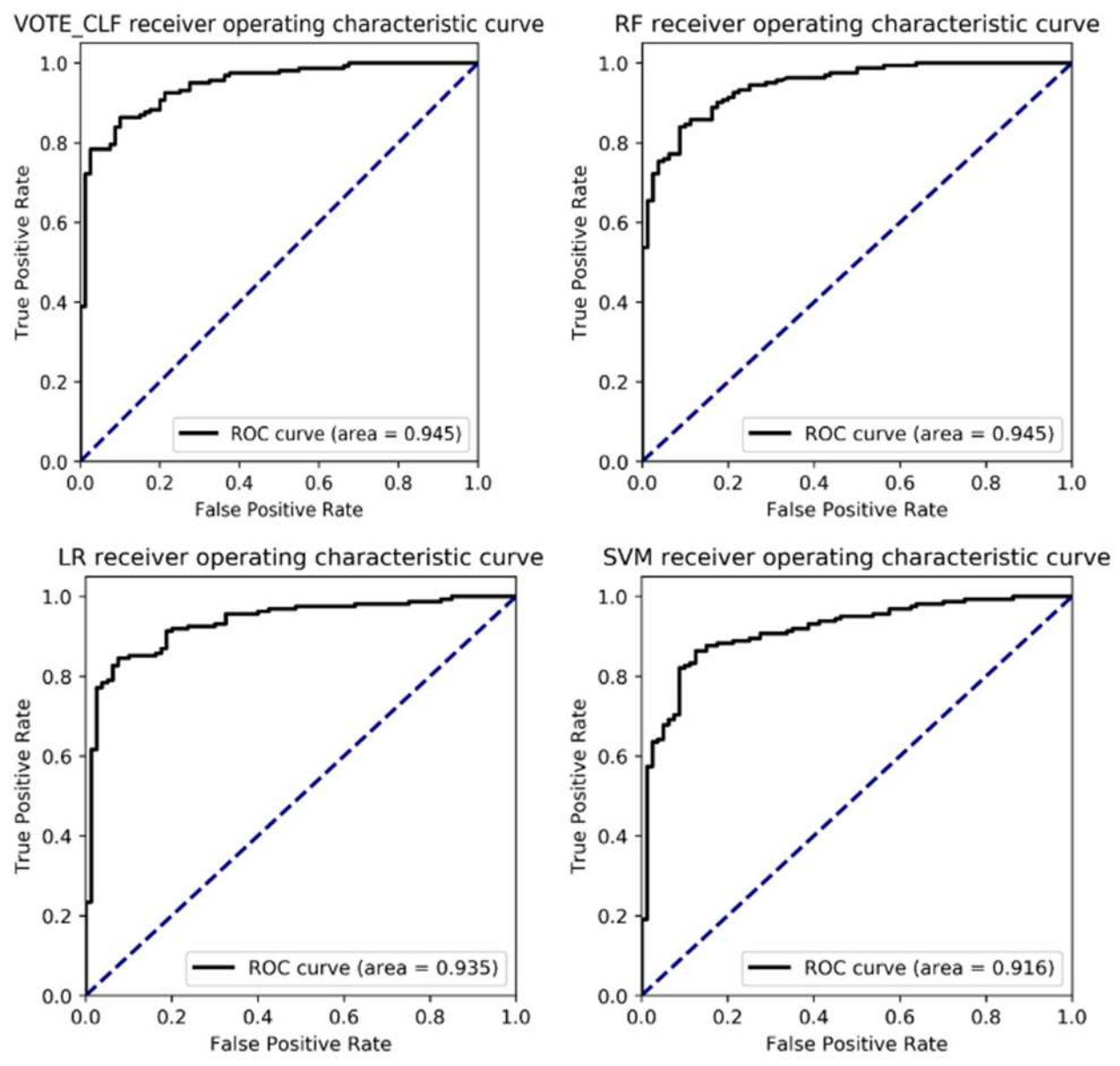

**Fig.4 Receiver operating characteristic curves for four models**

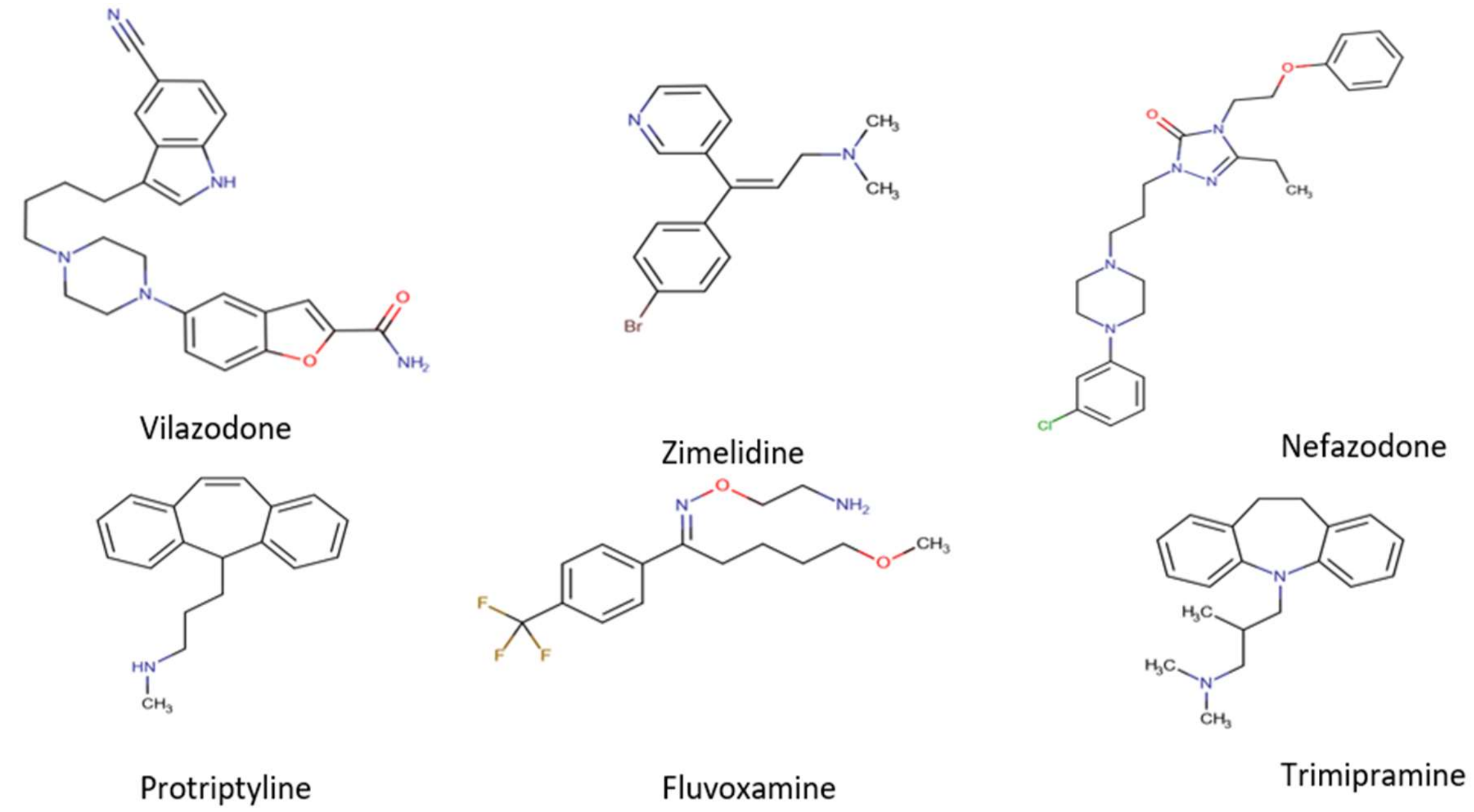

**Fig5. The correct predicted molecules in the RF model**

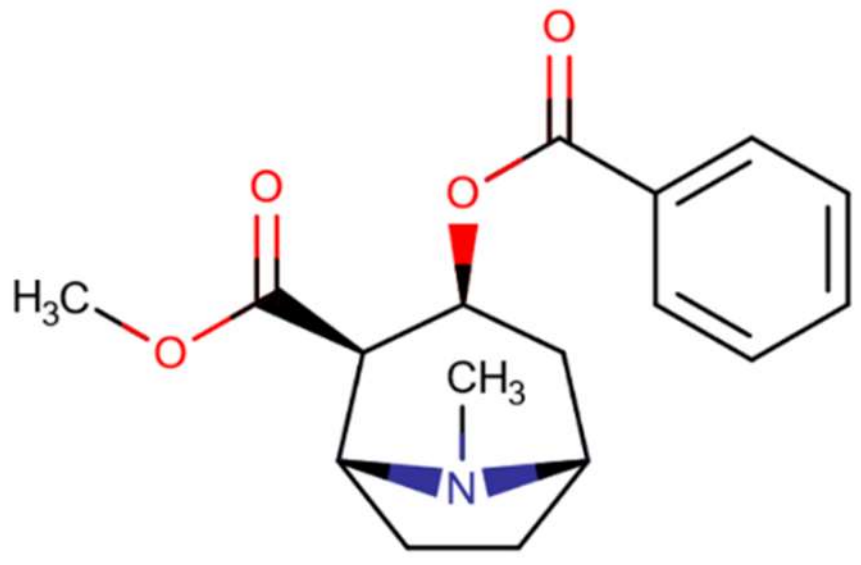

Cocaine

**Fig6. The wrong predicted molecules in the RF model**

## Funding

This article was sponsored by the National Natural Science Foundation of China (No. 11771048 and No. 81871854).

## Author Contributions Statement

All authors contributed to the work presented in this paper. Weikaixin Kong and Jinbing An designed the calculation process, wrote the manuscript, and prepared the figures. Weikaixin Kong and Wenyu Wang completed the selection of topics, the collection of background materials and data, and wrote relevant content. Weikaixin Kong completed the calculations and wrote relevant content. Jinbing An gave advice on the determination of analytical methods and the writing of the article. The manuscript was reviewed and approved by all authors.

## Acknowledgements

This study is the project content of the "1st Big Data Workshop" (March 2018) organized by Peking University's Department of Natural Sciences for Medicine.

## Competing interests

The authors declare that they have no competing interests.

# Supplementary material

**Supplementary Table 1 The training data set**

| CHEMBL ID | smiles | IC50/nM |
|---|---|---|
| CHEMBL283416 | CN1CC=C(c2c[nH]c3ccc([N+](=O)[O-])cc23)CC1 | 40 |
| CHEMBL1214367 | Cc1cccc(N2CCN(CCCNC(=O)c3cc(-c4ccc(Cl)cc4)n(C)c3C)CC2)c1C.Cl | 129 |
| CHEMBL2386162 | COc1ccc(OC(CC(C)C)C2CCNC2)c(C)n1 | 1.9 |
| CHEMBL362298 | O=C1OC2(CCC(N3CCC(Cc4ccccc4C(F)(F)F)CC3)CC2)c2ccc3c(c21)OCO3 | 180 |
| CHEMBL446084 | NCCC(Oc1ccc(C(F)(F)F)cc1)c1ccccc1 | 13 |
| CHEMBL260036 | CN(C)Cc1ncccc1Oc1ccc(Cl)cc1Cl | 8 |
| CHEMBL81525 | COC(=O)C1C2CCC(CC1c1ccc(Cl)cc1)N2 | 1.63 |
| CHEMBL1214116 | CCCn1c(-c2ccccc2)cc(C(=O)NCCCN2CCN(c3cccc(C(F)(F)F)c3)CC2)c1C.Cl | 381 |
| CHEMBL101250 | Cc1ccc(C2CC3CCC4C2C(C)(O)CN34)cc1 | 147 |
| CHEMBL3215853 | Cc1c(Cl)cccc1N1CCN(CCCNC(=O)c2nc(C)n(CC(C)C)c2C)CC1.Cl.Cl.Cl | 447.2 |
| CHEMBL303266 | CN1C2CCC1C(C(=O)NCCCNC(=O)C1C(c3ccc(Cl)cc3)CC3CCC1N3C)C(c1ccc(Cl)cc1)C2 | 441 |
| CHEMBL212878 | c1ccc(C(Cc2cccc3ccccc23)N2CCNCC2)cc1 | 12 |
| CHEMBL1088165 | Cc1cccc(N2CCN(CCCNC(=O)c3cc(-c4ccccc4)n(N4CCCC4)c3C)CC2)c1C | 92 |
| CHEMBL213035 | N#Cc1ccccc1CC(c1ccccc1)N1CCNCC1 | 87 |
| CHEMBL6467 | CSc1ccc(CC(C)N)cc1 | 74 |
| CHEMBL1080522 | Cc1[nH]c(-c2ccccc2)cc1C(=O)NCCN1CCN(c2cccc(Cl)c2Cl)CC1 | 87 |
| CHEMBL169471 | COC(=O)C1C2CCC(CC1c1ccc(Cl)c(Cl)c1)N2 | 1.24 |
| CHEMBL549965 | c1ccc(Oc2ccccc2OC(c2cccnc2)C2CCNCC2)cc1 | 76 |
| CHEMBL578610 | CC(NC(C)(C)C)C(=O)c1ccc(F)c(F)c1 | 100 |
| CHEMBL1080728 | CCCn1c(-c2ccccc2)cc(C(=O)NCCCN2CCN(c3cccc(Cl)c3)CC2)c1C | 93 |
| CHEMBL201978 | CCC(C(=O)c1ccc(Cl)c(Cl)c1)N1CCCC1 | 441 |
| CHEMBL1790051 | COC(=O)C1C(c2ccccc2)CC2CCC1N2C.Cl | 21 |
| CHEMBL2114227 | COC(=O)C1CC(c2ccc(Cl)c(Cl)c2)C2CC(=O)C1N2C | 290 |
| CHEMBL3323092 | CN1Cc2ccccc2C(c2coc3ccccc23)C1 | 67 |
| CHEMBL809 | CNC1CCC(c2ccc(Cl)c(Cl)c2)c2ccccc21 | 3 |
| CHEMBL481697 | CN(C)Cc1cc(CNS(C)(=O)=O)ccc1Oc1ccc(OC(F)(F)F)cc1 | 11 |
| CHEMBL245482 | Cc1cccc(CN(Cc2ccccn2)C2CCNC2)c1C | 8 |
| CHEMBL477580 | CC(C)C(=O)N(Cc1cc(Cl)cc(Cl)c1Cl)C1CCNC1 | 44 |
| CHEMBL3612828 | O=C(NCCCN1CCC(Cc2ccccc2)CC1)c1cc2ccccc2s1 | 449 |
| CHEMBL476954 | CC(C)C(=O)N(Cc1cccc(Cl)c1Cl)C1CCNC1 | 12 |
| CHEMBL377900 | Clc1ccc(Cl)c(CC(c2ccccc2)N2CCNCC2)c1 | 8.2 |
| CHEMBL3216097 | CCc1nc(C(=O)NCCN2CCN(c3cccc(Cl)c3Cl)CC2)c(C)n1-c1ccccc1.Cl.Cl.Cl.Cl | 245 |
| CHEMBL1642893 | NC1Cc2ccccc2C(c2ccc(Cl)c(Cl)c2)C1 | 108 |
| CHEMBL3216074 | Cc1cccc(N2CCN(CC(O)CNC(=O)c3nc(-c4ccccc4)n(-c4ccc(F)cc4)c3C)CC2)c1C.Cl.Cl | 150.1 |
| CHEMBL124222 | O=C1CN(CCOC(c2ccccc2)c2ccccc2)CCN1CCCc1ccccc1 | 10 |
| CHEMBL1812745 | COC(=O)C1C(c2ccc(Cl)c(C)c2)CC2CCC1N2C | 1 |
| CHEMBL563987 | Fc1cccc(OC(c2cccnc2)C2CCNCC2)c1F | 3 |
| CHEMBL481353 | CN(C)Cc1ccccc1Oc1ccc(OC(F)(F)F)cc1 | 7 |
| CHEMBL3215624 | CCCc1nc(C(=O)NCCCN2CCN(c3cccc(Cl)c3Cl)CC2)c(C)n1-c1ccccc1F.Cl.Cl.Cl.Cl | 37 |
| CHEMBL1173428 | CC1NC(C)(C)COC1(O)c1ccc(Cl)c(Cl)c1 | 360 |
| CHEMBL3407785 | CNC(=O)C12CC1C(n1cnc3c(NC)nc(C#Cc4ccc(Br)s4)nc31)C(O)C2O | 10 |
| CHEMBL3216289 | Cc1cccc(N2CCN(CC(O)CNC(=O)c3nc(C)n(-c4ccccc4F)c3C)CC2)c1C.Cl.Cl | 7.59 |
| CHEMBL257593 | COc1cc(Cl)ccc1Oc1ccncc1CN(C)C | 6 |
| CHEMBL481306 | CNCc1cc(S(N)(=O)=O)ccc1Cc1ccc(Cl)c(Cl)c1 | 6 |
| CHEMBL321067 | Cc1ccc(C2CC3CCC4C2C(O)(c2ccccc2)CN34)cc1 | 7.9 |
| CHEMBL3215864 | Cc1cccc(N2CCN(CCCNC(=O)c3nc(C)n(-c4ccc5c(c4)OCCO5)c3C)CC2)c1C.Cl.Cl | 6.08 |
| CHEMBL451165 | CNCc1cc(C(N)=O)ccc1Cc1ccc(Cl)c(Cl)c1 | 4 |
| CHEMBL1080726 | Cc1c(C(=O)NCCCN2CCN(c3cccc(Cl)c3)CC2)cc(-c2ccccc2)n1C | 11.1 |
| CHEMBL2337593 | Clc1ccccc1OC(c1ccccc1)C1CCNC1 | 3.981 |
| CHEMBL549825 | Clc1cccc(OC(c2cccnc2)C2CCCNC2)c1Cl | 4 |
| CHEMBL445 | CNCCC=C1c2ccccc2CCc2ccccc21 | 13 |
| CHEMBL259136 | CNCc1cnccc1Oc1ccc(Cl)cc1Cl | 39 |
| CHEMBL489186 | CSc1ccc(Oc2ccncc2CN(C)C)cc1C | 8 |
| CHEMBL450067 | CSc1ccc(Oc2ccc(NS(C)(=O)=O)cc2CN(C)C)cc1 | 5 |
| CHEMBL113136 | CN(C)CC1C2CCC(CC2)C1c1cccc(Cl)c1 | 303 |
| CHEMBL211412 | Fc1ccccc1C(Cc1ccccc1Cl)N1CCNCC1 | 30 |
| CHEMBL363507 | O=C1OC2(CCC(N3CCC(Cc4ccccc4F)CC3)CC2)c2ccc3c(c21)OCO3 | 410 |
| CHEMBL4093756 | CCOC(=O)C12CC1C(n1cnc3c(NC)nc(C#Cc4ccc(Cl)s4)nc31)C(O)C2O | 10 |
| CHEMBL513313 | CCCCC(=O)N(Cc1ccc(Cl)cc1Cl)C1CCNC1 | 33 |
| CHEMBL1644474 | Clc1cccc(OC2CC3CCC(C2)N3)c1 | 135 |
| CHEMBL3216967 | CCCc1nc(C(=O)NCC(O)CN2CCN(c3cccc(C)c3C)CC2)c(C)n1-c1ccc(OC)cc1.Cl.Cl | 6 |
| CHEMBL256293 | CCc1cc(Cl)ccc1Oc1cc(C)ncc1CN(C)C | 19 |
| CHEMBL562691 | Cc1ccnc(C(Oc2cccc(Cl)c2Cl)C2CCNCC2)c1 | 8 |
| CHEMBL3944260 | O=C(OC(c1ccccc1)C1CNC1)c1ccccc1 | 402 |
| CHEMBL2447952 | CCCc1nc(C(=O)NCC(O)CN2CCN(c3cccc(C)c3C)CC2)c(C)n1-c1ccc(F)cc1.Cl.Cl | 6.3 |
| CHEMBL637 | COc1ccc(C(CN(C)C)C2(O)CCCCC2)cc1 | 7.18 |
| CHEMBL1852659 | CCCc1nc(C(=O)NCCCN2CCN(c3cccc(Cl)c3Cl)CC2)c(C)n1-c1ccc(OC)cc1 | 84 |
| CHEMBL2337599 | N#Cc1cccc(OC(c2ccccc2)C2CCNC2)c1 | 0.631 |
| CHEMBL2338044 | Fc1ccc(F)c(OC(c2ccccc2)C2CCNC2)c1F | 63.1 |
| CHEMBL2219901 | COC(c1ccc(Cl)cc1)C1CNC1 | 42.9 |
| CHEMBL1214366 | Cc1cccc(N2CCN(CCCNC(=O)c3cc(-c4ccc(Cl)cc4)[nH]c3C)CC2)c1C.Cl | 19 |
| CHEMBL443922 | CN(C)Cc1ccccc1Oc1ccc(Cl)c(Cl)c1 | 11 |
| CHEMBL3407784 | CNC(=O)C12CC1C(n1cnc3c(NC)nc(C#Cc4ccc(Cl)s4)nc31)C(O)C2O | 10 |
| CHEMBL1852472 | CCCc1nc(C(=O)NCCN2CCN(c3cccc(Cl)c3Cl)CC2)c(C)n1-c1ccccc1 | 82 |
| CHEMBL1642897 | CNC1Cc2ccccc2C(c2ccc(Cl)c(Cl)c2)C1 | 107 |
| CHEMBL214259 | Clc1cccc(Cl)c1CC(c1ccccc1)N1CCNCC1 | 12 |
| CHEMBL402057 | CN(C)Cc1ccccc1Oc1ccc(Cl)cc1Cl | 4 |
| CHEMBL559456 | Clc1cccc(OC(c2cccnc2)C2CCNCC2)c1Cl | 11 |
| CHEMBL3612825 | O=C(NCCCN1CCC(Cc2ccccc2)CC1)c1cc(Cl)cc(Cl)c1 | 422 |
| CHEMBL214335 | COc1ccccc1CC(c1ccccc1)N1CCNCC1 | 18 |
| CHEMBL493057 | CN(C)Cc1ccccc1Sc1cccc(C(F)(F)F)c1 | 40 |
| CHEMBL2380974 | CN(Cc1ccccc1)CC1CCCC1c1ccc2[nH]cc(C#N)c2c1 | 140 |
| CHEMBL482295 | CN(C)Cc1cc(NS(C)(=O)=O)ccc1Sc1ccc(C(F)(F)F)cc1 | 3 |
| CHEMBL565396 | CC(NC(C)(C)C)C(=O)c1cccc(Br)c1 | 100 |
| CHEMBL469947 | C=CCN(C(=O)C1(c2cccs2)CC1CN)C1CC1 | 45 |
| CHEMBL592854 | Cc1cccc(C(=O)C(C)NC2CCCC2)c1 | 100 |
| CHEMBL2103774 | C#CC1(O)CCC2C3C(C)CC4=C(CCC(=O)C4)C3CCC21C | 90.7 |
| CHEMBL451120 | CSc1ccc(Oc2ccc(Br)cc2CN(C)C)cc1 | 2 |
| CHEMBL213632 | CS(=O)(=O)c1ccccc1CC(c1ccccc1)N1CCNCC1 | 400 |
| CHEMBL3216073 | CCCc1nc(C(=O)NCC(O)CN2CCN(c3cccc(Cl)c3Cl)CC2)c(C)n1-c1ccc(F)cc1.Cl.Cl.Cl.Cl | 65 |
| CHEMBL391806 | Clc1cccc(CN(Cc2ccccn2)C2CCNC2)c1Cl | 5 |
| CHEMBL256137 | CNCc1cnc(C)cc1Oc1ccc(Cl)cc1C | 6 |
| CHEMBL3769872 | CC(C)C(C)C1(c2ccc3ccccc3c2)CCC(C)(O)CC1 | 18 |
| CHEMBL3216082 | Cc1nc(C(=O)NCC(O)CN2CCN(c3cccc(Cl)c3Cl)CC2)c(C)n1-c1ccccc1F.Cl.Cl.Cl.Cl | 13 |
| CHEMBL2380976 | CCN(C)CC1CCCC1c1ccc2[nH]cc(C#N)c2c1 | 200 |
| CHEMBL91184 | COC(=O)C1C(c2ccc(Cl)c(Cl)c2)CC2CC(O)C1N2C | 15 |
| CHEMBL1213931 | Cc1c(C(=O)NCCCN2CCN(c3ccccc3Br)CC2)cc(-c2ccccc2)n1C | 225 |
| CHEMBL596775 | CC(NC1CCCC1)C(=O)c1cccc([N+](=O)[O-])c1 | 100 |
| CHEMBL1080727 | CCn1c(-c2ccccc2)cc(C(=O)NCCCN2CCN(c3cccc(Cl)c3)CC2)c1C | 25.8 |
| CHEMBL1214177 | Cc1c(C(=O)N(C)CCCN2CCN(c3cccc(Cl)c3Cl)CC2)cc(-c2ccccc2)n1C | 98 |
| CHEMBL3323183 | CN1Cc2cc(N3CCOCC3)ccc2C(c2ccc3ccsc3c2)C1 | 4.5 |
| CHEMBL103936 | COC(=O)c1ccccc1-c1ccc2ccccc2c1 | 10 |
| CHEMBL3217172 | Cc1c(Cl)cccc1N1CCN(CC(O)CNC(=O)c2nc(C)n(-c3cccc(Cl)c3)c2C)CC1.Cl.Cl.Cl.Cl | 14.2 |
| CHEMBL580143 | COc1cccc(C(=O)C(C)N2CCCCC2)c1 | 100 |
| CHEMBL3334794 | CN(C)CCC(c1cc2ccccc2s1)N1CCCC1 | 401 |
| CHEMBL11 | CN(C)CCCN1c2ccccc2CCc2ccccc21 | 6.8 |
| CHEMBL313664 | COC(=O)C1C2CCC(CC1c1ccc(I)cc1)O2 | 12 |
| CHEMBL1081468 | Cc1cccc(N2CCN(CCCNC(=O)c3cc(-c4ccccc4)[nH]c3C)CC2)c1C | 77 |
| CHEMBL1852372 | CCCc1nc(C(=O)NCCCN2CCN(c3cccc(Cl)c3Cl)CC2)c(C)n1-c1ccccc1 | 13 |
| CHEMBL3216290 | Cc1c(Cl)cccc1N1CCN(CC(O)CNC(=O)c2nc(C)n(-c3ccc(F)cc3)c2C)CC1.Cl.Cl.Cl.Cl | 20 |
| CHEMBL2337595 | Clc1cccc(OC(c2ccccc2)C2CCNC2)c1 | 2.512 |
| CHEMBL1213985 | Cc1ccc2cccc(N3CCN(CCCNC(=O)c4cc(-c5ccccc5)n(C)c4C)CC3)c2n1 | 300 |
| CHEMBL1852552 | Cc1c(Cl)cccc1N1CCN(CCCNC(=O)c2nc(C)n(-c3ccccc3F)c2C)CC1 | 6.17 |
| CHEMBL381045 | CN1C2C=C(c3ccc4c(c3)Cc3ccccc3-4)CC1CC2 | 14.5 |
| CHEMBL377189 | Clc1ccccc1CC(c1cncs1)N1CCNCC1 | 12 |
| CHEMBL3334793 | CN(C)CCC(c1cc2ccccc2s1)N1CCOCC1 | 449.2 |
| CHEMBL3216956 | CCCc1nc(C(=O)NCC(O)CN2CCN(c3cccc(C)c3C)CC2)c(C)n1-c1ccc(Cl)cc1.Cl.Cl.Cl | 347 |
| CHEMBL2219884 | Clc1ccc(C(Oc2ccccc2)C2CNC2)cc1Cl | 4.3 |
| CHEMBL1080555 | CCCn1c(-c2ccccc2)cc(C(=O)NCCCCN2CCN(c3cccc(C)c3C)CC2)c1C | 76 |
| CHEMBL629 | CN(C)CCC=C1c2ccccc2CCc2ccccc21 | 1.661 |
| CHEMBL550779 | Clc1cccc(OC(c2ccncc2)C2CCNCC2)c1Cl | 8 |
| CHEMBL505147 | CSc1ccc(Oc2ccc(S(N)(=O)=O)cc2CN(C)C)cc1 | 4 |
| CHEMBL4071830 | CNc1nc(C#Cc2ccc(Br)s2)nc2c1ncn2C1C(O)C(O)C2(C(=O)OC)CC12 | 10 |
| CHEMBL1214305 | Cc1[nH]c(-c2ccccc2)c(C)c1C(=O)NCCCN1CCN(c2cccc(Cl)c2Cl)CC1.Cl | 36 |
| CHEMBL1080745 | Cc1c(C(=O)NCCCN2CCN(c3cccc(Cl)c3)CC2)cc(-c2ccccc2)n1CC(C)C | 110 |
| CHEMBL260039 | CNCc1cnc(C)cc1Oc1ccc(F)cc1C | 231 |
| CHEMBL452928 | NCC1CCC(CCc2c(F)c(F)c(F)c(F)c2F)O1 | 225 |
| CHEMBL335423 | COC(=O)C1C(c2ccc(OC)c(OC)c2)CC2CCC1N2C | 23 |
| CHEMBL3681357 | Cc1cccc(N2CCN(CCCCNC(=O)c3nc(C)n(-c4ccccc4Cl)c3C)CC2)c1C | 30 |
| CHEMBL3217182 | CCc1nc(C(=O)NCCCN2CCN(c3cccc(Cl)c3Cl)CC2)c(C)n1-c1ccccc1.Cl.Cl.Cl.Cl | 28 |
| CHEMBL212763 | COc1cccc(CC(c2ccccc2)N2CCNCC2)c1 | 25 |
| CHEMBL4097442 | CCCCN(C)CCCC1(c2ccc(F)cc2)OCc2cc(C#N)ccc21 | 160 |
| CHEMBL4071311 | CN(C)CCCC1(c2cc(Cl)cc(Cl)c2)OCc2ccccc21 | 340 |
| CHEMBL485540 | CC1CN(c2ccc3ccccc3n2)CCN1CCCCN1C(=O)c2ccccc2C1=O | 8 |
| CHEMBL469222 | CC(C)CC(=O)N(Cc1ccc(Cl)cc1Cl)C1CCNC1 | 93 |
| CHEMBL3323094 | CN1Cc2ccccc2C(c2cccc3ccoc23)C1 | 31 |
| CHEMBL397392 | Cc1ccccc1CN(C1CCOCC1)C1CCNC1 | 29 |
| CHEMBL449015 | CSc1ccc(Sc2ccc(NS(C)(=O)=O)cc2CN(C)C)cc1 | 3 |
| CHEMBL1214547 | Cc1c(C(=O)NCCCN2CCN(c3cccc(Cl)c3Cl)CC2)cc(-c2ccncc2)n1C | 85 |
| CHEMBL258063 | CNCc1cc(C#N)ccc1Oc1ccc(Cl)cc1Cl | 28 |
| CHEMBL481526 | CCOc1ccc(Oc2ccccc2CN(C)C)cc1 | 30 |
| CHEMBL3215621 | CCCc1nc(C(=O)NCCCN2CCN(c3cccc(Cl)c3Cl)CC2)c(C)n1-c1ccc(OC)cc1.Cl.Cl.Cl.Cl | 84 |
| CHEMBL219281 | Clc1ccc(C(CCc2ccccc2)C2CCCCN2)cc1 | 210 |
| CHEMBL315172 | COC(=O)C1C(c2ccc3ccccc3c2)CC2C(O)CC1N2C | 34.2 |
| CHEMBL361836 | COc1ccc(F)c(CC2CCN(C3CCC4(CC3)OC(=O)c3c4ccc4c3OCCO4)CC2)c1 | 430 |
| CHEMBL210028 | Cc1ccccc1CC(c1ccccc1)N1CCNCC1 | 12 |
| CHEMBL489562 | CN(C)Cc1cc(NS(C)(=O)=O)ccc1Oc1ccc2c(c1)CCS2 | 7 |
| CHEMBL3681354 | CCc1c(C(=O)NCCCN2CCN(c3cccc(C)c3C)CC2)nc(-c2ccc(Cl)cc2Cl)n1-c1ccc(Br)cc1 | 298 |
| CHEMBL2012106 | COC(=O)C1C2CCC(CC1c1cccc(-c3ccco3)c1)O2 | 77 |
| CHEMBL3216085 | CCCc1nc(C(=O)NCCCN2CCN(c3cccc(C)c3C)CC2)c(C)n1-c1ccccc1.Cl.Cl | 5 |
| CHEMBL2338054 | Cc1cc(Cl)cc(-c2ccccn2)c1OC(c1ccccc1)C1CCNC1 | 6.31 |
| CHEMBL257482 | CN1C2CCC1C(COC(=O)CCCCC(=O)OCC1C(c3ccc(Cl)c(Cl)c3)CC3CCC1N3C)C(c1ccc(Cl)c(Cl)c1)C2 | 56 |
| CHEMBL378447 | CCOc1ccccc1CC(c1ccccc1F)N1CCNCC1 | 10 |
| CHEMBL210026 | Oc1cccc(CC(c2ccccc2)N2CCNCC2)c1 | 9.5 |
| CHEMBL3216508 | CCCc1nc(C(=O)NCCCN2CCN(c3cccc(Cl)c3C)CC2)c(C)n1-c1ccccc1.Cl.Cl.Cl | 3.9 |
| CHEMBL234927 | CN(C)C1CC=C(c2c[nH]c3ccc(F)cc23)CC1 | 14 |
| CHEMBL3215618 | CCCc1nc(C(=O)NCCCN2CCN(c3cccc(C)c3C)CC2)c(C)n1-c1ccc(F)cc1.Cl.Cl | 6.6 |
| CHEMBL3799072 | CN(C)CCCN1c2ccccc2CCc2ccc(C#CCOCCOc3ccc4ccccc4c3)cc21 | 440 |
| CHEMBL214003 | CCOc1ccccc1CC(c1ccccc1)N1CCCNCC1 | 300 |
| CHEMBL415 | CN(C)CCCN1c2ccccc2CCc2ccc(Cl)cc21 | 0.088 |
| CHEMBL213287 | Clc1ccccc1CC(c1ccccc1)N1CCNCC1 | 5.4 |
| CHEMBL1080671 | Cc1c(C(=O)NCCCN2CCN(c3cccc(Cl)c3Cl)CC2)cc(-c2ccccc2)n1Cc1ccccc1 | 125 |
| CHEMBL3216757 | Cc1c(Cl)cccc1N1CCN(CC(O)CNC(=O)c2nc(-c3ccccc3)n(-c3ccccc3)c2C)CC1.Cl.Cl.Cl | 370 |
| CHEMBL256817 | CN(C)Cc1ccccc1Oc1ccc(F)cc1Cl | 32 |
| CHEMBL1628227 | CN(C)CCC=C1c2ccccc2COc2ccccc21 | 42 |
| CHEMBL3215622 | COc1ccc(-n2c(-c3ccccc3)nc(C(=O)NCCCN3CCN(c4cccc(Cl)c4Cl)CC3)c2C)cc1.Cl.Cl.Cl.Cl | 76 |
| CHEMBL3217176 | Cc1cccc(N2CCN(CCCNC(=O)c3nc(C)n(-c4ccccc4)c3C)CC2)c1C.Cl.Cl | 6 |
| CHEMBL565845 | Cc1cc(C(=O)C(C)NC(C)(C)C)ccc1Br | 473 |
| CHEMBL1642912 | NCCC1Cc2ccccc2C(c2ccc(Cl)c(Cl)c2)C1 | 1 |
| CHEMBL517963 | CC(C)C(=O)N(Cc1ccccc1Cl)C1CCNC1 | 91 |
| CHEMBL214159 | CCOc1ccccc1CC(c1ccc(F)cc1)N1CCNCC1 | 17 |
| CHEMBL1173771 | CC1NC(C)(C)COC1(O)c1cccc(-c2ccc3ccccc3c2)c1 | 334 |
| CHEMBL3681361 | COc1cc(OC)cc(-n2c(C)nc(C(=O)NCCN3CCN(c4cccc(C)c4C)CC3)c2C)c1 | 164 |
| CHEMBL3799795 | CN(C)CCCN1c2ccccc2CCc2ccc(C#CCOCCOCCc3cccc4ccccc34)cc21 | 300 |
| CHEMBL1852665 | COc1ccccc1-n1c(C)nc(C(=O)NCC(O)CN2CCN(c3cccc(Cl)c3Cl)CC2)c1C | 12.2 |
| CHEMBL446481 | CSc1ccc(Oc2ccc(F)cc2CN(C)C)cc1 | 10 |
| CHEMBL1088313 | Cc1c(C(=O)NCCCN2CCN(c3cccc(Cl)c3Cl)CC2)cc(-c2ccccc2)n1N1CCCCC1 | 298 |
| CHEMBL3217183 | Cc1c(C(=O)NCCCN2CCN(c3cccc(Cl)c3Cl)CC2)nc(-c2ccccc2)n1-c1ccccc1.Cl.Cl.Cl.Cl | 106 |
| CHEMBL1642907 | CNCC1Cc2ccccc2C(c2ccc(Cl)c(Cl)c2)C1 | 54 |
| CHEMBL403251 | CCc1cc(Cl)ccc1Oc1cc(C)ncc1CNC | 20 |
| CHEMBL245880 | Clc1cccc(CN(C2CCOCC2)C2CCNC2)c1Cl | 10 |
| CHEMBL2430681 | O=C1NCC(c2ccc(Cl)cc2)C1c1ccc(Cl)cc1 | 99.2 |
| CHEMBL451500 | COc1ccc(F)c(CCCC2CCC(CCNCCCCCN3N=C(c4ccc(OC)c(OC)c4)C4CC=CCC4C3=O)O2)c1 | 194 |
| CHEMBL378319 | Clc1ccccc1CC(c1cccnc1)N1CCNCC1 | 14 |
| CHEMBL3216503 | Cc1c(Cl)cccc1N1CCN(CCCNC(=O)c2nc(C)n(-c3ccccc3F)c2C)CC1.Cl.Cl.Cl | 6.17 |
| CHEMBL2219829 | COC(c1ccc(Cl)c(Cl)c1)C1CNC1 | 271 |
| CHEMBL448689 | CSc1ccc(Oc2ccc(Cl)cc2CN(C)C)cc1 | 4 |
| CHEMBL4093087 | CCN(C)CCCC1(c2ccc(F)cc2)OCc2cc(C#N)ccc21 | 280 |
| CHEMBL1852779 | COc1cc(OC)cc(-n2c(C)nc(C(=O)NCCCN3CCN(c4cccc(C)c4C)CC3)c2C)c1 | 10 |
| CHEMBL185025 | COc1ccc(F)c(CC2CCN(C3CCC4(CC3)OC(=O)c3c4ccc4c3OCO4)CC2)c1 | 400 |
| CHEMBL4071867 | CNC(=O)C12CC1C(n1cnc3c(NC)nc(C#Cc4ccc(Cl)s4)cc31)C(O)C2O | 10 |
| CHEMBL492841 | CNCc1ccccc1Cc1ccc(SC)cc1 | 4 |
| CHEMBL1088164 | Cc1cccc(N2CCN(CCCNC(=O)c3cc(-c4ccccc4)n(CC(C)C)c3C)CC2)c1C | 350 |
| CHEMBL2380975 | CNCC1CCCC1c1ccc2[nH]cc(C#N)c2c1 | 4.3 |
| CHEMBL469946 | C=C(F)CN(CC)C(=O)C1(c2cccs2)CC1CN | 97 |
| CHEMBL1213933 | Cc1c(C(=O)NCCCN2CCN(c3ccc(F)cc3)CC2)cc(-c2ccccc2)n1C | 183 |
| CHEMBL3216974 | Cc1cccc(N2CCN(CCCNC(=O)c3nc(C)n(-c4cccc(Cl)c4)c3C)CC2)c1C.Cl.Cl.Cl | 4.88 |
| CHEMBL2337601 | O=[N+]([O-])c1ccccc1OC(c1ccccc1)C1CCNC1 | 63.1 |
| CHEMBL1852341 | Cc1nc(C(=O)NCCCN2CCN(c3cccc(Cl)c3Cl)CC2)c(C)n1-c1ccccc1 | 10.2 |
| CHEMBL432727 | CN1C2CCC1C(C(=O)NCCCCCCNC(=O)C1C(c3ccc(Cl)cc3)CC3CCC1N3C)C(c1ccc(Cl)cc1)C2 | 225 |
| CHEMBL258060 | CNCc1cc(NS(C)(=O)=O)ccc1Oc1ccc(Cl)cc1C | 16 |
| CHEMBL2012116 | COC(=O)C1C2CCC(CC1c1ccc(-c3ccccc3)s1)O2 | 78 |
| CHEMBL379536 | NC(=O)c1cccc(CC(c2ccccc2)N2CCNCC2)c1 | 210 |
| CHEMBL43048 | CNC(C)Cc1ccc2c(c1)OCO2 | 425 |
| CHEMBL1214427 | Cc1[nH]c(-c2ccc(Cl)cc2)cc1C(=O)NCCCN1CCN(c2cccc(Cl)c2Cl)CC1.Cl | 56 |
| CHEMBL402371 | CNCc1cnc(C)cc1Oc1ccc(Cl)cc1Br | 10 |
| CHEMBL3216282 | Cc1cccc(N2CCN(CC(O)CNC(=O)c3nc(C)n(-c4cccc(Cl)c4)c3C)CC2)c1C.Cl.Cl.Cl | 2.9 |
| CHEMBL1175 | CNCCC(Oc1cccc2ccccc12)c1cccs1 | 6 |
| CHEMBL3681359 | CCc1nc(C(=O)NCC(O)CN2CCN(c3cccc(C)c3C)CC2)c(C)n1-c1ccccc1 | 9 |
| CHEMBL522584 | CNCc1cc(S(=O)(=O)N(C)CCO)ccc1Oc1ccc(SC)cc1 | 5 |
| CHEMBL491400 | CSc1ccc(Oc2ccc(NS(C)(=O)=O)cc2CN(C)C)cc1C | 60 |
| CHEMBL399740 | CN1C2CCC1C(COC(=O)CCCCCCC(=O)OCC1C(c3ccc(Cl)c(Cl)c3)CC3CCC1N3C)C(c1ccc(Cl)c(Cl)c1)C2 | 31 |
| CHEMBL307314 | CN1C2CCC1C(C(=O)NCCCCCCCCNC(=O)C1C(c3ccc(Cl)cc3)CC3CCC1N3C)C(c1ccc(Cl)cc1)C2 | 99 |
| CHEMBL481527 | CN(C)Cc1cc(Br)ccc1Oc1ccc(C(F)(F)F)cc1 | 3 |
| CHEMBL402000 | CCc1cc(Cl)ccc1Oc1ccncc1CN(C)C | 11 |
| CHEMBL1644605 | N#Cc1cc(OC2CC3CCC(C2)N3)nc(-c2ccccc2)c1 | 208 |
| CHEMBL1214302 | Cc1[nH]c(-c2ccccc2)c(C)c1C(=O)NCCN1CCN(c2cccc(Cl)c2Cl)CC1.Cl | 162 |
| CHEMBL562444 | CCc1ccc(OC(c2cccnc2)C2CCNCC2)cc1 | 5 |
| CHEMBL2380973 | N#Cc1c[nH]c2ccc(C3CCCC3CN3CCCC3)cc12 | 300 |
| CHEMBL196468 | CN(C)CC1CC1c1c[nH]c2ccc(C#N)cc12 | 0.56 |
| CHEMBL3936578 | O=C(OC(c1ccc(Cl)cc1)C1CNC1)c1ccccc1 | 357 |
| CHEMBL1086626 | Cc1[nH]c(-c2ccccc2)cc1C(=O)NCCCN1CCN(c2cccc(Cl)c2Cl)CC1 | 137 |
| CHEMBL1080215 | Cc1[nH]c(-c2ccccc2)cc1C(=O)NCCCN1CCN(c2cccc(Cl)c2)CC1 | 92 |
| CHEMBL1214596 | Cc1cccc(N2CCN(CCCNC(=O)c3cc(C(C)(C)C)n(C)c3C)CC2)c1C | 252 |
| CHEMBL1080554 | Cc1cccc(N2CCN(CCCCNC(=O)c3cc(-c4ccccc4)n(C)c3C)CC2)c1C | 149 |
| CHEMBL1214304 | CCCn1c(C)c(C(=O)NCCN2CCN(c3cccc(Cl)c3Cl)CC2)c(C)c1-c1ccccc1.Cl | 78 |
| CHEMBL362363 | O=C1OC2(CCC(N3CCN(Cc4ccccc4I)CC3)CC2)c2ccc3c(c21)OCCO3 | 54 |
| CHEMBL209969 | NC(=O)c1ccccc1CC(c1ccccc1)N1CCNCC1 | 400 |
| CHEMBL363320 | O=C1OC2(CCC(N3CCC(Cc4ccccc4Br)CC3)CC2)c2ccc3c(c21)OCCO3 | 73 |
| CHEMBL1212957 | Cc1c(C(=O)NCCCN2CCN(c3cccc(Cl)c3Cl)CC2)cc(-c2cccnc2)n1C | 7 |
| CHEMBL3908637 | O=C(OC(c1ccc(Cl)c(Cl)c1)C1CNC1)c1ccccc1 | 69.9 |
| CHEMBL425961 | c1ccc(C(Cc2ccc3ccccc3c2)N2CCNCC2)cc1 | 17 |
| CHEMBL1644602 | Clc1cc(OC2CC3CCC(C2)N3)cc(-c2ccccc2)c1 | 452 |
| CHEMBL3217167 | Cc1nc(C(=O)NCC(O)CN2CCN(c3cccc(Cl)c3Cl)CC2)c(C)n1-c1ccc2c(c1)OCCO2.Cl.Cl.Cl.Cl | 12 |
| CHEMBL2338047 | Fc1cc(Cl)cc(Cl)c1OC(c1ccccc1)C1CCNC1 | 0.5012 |
| CHEMBL1271929 | CSc1nc(C)cc(C(=O)NCCCN2CCN(c3cccc(C)c3C)CC2)n1.Cl | 394 |
| CHEMBL371726 | CN1C2C=C(c3c[nH]c4ccccc34)CC1CC2 | 72.6 |
| CHEMBL454507 | COc1ccc(C2=NN(CCCCCNCCC(Oc3ccc(C(F)(F)F)cc3)c3ccccc3)C(=O)C3CC=CCC23)cc1OC | 42 |
| CHEMBL3334776 | Clc1ccc(C(CCN2CCOCC2)N2CCCC2)cc1Cl | 389.4 |
| CHEMBL507422 | CN(C)Cc1ccccc1Oc1ccc(C2CC2)cc1 | 17 |
| CHEMBL3216961 | Cc1c(Cl)cccc1N1CCN(CCCNC(=O)c2nc(C)n(C3CCCC3)c2C)CC1.Cl.Cl.Cl | 9.53 |
| CHEMBL1081644 | Cc1cccc(N2CCN(CCCNC(=O)c3cc(-c4ccccc4)n(C)c3C)CC2)c1C | 11.8 |
| CHEMBL566830 | CC(C(=O)c1cccc(Cl)c1)N(C)C | 100 |
| CHEMBL3215851 | CCCc1nc(C(=O)NCC(O)CN2CCN(c3cccc(C)c3C)CC2)c(C)n1-c1ccc2c(c1)OCCO2.Cl.Cl | 9.3 |
| CHEMBL1271698 | Cc1cc(C(=O)NCCCN2CCN(c3cccc(Cl)c3Cl)CC2)nc(C)n1.Cl | 289 |
| CHEMBL513248 | C#CCN(CC#C)C(=O)C1(c2cccs2)CC1CN | 26 |
| CHEMBL100941 | COC(=O)C1C2CCC(C2)CC1c1ccc(Cl)c(Cl)c1 | 33.4 |
| CHEMBL257423 | CN(C)Cc1ccccc1Oc1ccc(Cl)cc1 | 18 |
| CHEMBL86590 | CCOC(=O)C1C(c2ccc(Cl)c(Cl)c2)CC2CC(O)C1N2C | 97 |
| CHEMBL378418 | Brc1ccccc1CC(c1ccccc1)N1CCNCC1 | 15 |
| CHEMBL3681360 | CCCc1nc(C(=O)NCC(O)CN2CCN(c3cccc(C)c3C)CC2)cn1-c1ccc(Cl)cc1 | 347 |
| CHEMBL456639 | NCC1CCC(CCc2ccccc2)O1 | 129 |
| CHEMBL3799668 | CN(C)CCCN1c2ccccc2CCc2ccc(CCCOCCOc3cccc4ccccc34)cc21 | 280 |
| CHEMBL493667 | CC(c1ccc(Cl)c(Cl)c1)c1ccccc1CN(C)C | 40 |
| CHEMBL1852387 | Cc1cccc(N2CCN(CC(O)CNC(=O)c3nc(-c4ccccc4)n(-c4ccccc4)c3C)CC2)c1C | 19.1 |
| CHEMBL2219839 | Clc1ccc(C(OCc2ccccc2)C2CNC2)cc1 | 21.6 |
| CHEMBL589494 | Cc1cccc(C(=O)C(C)N2CCCCC2)c1 | 100 |
| CHEMBL493264 | CSc1ccc(Sc2ccccc2CN(C)C)cc1 | 22 |
| CHEMBL3217184 | COc1ccccc1-n1c(C)nc(C(=O)NCCCN2CCN(c3cccc(Cl)c3Cl)CC2)c1C.Cl.Cl.Cl.Cl | 15.8 |
| CHEMBL563852 | FC(F)(F)Oc1ccccc1OC(c1cccnc1)C1CCNCC1 | 95 |
| CHEMBL1812747 | Cc1cc(C2C(c3ccc(Cl)c(C)c3)CC3CCC2N3C)on1 | 115 |
| CHEMBL438889 | CNCc1cc(N2CCCS2(=O)=O)ccc1Oc1ccc(Cl)cc1Cl | 12 |
| CHEMBL3216081 | COc1ccccc1-n1c(C)nc(C(=O)NCCCN2CCN(c3cccc(C)c3C)CC2)c1C.Cl.Cl | 9 |
| CHEMBL491394 | CN(C)Cc1cnccc1Oc1ccc2c(c1)CCS2 | 3 |
| CHEMBL3216738 | Cc1nc(C(=O)NCCCN2CCN(c3cccc(Cl)c3Cl)CC2)c(C)n1-c1ccccc1C(F)(F)F.Cl.Cl.Cl.Cl | 42 |
| CHEMBL3323097 | CN1Cc2ccccc2C(c2cc3ccccc3s2)C1 | 35 |
| CHEMBL55097 | COC(=O)C1C(c2ccc(F)cc2)CC2CCC1N2CCCCCc1ccccc1 | 49.9 |
| CHEMBL512472 | O=C(C1CC1)N(Cc1ccc(Cl)cc1Cl)C1CCNC1 | 38 |
| CHEMBL430840 | O=C1CN2C3CCC2C1C(c1ccc(F)cc1)C3 | 251 |
| CHEMBL3681370 | Cc1ccc(-n2c(C)nc(C(=O)NCC(O)CN3CCN(c4cccc(C)c4C)CC3)c2C)cc1 | 3.89 |
| CHEMBL3216957 | Cc1c(Cl)cccc1N1CCN(CC(O)CNC(=O)c2nc(C)n(-c3ccc4c(c3)OCCO4)c2C)CC1.Cl.Cl.Cl | 9.1 |
| CHEMBL1852374 | Cc1cccc(N2CCN(CC(O)CNC(=O)c3nc(C)n(-c4ccccc4)c3C)CC2)c1C | 5.08 |
| CHEMBL245492 | Fc1cccc(CN(C2CCOCC2)C2CCNC2)c1C(F)(F)F | 24 |
| CHEMBL384687 | c1ccc(CCC(Cc2ccccc2)N2CCNCC2)cc1 | 23 |
| CHEMBL456223 | COc1ccc(F)cc1CCC1CCC(CN)O1 | 11.7 |
| CHEMBL214004 | c1ccc(C(Cc2ccncc2)N2CCNCC2)cc1 | 38 |
| CHEMBL3216517 | CCCc1nc(C(=O)NCCCN2CCN(c3cccc(C)c3C)CC2)cn1-c1ccc(Cl)cc1.Cl.Cl.Cl | 67.2 |
| CHEMBL67024 | Fc1ccc(CCNC2CCN(CCOC(c3ccccc3)c3ccccc3)CC2)cc1 | 470 |
| CHEMBL260038 | CNCc1cnc(C)cc1Oc1ccc(C)cc1OC | 18 |
| CHEMBL1852358 | Cc1c(Cl)cccc1N1CCN(CC(O)CNC(=O)c2nc(-c3ccccc3)n(-c3ccc(F)cc3)c2C)CC1 | 232 |
| CHEMBL234705 | CN(C)C1CC=C(c2c[nH]c3ccc(C#N)cc23)CC1 | 1.1 |
| CHEMBL4105127 | CCCOC(=O)C12CC1C(n1cnc3c(NC)nc(C#Cc4ccc(Cl)s4)nc31)C(O)C2O | 10 |
| CHEMBL1852507 | Cc1cccc(N2CCN(CC(O)CNC(=O)c3nc(C)n(-c4ccccc4F)c3C)CC2)c1C | 0.75 |
| CHEMBL1852359 | Cc1cccc(N2CCN(CCCNC(=O)c3nc(C)n(-c4ccccc4Cl)c3C)CC2)c1C | 17 |
| CHEMBL246519 | Fc1ccc(CN(C2CCOCC2)C2CCNC2)c(F)c1F | 15 |
| CHEMBL3798279 | CN(C)CCCN1c2ccccc2CCc2ccc(C#CCOCCOCCOc3cccc4ccccc34)cc21 | 420 |
| CHEMBL1644606 | N#Cc1ccc(-c2cccc(OC3CC4CCC(C3)N4)c2)cc1 | 105 |
| CHEMBL493058 | CN(C)Cc1ccccc1Sc1ccc(Cl)cc1 | 20 |
| CHEMBL3216288 | Cc1c(Cl)cccc1N1CCN(CC(O)CNC(=O)c2nc(C)n(-c3ccccc3F)c2C)CC1.Cl.Cl.Cl | 36 |
| CHEMBL520760 | CSc1ccc(Sc2ccc(S(N)(=O)=O)cc2CN(C)C)cc1 | 20 |
| CHEMBL211372 | CN1C2CCC1C1COC(=O)CCCCCCCCC(=O)Nc3ccc(cc3)C1C2 | 7.8 |
| CHEMBL1852418 | Cc1c(Cl)cccc1N1CCN(CCCNC(=O)c2nc(C)n(CC(C)C)c2C)CC1 | 447 |
| CHEMBL214237 | c1ccc(CC(c2ccccc2)N2CCNCC2)cc1 | 14 |
| CHEMBL103504 | CCCC1(O)CN2C3CCC2C1C(c1ccc(C)cc1)C3 | 41 |
| CHEMBL257592 | CCc1cc(Cl)ccc1Oc1ccncc1CNC | 46 |
| CHEMBL401819 | CNCc1cc(C(N)=O)ccc1Oc1ccc(Cl)cc1C | 7 |
| CHEMBL1088312 | Cc1c(C(=O)NCCCN2CCN(c3cccc(Cl)c3Cl)CC2)cc(-c2ccccc2)n1CC(C)C | 381 |
| CHEMBL3323175 | CN1Cc2cc(O)ccc2C(c2ccc3sccc3c2)C1 | 10 |
| CHEMBL1812746 | COC(=O)C1C2CCC(CC1c1ccc(Cl)c(C)c1)N2 | 6.2 |
| CHEMBL258271 | CNCc1cc(C(N)=O)ccc1Oc1ccc(Cl)cc1Cl | 37 |
| CHEMBL556888 | Clc1ccc(OC(c2ccncc2)C2CCNCC2)c(Cl)c1 | 7 |
| CHEMBL534942 | Cl.Clc1ccccc1CC(c1ccccc1)N1CCNCC1 | 13 |
| CHEMBL3216093 | Cc1c(Cl)cccc1N1CCN(CCCNC(=O)c2nc(C)n(-c3ccccc3Cl)c2C)CC1.Cl.Cl.Cl.Cl | 18 |
| CHEMBL3215852 | COc1ccccc1-n1c(C)nc(C(=O)NCC(O)CN2CCN(c3cccc(Cl)c3Cl)CC2)c1C.Cl.Cl.Cl.Cl | 12.2 |
| CHEMBL72 | CNCCCN1c2ccccc2CCc2ccccc21 | 64 |
| CHEMBL3216521 | Cc1cccc(N2CCN(CC(O)CNC(=O)c3nc(-c4ccccc4)n(-c4ccccc4)c3C)CC2)c1C.Cl.Cl | 19.1 |
| CHEMBL84680 | COC(=O)C1=C(c2ccc3ccccc3c2)CC2CCC1N2C | 109 |
| CHEMBL1214176 | CCCn1c(-c2ccccc2)cc(C(=O)N(C)CCCN2CCN(c3cccc(C)c3C)CC2)c1C | 163 |
| CHEMBL3334797 | CN(C)CCC(c1ccc(Cl)c(Cl)c1)N1CCOCC1 | 378.5 |
| CHEMBL363768 | COc1ccccc1CC1CCN(C2CCC3(CC2)OC(=O)c2c3ccc3c2OCO3)CC1 | 150 |
| CHEMBL102144 | COC(=O)c1ccccc1-c1ccc(-c2ccc(Br)cc2)cc1 | 10 |
| CHEMBL453493 | CN(C)Cc1ccccc1C(=O)c1ccc(Cl)c(Cl)c1 | 50 |
| CHEMBL1081036 | Cc1c(C(=O)NCCCN2CCN(c3cccc(Cl)c3Cl)CC2)cc(-c2ccccc2)n1C | 93 |
| CHEMBL213785 | Fc1ccc(C(Cc2ccccc2Cl)N2CCNCC2)cc1 | 13 |
| CHEMBL363328 | O=C1OC2(CCC(N3CCC(Cc4cc(F)ccc4F)CC3)CC2)c2ccc3c(c21)OCO3 | 110 |
| CHEMBL900 | Cc1ccccc1C(OCCN(C)C)c1ccccc1 | 458 |
| CHEMBL2338036 | COc1cc(Cl)ccc1OC(c1ccccc1)C1CCNC1 | 1.259 |
| CHEMBL1081303 | Cc1c(C(=O)NCCCN2CCN(c3cccc(Cl)c3)CC2)cc(-c2ccccc2)n1CC1CCCCC1 | 213 |
| CHEMBL247538 | COc1cccc(CN(C2CCOCC2)C2CCNC2)c1C | 11 |
| CHEMBL434728 | COc1ccccc1CC1CCN(C2CCC3(CC2)OC(=O)c2c3ccc3c2OCCO3)CC1 | 59 |
| CHEMBL455221 | COc1ccc(F)cc1CCCCC1CCC(CN)O1 | 84 |
| CHEMBL3215629 | COc1ccc(-n2cc(C(=O)NCCCN3CCN(c4cccc(C)c4C)CC3)nc2-c2ccccc2)cc1.Cl.Cl | 42.6 |
| CHEMBL101905 | C=C1CN2C3CCC2C1C(c1ccc(C)cc1)C3 | 33 |
| CHEMBL2337608 | Clc1cc(Cl)cc(OC(c2ccccc2)C2CCNC2)c1 | 0.7943 |
| CHEMBL566207 | Cc1ccc(C(=O)C(C)NC(C)(C)C)cc1 | 100 |
| CHEMBL723 | COc1ccccc1OCCNCC(O)COc1cccc2[nH]c3ccccc3c12 | 221 |
| CHEMBL3217175 | COc1ccccc1-n1c(C)nc(C(=O)NCC(O)CN2CCN(c3cccc(Cl)c3C)CC2)c1C.Cl.Cl.Cl | 19 |
| CHEMBL1852774 | CCCc1nc(C(=O)NCC(O)CN2CCN(c3cccc(Cl)c3Cl)CC2)c(C)n1-c1ccc(OC)cc1 | 51.8 |
| CHEMBL402345 | CNCc1cc(C(=O)N(C)C)ccc1Oc1ccc(Cl)cc1OC | 7 |
| CHEMBL323447 | CN(C)CC1C2CCC(C2)C1c1ccc2ccccc12 | 93 |
| CHEMBL2338053 | Fc1cc(F)c(F)c(OC(c2ccccc2)C2CCNC2)c1F | 5.012 |
| CHEMBL3217173 | Cc1cccc(N2CCN(CCCNC(=O)c3nc(C)n(-c4ccc(F)cc4)c3C)CC2)c1C.Cl.Cl | 21.6 |
| CHEMBL3216098 | Cc1nc(C(=O)NCCCN2CCN(c3cccc(Cl)c3Cl)CC2)c(C)n1-c1ccccc1.Cl.Cl.Cl.Cl | 10.2 |
| CHEMBL452376 | CN(C)Cc1ccccc1Oc1ccc(Br)cc1 | 12 |
| CHEMBL3215620 | COc1ccc(-n2c(C)nc(C(=O)NCCCN3CCN(c4cccc(Cl)c4Cl)CC3)c2C)cc1.Cl.Cl.Cl.Cl | 33 |
| CHEMBL403071 | CNCc1ccccc1Oc1ccc(Cl)cc1C | 18 |
| CHEMBL556268 | Cc1ccc(C(Oc2cccc(Cl)c2Cl)C2CCNCC2)nc1 | 16 |
| CHEMBL2337598 | Fc1ccccc1OC(c1ccccc1)C1CCNC1 | 19.95 |
| CHEMBL403660 | CN1C2CCC1C(COC(=O)CCCCCCCCCCC(=O)OCC1C(c3ccc(Cl)c(Cl)c3)CC3CCC1N3C)C(c1ccc(Cl)c(Cl)c1)C2 | 140 |
| CHEMBL407815 | CNCc1ccccc1Oc1ccc(Cl)cc1F | 73 |
| CHEMBL2012108 | COC(=O)C1C2CCC(CC1c1cccc(-c3cccn3C)c1)O2 | 142 |
| CHEMBL2219900 | CCOC(c1ccc(Cl)c(Cl)c1)C1CNC1 | 97.5 |
| CHEMBL443797 | CN(CCCc1ccccc1)C1CCN(CCOC(c2ccccc2)c2ccccc2)CC1 | 310 |
| CHEMBL3892951 | Cc1ccccc1OC(c1ccc(Cl)c(Cl)c1)C1CNC1 | 11.6 |
| CHEMBL560050 | Cc1cccnc1C(Oc1cccc(Cl)c1Cl)C1CCNCC1 | 11 |
| CHEMBL2064645 | CNC(=O)C12CC1C(n1cnc3c(NC)nc(C#Cc4ccc(F)c(F)c4)nc31)C(O)C2O | 10 |
| CHEMBL211527 | Clc1ccc(CC(c2ccccc2)N2CCNCC2)c(Cl)c1 | 11 |
| CHEMBL1201066 | COc1ccc(C(CN(C)C)C2(O)CCCCC2)cc1.Cl | 204 |
| CHEMBL1684047 | CC(N(C)C)C1(c2ccc(Cl)c(Cl)c2)CCCCC1 | 12 |
| CHEMBL549761 | CCOc1ccccc1OC(c1cccnc1)C1CCNCC1 | 80 |
| CHEMBL283185 | CN1CC=C(c2c[nH]c3ccc(Cl)cc23)CC1 | 240 |
| CHEMBL3216742 | Cc1c(Cl)cccc1N1CCN(CCCNC(=O)c2nc(C)n(-c3ccc(F)cc3)c2C)CC1.Cl.Cl.Cl | 39 |
| CHEMBL3612835 | COc1ccc2cc(C(=O)NCCN3CCC(Cc4ccccc4)CC3)ccc2c1 | 323 |
| CHEMBL3215862 | Cc1cccc(N2CCN(CCCNC(=O)c3nc(C)n(-c4ccc(Cl)cc4)c3C)CC2)c1C.Cl.Cl.Cl | 47 |
| CHEMBL1080711 | CCCn1c(-c2ccccc2)cc(C(=O)NCCCCN2CCN(c3cccc(Cl)c3Cl)CC2)c1C | 76 |
| CHEMBL3798035 | C#Cc1ccc2c(c1)N(CCCN(C)C)c1ccccc1CC2 | 40 |
| CHEMBL182288 | Oc1nc2cc3c(cc2c2c1C(N1CCN(Cc4ccccc4I)CC1)CCC2)OCO3 | 23 |
| CHEMBL490383 | CNCc1cc(NS(C)(=O)=O)ccc1Oc1ccc(SC)c(F)c1 | 16 |
| CHEMBL1087896 | Cc1cccc(N2CCN(CCCNC(=O)c3cc(-c4ccccc4)n(CC4CCCCC4)c3C)CC2)c1C | 352 |
| CHEMBL403078 | CNCc1cnc(C)cc1Oc1ccc(Cl)cc1OC | 4 |
| CHEMBL3323090 | CN1Cc2ccccc2C(c2ccc3ccoc3c2)C1 | 250 |
| CHEMBL395931 | Clc1cccc(CN(CC2CC2)C2CCNC2)c1Cl | 4 |
| CHEMBL551443 | CCc1cccc(OC(c2cccnc2)C2CCNCC2)c1 | 7 |
| CHEMBL1644481 | FC(F)(F)c1cccc(OC2CC3CCC(C2)N3)c1 | 99 |
| CHEMBL379181 | CN1CCCCCCCCCCCC(=O)OCC2C(CC3CCC2N3C)c2ccc1cc2 | 450 |
| CHEMBL1214594 | CCCn1c(-c2ccncc2)cc(C(=O)NCCCN2CCN(c3cccc(Cl)c3Cl)CC2)c1C | 78 |
| CHEMBL173594 | CC=Cc1ccc(C2CC3CCC(C2C(=O)OC)N3C)cc1 | 1.3 |
| CHEMBL596175 | COc1ccc(C2=NN(CCCCCNCCC(Oc3cccc4ccccc34)c3cccs3)C(=O)C3CC=CCC23)cc1OC | 442 |
| CHEMBL492815 | CN(C)Cc1ccccc1Cc1ccc(OC(F)(F)F)cc1 | 45 |
| CHEMBL490382 | CSc1ccc(Oc2ccc(NS(C)(=O)=O)cc2CN(C)C)cc1F | 10 |
| CHEMBL3216284 | CCCc1nc(C(=O)NCCCN2CCN(c3cccc(Cl)c3Cl)CC2)c(C)n1-c1ccccc1OC.Cl.Cl.Cl.Cl | 84 |
| CHEMBL386224 | CCOc1ccccc1CC(c1ccccc1)N1CCNCC1 | 13 |
| CHEMBL3216959 | Cc1c(Cl)cccc1N1CCN(CCCNC(=O)c2nc(-c3ccccc3)n(-c3ccc(F)cc3)c2C)CC1.Cl.Cl.Cl | 281.9 |
| CHEMBL124670 | COC(=O)C1C(OC(c2ccc(F)cc2)c2ccc(F)cc2)CC2CCC1N2CCCc1ccc(Br)cc1 | 234 |
| CHEMBL479999 | CN(C)Cc1cc(C(N)=O)ccc1Cc1ccc(Cl)c(Cl)c1 | 3 |
| CHEMBL4071385 | c1ccc(CC2CCN(CCCn3nnnc3-c3cccc4ccccc34)CC2)cc1 | 310 |
| CHEMBL391143 | Cc1c(CN(C2CCOCC2)C2CCNC2)ccc1C(F)(F)F | 13 |
| CHEMBL512851 | C=CCN(CC=C)C(=O)C1(c2cccs2)CC1CN | 32 |
| CHEMBL2012107 | COC(=O)C1C2CCC(CC1c1cccc(-c3cccs3)c1)O2 | 287 |
| CHEMBL3323099 | CN1Cc2ccccc2C(c2cccc3sccc23)C1 | 130 |
| CHEMBL3216975 | CCCc1nc(C(=O)NCCCN2CCN(c3cccc(Cl)c3C)CC2)c(C)n1-c1ccc2c(c1)OCCO2.Cl.Cl.Cl | 14 |
| CHEMBL245687 | c1ccc(-c2ccccc2CN(C2CCOCC2)C2CCNC2)cc1 | 400 |
| CHEMBL1081035 | CCCn1c(-c2ccccc2)cc(C(=O)NCCN2CCN(c3cccc(Cl)c3Cl)CC2)c1C | 32 |
| CHEMBL256816 | CNCc1ccccc1Oc1ccc(Cl)cc1Cl | 56 |
| CHEMBL4083724 | c1ccc(CCCN2CCN(CCCCn3nnnc3C(c3ccccc3)c3ccccc3)CC2)cc1 | 158.7 |
| CHEMBL150932 | c1ccc(C(Oc2cccc3ccccc23)C2CCNCC2)cc1 | 15 |
| CHEMBL491393 | CSc1ccc(Oc2ccncc2CN(C)C)cc1Cl | 5 |
| CHEMBL30008 | Fc1ccc(C(c2ccc(F)cc2)N2CCN(CC=Cc3ccccc3)CC2)cc1 | 499 |
| CHEMBL481941 | CN(C)Cc1ccccc1Oc1ccc(C#N)cc1 | 60 |
| CHEMBL518580 | CC(C)C(=O)N(Cc1ccc(F)cc1Cl)C1CCNC1 | 161 |
| CHEMBL479409 | CN(C)Cc1cc(S(N)(=O)=O)ccc1Oc1ccc(OC(F)(F)F)cc1 | 9 |
| CHEMBL1852587 | Cc1cccc(N2CCN(CCCNC(=O)c3nc(C)n(CC(C)C)c3C)CC2)c1C | 436 |
| CHEMBL3217141 | Cc1[nH]c(-c2ccccn2)cc1C(=O)NCCCN1CCN(c2cccc(Cl)c2Cl)CC1.Cl.Cl.Cl.Cl | 28 |
| CHEMBL401878 | CNCc1cc(C#N)ccc1Oc1ccc(Cl)cc1OC | 5 |
| CHEMBL184713 | O=C1OC2(CCC(N3CCC(Cc4cc(F)ccc4F)CC3)CC2)c2ccc3c(c21)OCO3 | 250 |
| CHEMBL3216501 | Cc1cccc(N2CCN(CCCNC(=O)c3cn(-c4ccccc4)c(-c4ccccc4)n3)CC2)c1C.Cl.Cl | 50.5 |
| CHEMBL3216745 | Cc1cccc(N2CCN(CCCNC(=O)c3nc(C(C)C)n(-c4ccccc4)c3C)CC2)c1C.Cl.Cl | 16 |
| CHEMBL1644473 | Clc1ccc(OC2CC3CCC(C2)N3)cc1 | 43 |
| CHEMBL1172 | Clc1ccc2c(c1)CCc1cccnc1C2=C1CCNCC1 | 230 |
| CHEMBL1086754 | Cc1c(C(=O)NCCCN2CCN(c3cccc(Cl)c3Cl)CC2)cc(-c2ccccc2)n1C | 61.7 |
| CHEMBL255957 | CNCc1ccccc1Oc1ccc(OC)cc1Cl | 22 |
| CHEMBL513139 | CC(C)C(=O)N(Cc1ccc(Cl)cc1Cl)C1CCNC1 | 13 |
| CHEMBL569465 | CCCC(NC(C)(C)C)C(=O)c1cccc(Cl)c1 | 100 |
| CHEMBL214395 | CCOc1ccccc1CC(c1cccc(F)c1)N1CCNCC1 | 28 |
| CHEMBL185540 | O=C1OC2(CCC(N3CCC(Cc4ccccc4Br)CC3)CC2)c2ccc3c(c21)OCO3 | 45 |
| CHEMBL257481 | CNCc1cc(C(=O)NC)ccc1Oc1ccc(Cl)cc1OC | 13 |
| CHEMBL2012101 | COC(=O)C1C2CCC(CC1c1ccc(-c3cccs3)cc1)O2 | 35 |
| CHEMBL562108 | Clc1ccc(OC(c2ccccn2)C2CCNCC2)c(Cl)c1 | 7 |
| CHEMBL3216740 | CCCc1nc(C(=O)NCCCN2CCN(c3cccc(Cl)c3C)CC2)c(C)n1-c1ccc(OC)cc1.Cl.Cl.Cl | 10.2 |
| CHEMBL132683 | COC(=O)C1C(c2ccc(O)c(O)c2)CC2CCC1N2C | 308 |
| CHEMBL256136 | CN(C)Cc1cnccc1Oc1ccc(Cl)cc1Br | 4 |
| CHEMBL534723 | CCOc1ccccc1CC(c1ccccc1)N1CCNCC1.Cl | 13 |
| CHEMBL3681358 | Cc1cccc(N2CCN(CC(O)CNC(=O)c3nc(C(C)C)n(-c4ccccc4)c3C)CC2)c1C | 15 |
| CHEMBL2219828 | CCOC(c1ccc(Cl)cc1)C1CNC1 | 120 |
| CHEMBL402851 | CN(C)Cc1ccccc1Oc1ccccc1Cl | 282 |
| CHEMBL3216083 | CCCc1nc(C(=O)NCCCN2CCN(c3cccc(Cl)c3Cl)CC2)c(C)n1-c1ccc2c(c1)OCCO2.Cl.Cl.Cl.Cl | 49 |
| CHEMBL316342 | COC(=O)C1C2CCC(CC1c1ccc(Cl)c(Cl)c1)O2 | 64.5 |
| CHEMBL3917777 | COC(c1ccccc1C)C1CNC1 | 112 |
| CHEMBL229381 | Cc1ccc(C2CC3CCC(C2c2ncc(-c4ccc(Cl)cc4)s2)N3C)cc1 | 377 |
| CHEMBL539470 | CN(C)CC1CCCC1c1c[nH]c2ccc(C#N)cc12 | 1.1 |
| CHEMBL522625 | CSc1ccc(Oc2ccc(S(N)(=O)=O)cc2CN(C)C)cc1C | 5 |
| CHEMBL3216499 | Cc1nc(C(=O)NCCCN2CCN(c3cccc(Cl)c3Cl)CC2)c(C)n1-c1ccccc1Cl.Cl.Cl.Cl.Cl.Cl | 82 |
| CHEMBL3681366 | Cc1cccc(N2CCN(CCCNC(=O)c3nc(C)n(-c4ccccn4)c3C)CC2)c1C | 12 |
| CHEMBL3215627 | COc1ccc(-n2c(C)nc(C(=O)NCC(O)CN3CCN(c4cccc(C)c4C)CC3)c2C)cc1OC.Cl.Cl | 70.3 |
| CHEMBL403090 | CNCc1cc(N2CCCS2(=O)=O)ccc1Oc1ccc(Cl)cc1C | 10 |
| CHEMBL247341 | Cc1c(Cl)cccc1CN(C1CCOCC1)C1CCNC1 | 11 |
| CHEMBL1086755 | CCn1c(-c2ccccc2)cc(C(=O)NCCCN2CCN(c3cccc(Cl)c3Cl)CC2)c1C | 182 |
| CHEMBL450769 | COc1ccc(F)cc1CC1CCC(CCN)O1 | 23 |
| CHEMBL491595 | CNCc1cc(NS(C)(=O)=O)ccc1Oc1ccc(SC)cc1 | 12 |
| CHEMBL1852361 | CCCc1nc(C(=O)NCCCN2CCN(c3cccc(C)c3C)CC2)cn1-c1ccc(Cl)cc1 | 67.2 |
| CHEMBL3703749 | Cc1ccc(C2CNCc3cc(-c4cccnn4)ccc32)cc1 | 27.9 |
| CHEMBL599230 | CC(C(=O)c1cccc(Br)c1)N1CCCCC1 | 100 |
| CHEMBL1214303 | Cc1c(C(=O)NCCN2CCN(c3cccc(Cl)c3Cl)CC2)c(C)n(C)c1-c1ccccc1.Cl | 78 |
| CHEMBL3216965 | COc1ccc(-n2c(C)nc(C(=O)NCC(O)CN3CCN(c4cccc(Cl)c4C)CC3)c2C)cc1.Cl.Cl.Cl | 24 |
| CHEMBL485541 | CC1CN(c2ccc3ccccc3n2)CCN1CCCCN1C(=O)CCCC1=O | 50 |
| CHEMBL3215850 | Cc1c(Cl)cccc1N1CCN(CCCNC(=O)c2nc(-c3ccccc3)n(-c3ccccc3)c2Cn2cncn2)CC1.Cl.Cl.Cl | 181.7 |
| CHEMBL466963 | CCCC(=O)N(Cc1ccc(Cl)cc1Cl)C1CCNC1 | 72 |
| CHEMBL3216286 | CCCc1nc(C(=O)NCCCN2CCN(c3cccc(Cl)c3C)CC2)c(C)n1-c1ccccc1OC.Cl.Cl.Cl | 55 |
| CHEMBL52440 | COc1ccc2c(c1)C13CCCCC1C(C2)N(C)CC3 | 2.71 |
| CHEMBL1201353 | CN(C)CCC(c1ccc(Cl)cc1)c1ccccn1 | 25 |
| CHEMBL2380979 | CN(C)CC1CC1Cc1ccc2[nH]cc(C#N)c2c1 | 97 |
| CHEMBL258449 | CNCc1cnccc1Oc1ccc(Cl)cc1OC | 14 |
| CHEMBL219649 | CCCC(c1ccc(Cl)c(Cl)c1)C1CCCCN1 | 180 |
| CHEMBL456638 | NCC1CCC(c2cccc3ccccc23)O1 | 24 |
| CHEMBL401596 | Cc1cc(Cl)ccc1Oc1ccncc1CN(C)C | 3 |
| CHEMBL3216498 | CCCc1nc(C(=O)NCCCN2CCN(c3cccc(Cl)c3C)CC2)c(C)n1-c1ccc(F)cc1.Cl.Cl.Cl | 28 |
| CHEMBL87567 | COC(=O)C1C(c2ccc3ccccc3c2)CC2CCC1N2C | 2.19 |
| CHEMBL2337604 | Clc1ccc(OC(c2ccccc2)C2CCNC2)c(Cl)c1 | 3.162 |
| CHEMBL513884 | O=C(C1CCOCC1)N(Cc1cccc(Cl)c1Cl)C1CCNC1 | 123 |
| CHEMBL1214426 | CCCn1c(-c2ccc(Cl)cc2)cc(C(=O)NCCCN2CCN(c3cccc(C)c3C)CC2)c1C.Cl | 131 |
| CHEMBL621 | O=c1n(CCCN2CCN(c3cccc(Cl)c3)CC2)nc2ccccn12 | 192 |
| CHEMBL500629 | CN(C)Cc1cc(S(N)(=O)=O)ccc1Sc1ccc(C(F)(F)F)cc1 | 26 |
| CHEMBL6966 | COc1ccc(CCN(C)CCCC(C#N)(c2ccc(OC)c(OC)c2)C(C)C)cc1OC | 240 |
| CHEMBL3323179 | CN(C)Cc1ccc2c(c1)CN(C)CC2c1ccc2sccc2c1 | 4.4 |
| CHEMBL4085773 | CN(C)CCCC1(c2ccc(F)cc2)OCc2ccc3ccccc3c21 | 12 |
| CHEMBL41 | CNCCC(Oc1ccc(C(F)(F)F)cc1)c1ccccc1 | 7.2 |
| CHEMBL2337587 | COc1ccccc1OC(c1ccccc1)C1CCNC1 | 100 |
| CHEMBL376952 | c1ccc(C(Cc2ccccn2)N2CCNCC2)cc1 | 230 |
| CHEMBL1812750 | COC(=O)C1C2CCC(CC1c1ccc(C)c(F)c1)N2 | 110 |
| CHEMBL210126 | CCc1ccccc1CC(c1ccccc1)N1CCNCC1 | 16 |
| CHEMBL3216302 | CCCc1nc(C(=O)NCCN2CCN(c3cccc(Cl)c3Cl)CC2)c(C)n1-c1ccccc1.Cl.Cl.Cl.Cl | 13 |
| CHEMBL256819 | CN(C)Cc1ccccc1Oc1ccc(Cl)cc1F | 6 |
| CHEMBL256970 | Cc1cc(Oc2ccc(Cl)cc2C)c(CN(C)C)cn1 | 6 |
| CHEMBL599846 | CC(NC1CCCC1)C(=O)c1cccc(Br)c1 | 100 |
| CHEMBL367408 | C=Cc1ccc(C2CC3CCC(N3)C2C(=O)CC)cc1 | 0.32 |
| CHEMBL407803 | CN1C2CCC1C(C(=O)NCCCCNC(=O)C1C(c3ccc(Cl)cc3)CC3CCC1N3C)C(c1ccc(Cl)cc1)C2 | 176 |
| CHEMBL551646 | Clc1cc(Cl)cc(OC(c2cccnc2)C2CCNCC2)c1 | 10 |
| CHEMBL2337606 | Clc1cccc(Cl)c1OC(c1ccccc1)C1CCNC1 | 39.81 |
| CHEMBL560049 | Clc1cc(Cl)c(Cl)c(OC(c2cccnc2)C2CCNCC2)c1 | 10 |
| CHEMBL1214175 | CCCn1c(-c2ccccc2)cc(C(=O)NCCCN2CCN(c3cccc4cccnc34)CC2)c1C | 82 |
| CHEMBL125954 | COC(=O)C1C(OC(c2ccc(F)cc2)c2ccc(F)cc2)CC2CCC1N2CCCc1ccccc1 | 19.7 |
| CHEMBL559081 | Cc1ccc(C(Oc2cccc(Cl)c2Cl)C2CCNCC2)cn1 | 14 |
| CHEMBL505431 | COc1ccc(F)cc1CCC1CCC(CCNCCCCCN2N=C(c3ccc(OC)c(OC)c3)C3CC=CCC3C2=O)O1 | 127 |
| CHEMBL2430680 | O=C1NCC(c2ccc(Cl)c(Cl)c2)C1c1ccc(Cl)cc1 | 85.6 |
| CHEMBL202479 | CCCC(C(=O)c1ccc(I)cc1)N1CCCC1 | 197 |
| CHEMBL540978 | Clc1ccc(OC(c2cccnc2)C2CCNCC2)c(Cl)c1 | 5 |
| CHEMBL252923 | CCN(CC)C(=O)C1(c2ccccc2)CC1CN | 22 |
| CHEMBL482293 | CN(C)Cc1cc(S(N)(=O)=O)ccc1Sc1cccc(C(F)(F)F)c1 | 30 |
| CHEMBL493027 | CN(C)Cc1ccccc1Sc1ccc(C(F)(F)F)cc1 | 28 |
| CHEMBL477584 | CC(C)C(=O)N(Cc1ccc(C(F)(F)F)cc1F)C1CCNC1 | 24 |
| CHEMBL3216758 | Cc1nc(C(=O)NCCN2CCN(c3cccc(Cl)c3Cl)CC2)c(C)n1-c1ccccc1.Cl.Cl.Cl.Cl | 129 |
| CHEMBL559654 | Clc1ccc(Cl)c(OC(c2cccnc2)C2CCNCC2)c1 | 7 |
| CHEMBL1644476 | Brc1cccc(OC2CC3CCC(C2)N3)c1 | 136 |
| CHEMBL396537 | CN(C)C1CC=C(c2c[nH]c3ccc([N+](=O)[O-])cc23)CC1 | 1 |
| CHEMBL3216755 | Cc1c(Cl)cccc1N1CCN(CCCNC(=O)c2nc(C)n(-c3ccc4c(c3)OCCO4)c2C)CC1.Cl.Cl.Cl | 2.7 |
| CHEMBL3334799 | CN(C)CCC(c1ccc(Cl)c(Cl)c1)N1CCNCC1 | 399.3 |
| CHEMBL3323095 | CN1Cc2ccccc2C(c2ccc3sccc3c2)C1 | 69 |
| CHEMBL2012078 | COC(=O)C1=C(c2ccc(-c3ccco3)cc2)CC2CCC1O2 | 248 |
| CHEMBL234928 | CN(C)C1CC=C(c2c[nH]c3ccc(Cl)cc23)CC1 | 19 |
| CHEMBL3323089 | CN1Cc2ccccc2C(c2ccc3occc3c2)C1 | 24 |
| CHEMBL476116 | CC(C)C(=O)N(Cc1ccccc1-c1ccccc1)C1CCNC1 | 400 |
| CHEMBL1852466 | COc1ccccc1-n1c(C)nc(C(=O)NCCCN2CCN(c3cccc(Cl)c3C)CC2)c1C | 16.8 |
| CHEMBL87031 | COC(=O)C1C(c2ccc(Cl)c(Cl)c2)CC2CCC1N2C | 27 |
| CHEMBL474922 | COc1ccc(Oc2ccccc2CN(C)C)cc1 | 25 |
| CHEMBL4070493 | CN(C)CCCC1(c2ccc3ccccc3c2)OCc2cc(C#N)ccc21 | 34 |
| CHEMBL493056 | CNCc1ccccc1Sc1ccc(C(F)(F)F)cc1 | 70 |
| CHEMBL3216287 | COc1ccc(-n2c(C)nc(C(=O)NCCCN3CCN(c4cccc(Cl)c4C)CC3)c2C)cc1.Cl.Cl.Cl | 12.4 |
| CHEMBL506 | COc1cc(NC(C)CCCN)c2ncccc2c1 | 16 |
| CHEMBL511163 | NCC1CCC(CCc2cccc3ccccc23)O1 | 23 |
| CHEMBL1113 | Clc1ccc2c(c1)C(N1CCNCC1)=Nc1ccccc1O2 | 34 |
| CHEMBL576063 | CC(NC(C)(C)C)C(=O)c1cccc(F)c1 | 100 |
| CHEMBL401877 | CNCc1cc(C(N)=O)ccc1Oc1ccc(Cl)cc1OC | 12 |
| CHEMBL1200623 | CCC1(O)CCC2C3CCC4=CCCCC4C3CCC21C | 81 |
| CHEMBL1087773 | Cc1cccc(N2CCN(CCCNC(=O)c3cc(-c4ccccc4)n(-c4ccc(F)cc4)c3C)CC2)c1C | 53.6 |
| CHEMBL1201151 | C#CC1(O)CCC2C3CCc4cc(OC)ccc4C3CCC21C | 41 |
| CHEMBL565844 | Cc1ccc(C(=O)C(C)NC(C)(C)C)cc1Cl | 100 |
| CHEMBL3217171 | CCCc1nc(C(=O)NCCN2CCN(c3cccc(C)c3C)CC2)c(C)n1-c1ccccc1OC.Cl.Cl | 28 |
| CHEMBL566618 | CC(NC1CCCC1)C(=O)c1cccc(Cl)c1 | 100 |
| CHEMBL2338041 | Clc1cc(Cl)c(Cl)c(OC(c2ccccc2)C2CCNC2)c1 | 1.585 |
| CHEMBL1852617 | Cc1cccc(N2CCN(CCCNC(=O)c3nc(C(C)C)n(-c4ccccc4)c3C)CC2)c1C | 16 |
| CHEMBL1080672 | Cc1c(C(=O)NCCCN2CCN(c3cccc(Cl)c3Cl)CC2)cc(-c2ccccc2)n1CC1CCCCC1 | 412 |
| CHEMBL3612829 | O=C(NCCCN1CCC(Cc2ccccc2)CC1)c1ccc(-c2ccccc2)cc1 | 56 |
| CHEMBL528995 | Clc1ccc(C23CNCC2C3)cc1Cl | 14 |
| CHEMBL2012103 | COC(=O)C1C2CCC(CC1c1ccc(-c3nccs3)cc1)O2 | 128 |
| CHEMBL403080 | COc1cc(Cl)ccc1Oc1cc(C)ncc1CN(C)C | 13 |
| CHEMBL3797849 | CN(C)CCCN1c2ccccc2CCc2ccc(C#CCOCc3cccc4ccccc34)cc21 | 340 |
| CHEMBL184252 | O=C1OC2(CCC(N3CCC(Cc4cc(Cl)ccc4Cl)CC3)CC2)c2ccc3c(c21)OCO3 | 92 |
| CHEMBL481695 | CN(C)Cc1cc(C(N)=O)ccc1Oc1ccc(OC(F)(F)F)cc1 | 3 |
| CHEMBL2012079 | COC(=O)C1=C(c2ccc(-c3cccs3)cc2)CC2CCC1O2 | 242 |
| CHEMBL598026 | CC(NC(C)(C)C)C(=O)c1cccc([N+](=O)[O-])c1 | 100 |
| CHEMBL182431 | O=C1OC2(CCC(N3CCC(Cc4ccccc4F)CC3)CC2)c2ccc3c(c21)OCCO3 | 480 |
| CHEMBL456221 | COc1ccc(F)cc1C1CCC(CN)O1 | 122 |
| CHEMBL184910 | O=C1OC2(CCC(N3CCN(Cc4ccccc4I)CC3)CC2)c2ccc3c(c21)OCO3 | 48 |
| CHEMBL2337600 | FC(F)(F)c1ccc(OC(c2ccccc2)C2CCNC2)cc1 | 1.995 |
| CHEMBL171106 | COC(=O)C1C(c2ccc(Br)c2)CC2CCC1N2C | 20.4 |
| CHEMBL3216507 | COc1ccc(-n2c(-c3ccccc3)nc(C(=O)NCC(O)CN3CCN(c4cccc(Cl)c4C)CC3)c2C)cc1.Cl.Cl.Cl | 443 |
| CHEMBL3215863 | CCCc1nc(C(=O)NCCCN2CCN(c3cccc(Cl)c3C)CC2)cn1-c1ccc(Cl)cc1.Cl.Cl.Cl | 66 |
| CHEMBL3215848 | Cc1c(C(=O)NCCCN2CCN(c3cccc(Cl)c3Cl)CC2)nc(-c2ccccc2)n1-c1ccccc1F.Cl.Cl.Cl | 200 |
| CHEMBL2380978 | CCNCC1CCCC1c1ccc2[nH]cc(C#N)c2c1 | 260 |
| CHEMBL1852666 | Cc1nc(C(=O)NCC(O)CN2CCN(c3cccc(Cl)c3Cl)CC2)c(C)n1-c1ccc2c(c1)OCCO2 | 12 |
| CHEMBL3703750 | O=c1ccc(-c2cc3c(cc2F)C(c2ccc(Cl)c(Cl)c2)CNC3)n[nH]1 | 39 |
| CHEMBL552046 | Clc1cccc(OC(c2ccccn2)C2CCNCC2)c1Cl | 6 |
| CHEMBL3681356 | COc1ccccc1-n1c(C)nc(C(=O)NCC(O)CN2CCN(c3cccc(C)c3C)CC2)c1C | 12 |
| CHEMBL1080683 | Cc1c(C(=O)NCCN2CCN(c3cccc(Cl)c3Cl)CC2)cc(-c2ccccc2)n1C | 57 |
| CHEMBL493263 | CN(C)Cc1ccccc1Sc1cccc(Cl)c1 | 12 |
| CHEMBL493040 | CNCc1ccccc1Cc1ccc(Cl)c(Cl)c1 | 30 |
| CHEMBL549 | CN(C)CCCC1(c2ccc(F)cc2)OCc2cc(C#N)ccc21 | 0.902 |
| CHEMBL3215846 | Cc1c(C(=O)NCC(O)CN2CCN(c3cccc(Cl)c3Cl)CC2)nc(-c2ccccc2)n1-c1ccc(F)cc1.Cl.Cl.Cl.Cl | 195 |
| CHEMBL1644610 | N#Cc1cc(OC2CC3CCC(C2)N3)nc(-c2ccccc2C#N)c1 | 48 |
| CHEMBL214056 | Clc1cccc(CC(c2ccccc2)N2CCNCC2)c1Cl | 3.9 |
| CHEMBL564 | CN(C)CCCN1c2ccccc2Sc2ccccc21 | 86 |
| CHEMBL2337602 | Clc1cccc(OC(c2ccccc2)C2CCNC2)c1Cl | 10 |
| CHEMBL210116 | CCOc1ccccc1CC(c1nccs1)N1CCNCC1 | 39 |
| CHEMBL1214597 | Cc1c(C(=O)NCCN2CCN(c3cccc(Cl)c3Cl)CC2)cc(C(C)(C)C)n1C.Cl | 45 |
| CHEMBL1852517 | CCCc1nc(C(=O)NCCCN2CCN(c3cccc(Cl)c3C)CC2)c(C)n1-c1ccc(OC)cc1 | 10.2 |
| CHEMBL396540 | CN(C)C1CC=C(c2c[nH]c3ccc(Br)cc23)CC1 | 5.3 |
| CHEMBL442272 | FC(F)(F)c1ccccc1CN(C1CCOCC1)C1CCNC1 | 21 |
| CHEMBL566418 | CC(NC1CCC1)C(=O)c1cccc(Cl)c1 | 185 |
| CHEMBL567481 | CC(CNC(C)(C)C)C(=O)c1cccc(Cl)c1 | 100 |
| CHEMBL121460 | CN(CCOC(c1ccc(F)cc1)c1ccc(F)cc1)CCN(C)CCOC(c1ccc(F)cc1)c1ccc(F)cc1 | 10 |
| CHEMBL472253 | C#CCN(CC)C(=O)C1(c2cccs2)CC1CN | 81 |
| CHEMBL322067 | CN(C)CC1C2CCC(CC2)C1c1ccc(Cl)c(Cl)c1 | 270 |
| CHEMBL213229 | CCOc1ccccc1CC(c1cccnc1)N1CCNCC1 | 46 |
| CHEMBL3215630 | Cc1nc(C(=O)NCCCN2CCN(c3cccc(Cl)c3Cl)CC2)c(C)n1-c1ccc2c(c1)OCCO2.Cl.Cl.Cl.Cl | 9.17 |
| CHEMBL433956 | O=C1OC2(CCC(N3CCC(Cc4ccccc4Cl)CC3)CC2)c2ccc3c(c21)OCO3 | 140 |
| CHEMBL477586 | CC(C)C(=O)N(Cc1ccc(F)cc1F)C1CCNC1 | 400 |
| CHEMBL511 | COc1ccc(CN(CCN(C)C)c2ccccn2)cc1 | 47 |
| CHEMBL3323178 | CN1Cc2cc(CN)ccc2C(c2ccc3sccc3c2)C1 | 28 |
| CHEMBL520024 | CC1CN(c2ccc3ccccc3n2)CCN1CCCN1C(=O)c2ccccc2C1=O | 80 |
| CHEMBL491268 | CN(C)Cc1ccccc1Cc1ccc(Cl)c(Cl)c1 | 5 |
| CHEMBL1080747 | Cc1c(C(=O)NCCCN2CCN(c3cccc(Cl)c3)CC2)cc(-c2ccccc2)n1-c1ccc(F)cc1 | 80.2 |
| CHEMBL464422 | COc1ccc(F)cc1CCC1CCC(CCN)O1 | 2.5 |
| CHEMBL3323174 | COc1ccc2c(c1)CN(C)CC2c1ccc2sccc2c1 | 5.3 |
| CHEMBL550696 | Clc1cccc(OC(c2cccnc2)C2CCNCC2)c1 | 3 |
| CHEMBL184571 | COc1cc2nc(O)c3c(c2cc1OC)CCCC3N1CCC(Cc2cc(F)ccc2F)CC1 | 5.5 |
| CHEMBL1642901 | CN(C)C1Cc2ccccc2C(c2ccc(Cl)c(Cl)c2)C1 | 20 |
| CHEMBL1118 | CN(C)CC(c1ccc(O)cc1)C1(O)CCCCC1 | 17 |
| CHEMBL3216963 | CCCc1nc(C(=O)NCCCN2CCN(c3cccc(C)c3C)CC2)c(C)n1-c1ccc(OC)cc1.Cl.Cl | 13 |
| CHEMBL1852370 | Cc1nc(C(=O)NCCCN2CCN(c3cccc(Cl)c3Cl)CC2)c(C)n1-c1ccccc1F | 13 |
| CHEMBL599028 | CC(C(=O)c1cccc(F)c1)N1CCCCC1 | 100 |
| CHEMBL214273 | FC(F)(F)c1cccc(CC(c2ccccc2)N2CCNCC2)c1 | 13 |
| CHEMBL3216518 | CCCc1nc(C(=O)NCCCN2CCN(c3cccc(C)c3C)CC2)c(C)n1-c1ccc2c(c1)OCCO2.Cl.Cl | 5 |
| CHEMBL380056 | CN1CCCCCCCCCC(=O)OCC2C(CC3CCC2N3C)c2ccc1cc2 | 380 |
| CHEMBL1086756 | CCCn1c(-c2ccccc2)cc(C(=O)NCCCN2CCN(c3cccc(Cl)c3Cl)CC2)c1C | 420 |
| CHEMBL1088314 | Cc1c(C(=O)NCCCN2CCN(c3cccc(Cl)c3Cl)CC2)cc(-c2ccccc2)n1-c1ccc(F)cc1 | 266 |
| CHEMBL1214546 | CCCn1c(-c2cccnc2)cc(C(=O)NCCCN2CCN(c3cccc(Cl)c3Cl)CC2)c1C | 133 |
| CHEMBL1644483 | N#Cc1cccc(OC2CC3CCC(C2)N3)c1 | 25 |
| CHEMBL213496 | CCOc1ccccc1CC(c1cncs1)N1CCNCC1 | 40 |
| CHEMBL88792 | COC(=O)C1C2CCC(CC1c1ccc(Br)cc1)O2 | 30 |
| CHEMBL3215847 | Cc1cccc(N2CCN(CC(O)CNC(=O)c3nc(C)n(-c4ccc5c(c4)OCCO5)c3C)CC2)c1C.Cl.Cl | 12.6 |
| CHEMBL3216076 | Cc1cccc(N2CCN(CC(O)CNC(=O)c3nc(C)n(-c4ccccc4)c3C)CC2)c1C.Cl.Cl | 0.64 |
| CHEMBL2338050 | Fc1ccc(F)c(OC(c2ccccc2)C2CCNC2)c1Cl | 5.012 |
| CHEMBL3770663 | CNCC1(c2ccc3ccccc3c2)CCC(C)(O)CC1 | 11 |
| CHEMBL256046 | CN1C2CCC1C(COC(=O)CCC(=O)OCC1C(c3ccc(Cl)c(Cl)c3)CC3CCC1N3C)C(c1ccc(Cl)c(Cl)c1)C2 | 106 |
| CHEMBL599233 | CC(C(=O)c1cccc([N+](=O)[O-])c1)N1CCCCC1 | 100 |
| CHEMBL473966 | CNCCC1CCC(CCc2cc(F)ccc2OC)O1 | 18.4 |
| CHEMBL256135 | CNCc1cnccc1Oc1ccc(Cl)cc1Br | 23 |
| CHEMBL214240 | Clc1cccc(CC(c2ccccc2)N2CCNCC2)c1 | 9.4 |
| CHEMBL133070 | COC(=O)C1C(c2ccc(C(C)=O)c(C(C)=O)c2)CC2CCC1N2C | 330 |
| CHEMBL3215638 | Cc1cccc(N2CCN(CCCNC(=O)c3nc(-c4ccccc4)n(-c4ccc5c(c4)OCCO5)c3C)CC2)c1C.Cl.Cl | 185 |
| CHEMBL86654 | COC(=O)C1C(c2ccc3ccccc3c2)CC2CC(O)C1N2C | 94 |
| CHEMBL781 | OC1(c2ccc(Cl)cc2)c2ccccc2C2=NCCN21 | 54 |
| CHEMBL386223 | N#Cc1cccc(CC(c2ccccc2)N2CCNCC2)c1 | 4 |
| CHEMBL479411 | CNCc1cc(NS(C)(=O)=O)ccc1Sc1ccc(SC)cc1 | 10 |
| CHEMBL578611 | CC(NC(C)(C)C)C(=O)c1cc(Cl)cc(Cl)c1 | 100 |
| CHEMBL3216285 | Cc1cccc(N2CCN(CC(O)CNC(=O)c3nc(C)n(-c4ccc(F)cc4)c3C)CC2)c1C.Cl.Cl | 16.6 |
| CHEMBL1852777 | Cc1c(C(=O)NCC(O)CN2CCN(c3cccc(Cl)c3Cl)CC2)nc(-c2ccccc2)n1-c1ccccc1 | 71.1 |
| CHEMBL550289 | CCc1ccccc1OC(c1cccnc1)C1CCNCC1 | 61 |
| CHEMBL562244 | c1ccc(-c2ccccc2OC(c2cccnc2)C2CCNCC2)cc1 | 109 |
| CHEMBL563404 | c1ccc(-c2ccccc2OC(c2ccccn2)C2CCNCC2)cc1 | 400 |
| CHEMBL3215631 | CCCc1nc(C(=O)NCCCN2CCN(c3cccc(C)c3C)CC2)cn1-c1ccccc1.Cl.Cl | 16.8 |
| CHEMBL227757 | COC(=O)C1C2CCC(CC1c1ccc(Cl)c(Cl)c1)S2 | 3 |
| CHEMBL641 | CNCCC(Oc1ccccc1C)c1ccccc1 | 190 |
| CHEMBL256511 | CNCc1cnccc1Oc1ccc(Cl)cc1C | 14 |
| CHEMBL27025 | CN1CC=C(c2c[nH]c3ccc(F)cc23)CC1 | 160 |
| CHEMBL481696 | CNC(=O)c1ccc(Oc2ccc(OC(F)(F)F)cc2)c(CN(C)C)c1 | 10 |
| CHEMBL446281 | CC(C)(C)C(=O)N(Cc1cccc(Cl)c1Cl)C1CCNC1 | 26 |
| CHEMBL3216291 | COc1ccc(-n2c(-c3ccccc3)nc(C(=O)NCCCN3CCN(c4cccc(C)c4C)CC3)c2C)cc1.Cl.Cl | 12 |
| CHEMBL3216300 | CCCc1nc(C(=O)NCCN2CCN(c3cccc(Cl)c3Cl)CC2)c(C)n1-c1ccccc1.Cl.Cl.Cl.Cl | 82 |
| CHEMBL541226 | Clc1cccc(OC(c2ccccc2)C2CCNCC2)c1Cl | 8 |
| CHEMBL3681362 | Cc1cccc(N2CCN(CCNC(=O)c3nc(C(C)C)n(-c4ccccc4)c3C)CC2)c1C | 76 |
| CHEMBL102018 | Cc1ccc(C2CC3CCC4C2C(=O)CN34)cc1 | 68 |
| CHEMBL383706 | CNC1CCC(c2ccc(Cl)c(Cl)c2)c2ccc(S(N)(=O)=O)cc21 | 1 |
| CHEMBL491609 | CSc1ccc(Oc2ccc(S(N)(=O)=O)cc2CN(C)C)cc1F | 9 |
| CHEMBL2338032 | Fc1cccc(F)c1OC(c1ccccc1)C1CCNC1 | 158.49 |
| CHEMBL502129 | CN(C)Cc1cc(C(=O)N(C)C)ccc1Oc1ccc(OC(F)(F)F)cc1 | 100 |
| CHEMBL3216078 | CCCc1nc(C(=O)NCCCN2CCN(c3cccc(C)c3C)CC2)cn1-c1ccc(F)cc1.Cl.Cl | 188 |
| CHEMBL691 | C#CC1(O)CCC2C3CCc4cc(O)ccc4C3CCC21C | 54 |
| CHEMBL272493 | CNCc1cc(C(=O)N(C)C)ccc1Oc1ccc(Cl)cc1C | 17 |
| CHEMBL260037 | CCOc1cc(Cl)ccc1Oc1ccncc1CNC | 24 |
| CHEMBL1213856 | Cc1c(C(=O)NCCCN2CCN(c3ccc(Cl)cc3Cl)CC2)cc(-c2ccccc2)n1C | 23 |
| CHEMBL4093518 | CN(C)CCCC1(c2ccc(F)cc2)OCc2c1ccc1ccccc21 | 4 |
| CHEMBL1213674 | Cc1c(C(=O)NCCCN2CCN(c3cccc(Cl)c3Cl)CC2)cc(C(C)(C)C)n1C | 375 |
| CHEMBL210027 | FC(F)(F)c1ccccc1CC(c1ccccc1)N1CCNCC1 | 10 |
| CHEMBL3216094 | COc1cc(OC)cc(-n2c(C)nc(C(=O)NCCCN3CCN(c4cccc(C)c4C)CC3)c2C)c1.Cl.Cl | 10 |
| CHEMBL1214113 | CCCn1c(-c2ccccc2)cc(C(=O)NCCCN2CCN(c3cccc(OC)c3)CC2)c1C | 322 |
| CHEMBL213754 | CCCOc1ccccc1CC(c1ccccc1)N1CCNCC1 | 22 |
| CHEMBL271707 | CN1C2CCC1C(COC(=O)CCCCCCCCC(=O)OCC1C(c3ccc(Cl)c(Cl)c3)CC3CCC1N3C)C(c1ccc(Cl)c(Cl)c1)C2 | 17 |
| CHEMBL3215628 | COc1ccc(-n2c(C)nc(C(=O)NCCCN3CCN(c4cccc(C)c4C)CC3)c2C)cc1.Cl.Cl | 31.1 |
| CHEMBL1852503 | Cc1nc(C(=O)NCCCN2CCN(c3cccc(Cl)c3Cl)CC2)c(C)n1-c1ccc2c(c1)OCO2 | 9.17 |
| CHEMBL551573 | Clc1cccc(OC(c2ccccn2)C2CCNC2)c1Cl | 18 |
| CHEMBL394356 | COc1ccc2[nH]cc(C3=CCC(N(C)C)CC3)c2c1 | 150 |
| CHEMBL448900 | CC(C)CC(=O)N(Cc1cccc(Cl)c1Cl)C1CCNC1 | 17 |
| CHEMBL564631 | CSc1ccc(OC(c2cccnc2)C2CCNCC2)cc1 | 7 |
| CHEMBL482292 | CN(C)Cc1cc(S(N)(=O)=O)ccc1Cc1ccc(Cl)c(Cl)c1 | 60 |
| CHEMBL2219885 | c1ccc(COC(c2ccccc2)C2CNC2)cc1 | 82.9 |
| CHEMBL3797621 | CN(C)CCCN1c2ccccc2CCc2ccc(C#CCO)cc21 | 60 |
| CHEMBL185395 | O=C1OC2(CCC(N3CCC(Cc4ccccc4Cl)CC3)CC2)c2ccc3c(c21)OCCO3 | 100 |
| CHEMBL1213676 | Cc1c(Cl)cccc1N1CCN(CCNC(=O)c2cc(C(C)(C)C)n(C)c2C)CC1 | 54 |
| CHEMBL3216754 | CCCc1nc(C(=O)NCCCN2CCN(c3cccc(C)c3C)CC2)c(C)n1-c1ccccc1Cl.Cl.Cl.Cl | 7.3 |
| CHEMBL1411979 | CN(C)CCN(Cc1cccs1)c1ccccn1 | 200 |
| CHEMBL3681369 | Cc1ccc(-n2c(C)nc(C(=O)NCCCN3CCN(c4cccc(C)c4C)CC3)c2C)cc1 | 5.95 |
| CHEMBL566832 | Cc1cccc(C(=O)C(C)NC(C)(C)C)c1 | 100 |
| CHEMBL490 | Fc1ccc(C2CCNCC2COc2ccc3c(c2)OCO3)cc1 | 0.08 |
| CHEMBL231809 | CN(C)C1CC=C(c2c[nH]c3ccccc23)CC1 | 3.1 |
| CHEMBL565386 | CC(NC(C)(C)C)C(=O)c1cccs1 | 100 |
| CHEMBL3323091 | CN1Cc2ccccc2C(c2cc3ccccc3o2)C1 | 110 |
| CHEMBL3217169 | Cc1cccc(N2CCN(CCCNC(=O)c3nc(-c4ccccc4)n(-c4ccccc4)c3C)CC2)c1C.Cl.Cl | 20.5 |
| CHEMBL461268 | NCC1CCC(c2ccc(Br)cc2)O1 | 50 |
| CHEMBL3323101 | CN1Cc2ccccc2C(c2ccc3[nH]ccc3c2)C1 | 170 |
| CHEMBL187211 | COC(=O)C1=C(c2ccc(I)cc2)CC2CCC1O2 | 4800 |
| CHEMBL4076722 | [O-][S+](Cc1csc(Br)c1)C(c1ccccc1)c1ccccc1 | 44700 |
| CHEMBL195437 | CCCCCCCCc1ccc(O)cc1 | 6973 |
| CHEMBL387117 | CC(C)CC(c1ccc(Cl)c(Cl)c1)C1CCCCN1C | 2400 |
| CHEMBL229436 | Cc1ccc(C2CC3CCC(C2c2ncc(-c4ccc([N+](=O)[O-])cc4)s2)N3C)cc1 | 3170 |
| CHEMBL1173430 | CC1NC(C)(C)COC1(O)c1cc(Cl)cc(Cl)c1 | 1.00E+05 |
| CHEMBL226984 | Cc1ccc(C2CC3CCC(C2c2ncc(-c4cccc([N+](=O)[O-])c4)s2)N3C)cc1 | 2380 |
| CHEMBL3800215 | CN(C)CCCN1c2ccccc2CCc2ccc(C#CCOCCCOCCCN3c4ccccc4CCc4ccccc43)cc21 | 4100 |
| CHEMBL567479 | CC(C)(C)NC(CCC1CCCCC1)C(=O)c1cccc(Cl)c1 | 10000 |
| CHEMBL1381098 | S=C=Nc1cccc2ccccc12 | 14293 |
| CHEMBL1643651 | Clc1ccc(C2=C(c3cc(-c4ccccc4)no3)C3CCC(C2)S3)cc1Cl | 10000 |
| CHEMBL4060900 | C(=Cc1ccccc1)CN1CCN(CCCn2nnnc2C(c2ccccc2)c2ccccc2)CC1 | 7500 |
| CHEMBL1213675 | Cc1c(Cl)cccc1N1CCN(CCNC(=O)c2cc(-c3ccccc3)n(C)c2C)CC1 | 7096 |
| CHEMBL218849 | CCCCC(c1ccc(Cl)cc1)C1CCCCN1 | 4000 |
| CHEMBL121027 | CN(CCOC(c1ccccc1)c1ccccc1)CCN(C)CCc1ccc(F)cc1 | 1050 |
| CHEMBL219540 | CC(=O)C(c1ccccc1)C1CCCCN1 | 9700 |
| CHEMBL1255834 | CC(N)Cc1c[nH]c2ccc(OCc3cccs3)cc12 | 4785 |
| CHEMBL4074999 | c1ccc(CN2CCN(CCCn3nnnc3C(c3ccccc3)c3ccccc3)CC2)cc1 | 5050 |
| CHEMBL390743 | CN1C2CCC1C(c1ncc(-c3ccc(F)cc3)s1)C(c1ccc(Cl)cc1)C2 | 1010 |
| CHEMBL374769 | CC(C)(C)CC(c1ccc(Cl)cc1)C1CCCCN1 | 8300 |
| CHEMBL496 | Oc1c(Cl)cc(Cl)c(Cl)c1Cc1c(O)c(Cl)cc(Cl)c1Cl | 4349.4 |
| CHEMBL89906 | COC(=O)C1C(OC(c2ccc(F)cc2)c2ccc(F)cc2)CC2C(O)CC1N2C | 4850 |
| CHEMBL122859 | O=C1CN(CCc2ccccc2)CCN1CCOC(c1ccc(F)cc1)c1ccc(F)cc1 | 3940 |
| CHEMBL219116 | COC(=O)C(c1ccc(Cl)cc1)C1CCCCN1 | 1.00E+07 |
| CHEMBL448912 | NCCC1CCC(Cc2ccccc2)O1 | 2260 |
| CHEMBL3612834 | O=C(NCCCN1CCC(Cc2ccccc2)CC1)c1ccc2cc(Br)ccc2c1 | 1914 |
| CHEMBL1812742 | Cc1ccc(C2CC3CCC(C2c2cc(-c4ccc(Cl)cc4)no2)N3C)cc1 | 1.00E+05 |
| CHEMBL4063974 | Fc1ccc(-c2nnnn2CCCN2CCC(Cc3ccccc3)CC2)cc1 | 2000 |
| CHEMBL103138 | COC(=O)c1ccccc1-c1ccc(F)cc1 | 33000 |
| CHEMBL299890 | COC(=O)C1C(c2ccc(F)cc2)CC2CCC1N2Cc1ccccc1 | 1073 |
| CHEMBL451837 | NCCC1CCC(c2ccco2)O1 | 10000 |
| CHEMBL1643658 | Fc1ccc(C2CC3CCC(S3)C2c2cc(-c3ccccc3)no2)cc1 | 10000 |
| CHEMBL81 | O=C(c1ccc(OCCN2CCCCC2)cc1)c1c(-c2ccc(O)cc2)sc2cc(O)ccc12 | 3576 |
| CHEMBL1643662 | Cc1cc(C2C3CCC(CC2c2ccccc2)S3)on1 | 10000 |
| CHEMBL480988 | COc1ccc(C2=NN(CCCCCBr)C(=O)C3CC=CCC23)cc1OC | 1.00E+05 |
| CHEMBL1644608 | N#Cc1ccccc1-c1cccc(OC2CC3CCC(C2)N3)c1 | 2550 |
| CHEMBL568140 | CC(NC(C)(C)C)C(=O)c1ccc(Cl)c(Cl)c1 | 10000 |
| CHEMBL332506 | COC(=O)C1C(OC2c3ccccc3-c3ccccc32)CC2CCC1N2C | 3440 |
| CHEMBL566050 | CCCC(NC(C)(C)C)C(=O)c1ccc(Cl)c(Cl)c1 | 1580 |
| CHEMBL385359 | CN1C2CCC1C1COC(=O)CCCCCCC(=O)Nc3ccc(cc3)C1C2 | 2090 |
| CHEMBL1173353 | CC1NC(C)(C)COC1(O)c1ccc(F)cc1 | 1.00E+05 |
| CHEMBL193482 | CC12CCC3c4ccc(O)cc4CCC3C1CC(O)C2O | 10336 |
| CHEMBL2430700 | O=C1NCC(c2ccc(Cl)cc2)C1c1ccc2ccccc2c1 | 6451 |
| CHEMBL1643646 | Cc1cc(C2=C(c3ccc(Cl)c(Cl)c3)CC3CCC2S3)on1 | 10000 |
| CHEMBL125871 | COC(=O)C1C(OC(c2ccc(F)cc2)c2ccc(F)cc2)CC2CCC1N2CCCc1ccc(I)cc1 | 2200 |
| CHEMBL565594 | CC(C)CC(NC(C)(C)C)C(=O)c1cccc(Cl)c1 | 10000 |
| CHEMBL30713 | CC(N)Cc1c[nH]c2ccccc12 | 2344.23 |
| CHEMBL491889 | CCN(CC)CCCCCCn1c2ccccc2c(=O)c2ccccc21 | 3500 |
| CHEMBL4099773 | c1ccc(CCN2CCN(CCCn3nnnc3C(c3ccccc3)c3ccccc3)CC2)cc1 | 1121 |
| CHEMBL523652 | CCN(CC)CCCCn1c2ccccc2c(=O)c2ccccc21 | 10000 |
| CHEMBL1214489 | CCCn1c(-c2ccc(Cl)cc2)cc(C(=O)NCCCN2CCN(c3cccc(Cl)c3Cl)CC2)c1C.Cl | 1159 |
| CHEMBL204288 | CCCC(C(=O)c1ccc(Cl)c(Cl)c1)N1CCCCC1 | 10000 |
| CHEMBL1643669 | Clc1ccc(C2CC3CCC(S3)C2c2cc(-c3ccccc3)no2)cc1Cl | 10000 |
| CHEMBL2096861 | COC(=O)C1C(c2ccc(C)cc2)CC2CCC1N2C | 10000 |
| CHEMBL569700 | CC(NC1CC1)C(=O)c1cccc(Cl)c1 | 3180 |
| CHEMBL457966 | NCCC1CCC(CCc2ccc(O)cc2)O1 | 4041 |
| CHEMBL59 | NCCc1ccc(O)c(O)c1 | 30000 |
| CHEMBL314887 | COC(=O)C1C(c2ccc(F)cc2)CC2C(O)CC1N2C | 20000 |
| CHEMBL203403 | CCCC(C(=O)c1ccccc1C)N1CCCC1 | 2020 |
| CHEMBL3775405 | Cc1cccc(C)c1N1CCN(CCCN2CCC(Cc3ccccc3)CC2)CC1 | 2000 |
| CHEMBL4085259 | c1ccc(CCN2CCN(CCn3nnnc3C(c3ccccc3)c3ccccc3)CC2)cc1 | 4830 |
| CHEMBL4060019 | Clc1cccc(-c2nnnn2CCCN2CCC(Cc3ccccc3)CC2)c1Cl | 3960 |
| CHEMBL3613151 | O=C(NCCN1CCC(Cc2ccccc2)CC1)C(c1ccccc1)c1ccccc1 | 10000 |
| CHEMBL219334 | CCC(CC)C(c1ccc(Cl)cc1)C1CCCCN1 | 9400 |
| CHEMBL218894 | COc1ccc(C(CC(C)C)C2CCCCN2)cc1 | 3500 |
| CHEMBL2096856 | COC(=O)C1C(c2ccccc2)CC2CCC1N2C | 10000 |
| CHEMBL227232 | CN1C2CCC1C(c1ncc(-c3ccc([N+](=O)[O-])cc3)s1)C(c1ccc(Cl)cc1)C2 | 3600 |
| CHEMBL311469 | CC(=O)Nc1ccc2c(c1)Cc1ccccc1-2 | 6053 |
| CHEMBL89756 | COC(=O)C1=C(c2ccccc2)CC2C(O)CC1N2C | 1.00E+05 |
| CHEMBL4101722 | C(=Cc1ccccc1)CN1CCN(CCn2nnnc2C(c2ccccc2)c2ccccc2)CC1 | 10000 |
| CHEMBL426178 | Cc1ccc(C2CC3CCC(C2c2ccc(C)cc2)N3C)cc1 | 10000 |
| CHEMBL2012081 | COC(=O)C1=C(c2ccc(-c3nccs3)cc2)CC2CCC1O2 | 3000 |
| CHEMBL373773 | Clc1ccc(C(Cc2ccccc2)C2CCCCN2)cc1 | 1100 |
| CHEMBL1643654 | Cc1cc(C2C3CCC(CC2c2ccc(Cl)cc2)S3)on1 | 7000 |
| CHEMBL2012088 | COC(=O)C1=C(c2nc3ccccc3o2)CC2CCC1O2 | 3000 |
| CHEMBL926 | CC(CCc1ccc(O)cc1)NCCc1ccc(O)c(O)c1 | 1895 |
| CHEMBL4088715 | [O-][S+](Cc1ccsc1Br)C(c1ccccc1)c1ccccc1 | 59600 |
| CHEMBL4095430 | c1ccc(CN2CCN(CCn3nnnc3C(c3ccccc3)c3ccccc3)CC2)cc1 | 10000 |
| CHEMBL4074032 | CCCN(C)CCCC1(c2ccc(F)cc2)OCc2cc(C#N)ccc21 | 2400 |
| CHEMBL3775085 | Brc1ccc(N2CCN(CCN3CCC(Cc4ccccc4)CC3)CC2)cc1 | 4920 |
| CHEMBL607430 | COC(=O)C1C(OC(c2ccc(F)cc2)c2ccc(F)cc2)CC2CCC1N2C | 5500 |
| CHEMBL4081753 | c1ccc(CCN2CCN(CCCCn3nnnc3C(c3ccccc3)c3ccccc3)CC2)cc1 | 8000 |
| CHEMBL3705034 | O=C(O)COc1cc(CC2CCN(CCc3ccc4occ(=O)c4c3)CC2)ccc1Br | 10000 |
| CHEMBL1419 | CC(C)CC(N(C)C)C1(c2ccc(Cl)cc2)CCC1 | 2085 |
| CHEMBL1643645 | Cc1cc(C2=C(c3ccc(Br)cc3)CC3CCC2S3)on1 | 10000 |
| CHEMBL1173597 | CCCC1NC(C)(C)COC1(O)c1cccc(Cl)c1 | 4130 |
| CHEMBL123830 | O=C(Cc1ccc(F)cc1)NCCNCCOC(c1ccc(F)cc1)c1ccc(F)cc1 | 1900 |
| CHEMBL2012098 | COC(=O)C1=C(c2ccc(-c3ccccc3)n2C)CC2CCC1O2 | 3000 |
| CHEMBL2012099 | O=C1C2=C(CC3CCC2O3)c2cc3ccccc3n21 | 3000 |
| CHEMBL1272041 | COc1c(C)nc(C)nc1C(=O)NCCN1CCN(c2cccc(C)c2C)CC1.Cl | 2096 |
| CHEMBL217975 | COc1cccc(C(CC(C)C)C2CCCCN2)c1 | 2700 |
| CHEMBL418971 | CC(C)(c1ccc(O)cc1)c1ccc(O)cc1 | 20325 |
| CHEMBL204490 | CCCC(C(=O)c1ccc(C(F)(F)F)cc1)N1CCCC1 | 1030 |
| CHEMBL202420 | CCCC(C(=O)c1ccc(Br)cc1)N1CCCC1 | 1050 |
| CHEMBL2012093 | COC(=O)C1=C(c2ccco2)CC2CCC1O2 | 3000 |
| CHEMBL1170441 | Cc1ccc(C2(O)OCC(C)(C)NC2C)cc1 | 1.00E+05 |
| CHEMBL388760 | CN1C2CCC1C(c1ncc(-c3cccc([N+](=O)[O-])c3)s1)C(c1ccc(Cl)cc1)C2 | 1610 |
| CHEMBL338566 | COC(=O)C1C(OC(c2ccccc2)c2ccc(F)cc2)CC2CCC1N2C | 4930 |
| CHEMBL1213932 | COc1cccc(N2CCN(CCCNC(=O)c3cc(-c4ccccc4)n(C)c3C)CC2)c1OC | 1259 |
| CHEMBL388612 | CN1C2CCC1C(c1ncc(-c3ccc(Cl)c(Cl)c3)s1)C(c1ccc(Cl)cc1)C2 | 5300 |
| CHEMBL566886 | CCC(NC(C)(C)C)C(=O)c1cccc(Cl)c1 | 16000 |
| CHEMBL3612830 | O=C(NCCCN1CCC(Cc2ccccc2)CC1)C(c1ccc(Cl)cc1)c1ccc(Cl)cc1 | 7670 |
| CHEMBL186501 | COC(=O)C1=C(c2ccccc2)CC2CCC1S2 | 10000 |
| CHEMBL226933 | COc1ccc(-c2cnc(C3C(c4ccc(C)cc4)CC4CCC3N4C)s2)cc1 | 6800 |
| CHEMBL202409 | Cc1ccc(C(=O)C(CC(C)C)N2CCCC2)cc1 | 2040 |
| CHEMBL377498 | CCC(CC(=O)c1ccc(Cl)c(Cl)c1)N1CCCC1 | 1780 |
| CHEMBL715 | Cc1cc2c(s1)Nc1ccccc1N=C2N1CCN(C)CC1 | 1033 |
| CHEMBL588119 | CCCCCCCCNC(C)C(O)c1ccc(SC(C)C)cc1 | 1253.2 |
| CHEMBL314138 | COC(=O)C1=C(c2ccc(F)cc2)CC2CC(O)C1N2C | 10000 |
| CHEMBL1173280 | CC1NC(C)(C)COC1(O)c1cccc([N+](=O)[O-])c1 | 1.00E+05 |
| CHEMBL707 | COc1cc2nc(N3CCN(C(=O)C4COc5ccccc5O4)CC3)nc(N)c2cc1OC | 2688 |
| CHEMBL2012089 | COC(=O)C1=C(c2nc3ccccc3s2)CC2CCC1O2 | 3000 |
| CHEMBL288826 | O=C(NCCCN1CCC(Cc2ccccc2)CC1)C(c1ccccc1)c1ccccc1 | 10000 |
| CHEMBL123252 | CN(CCCc1ccccc1)CCCN(C)CCOC(c1ccccc1)c1ccccc1 | 1700 |
| CHEMBL442843 | COc1ccc(F)cc1C1CCC(CN)O1 | 2071 |
| CHEMBL388605 | COc1ccc(C2C(c3ccc(C)cc3)CC3CCC2N3C)cc1OC | 13700 |
| CHEMBL491408 | Cc1cc(Oc2ccc(S(N)(=O)=O)cc2CN(C)C)ccc1[S+](C)[O-] | 10000 |
| CHEMBL125140 | COC(=O)C1C(OC(c2ccccc2)c2ccc(C)cc2)CC2CCC1N2C | 13200 |
| CHEMBL305660 | CC(C)(C)c1ccc(C(=O)CCCN2CCC(OC(c3ccccc3)c3ccccc3)CC2)cc1 | 1228.7 |
| CHEMBL103831 | COC(=O)c1ccccc1-c1ccccc1 | 70000 |
| CHEMBL4082767 | C(=Cc1ccccc1)CN1CCN(CCCCn2nnnc2C(c2ccccc2)c2ccccc2)CC1 | 6290 |
| CHEMBL381046 | CN1C2C=C(c3ccccc3-c3ccccc3)CC1CC2 | 1710 |
| CHEMBL1672023 | NC(Cc1ccc(Cl)cc1)Cc1ccc(Cl)cc1 | 1900 |
| CHEMBL388850 | CN1C2CCC1C(c1ccccc1)C(c1ccccc1)C2 | 21000 |
| CHEMBL101986 | COC(=O)c1ccccc1-c1ccc(Cl)cc1 | 46000 |
| CHEMBL4075022 | [O-][S+](Cc1cc(Br)cs1)C(c1ccccc1)c1ccccc1 | 70900 |
| CHEMBL566208 | CCC(NC(C)(C)C)C(=O)c1ccc(Cl)c(Cl)c1 | 1090 |
| CHEMBL1479 | C#CC1(O)CCC2C3CCC4=Cc5oncc5CC4(C)C3CCC21C | 15420 |
| CHEMBL124278 | COC(=O)C1C(OC(c2ccc(Br)cc2)c2ccc(Br)cc2)CC2CCC1N2C | 2430 |
| CHEMBL313041 | COc1ccc(C2=NNC(=O)C3CC=CCC23)cc1OC | 1.00E+05 |
| CHEMBL218847 | CCCC(c1ccc(Cl)cc1)C1CCCCN1 | 1300 |
| CHEMBL2012094 | COC(=O)C1=C(c2cccn2C)CC2CCC1O2 | 3000 |
| CHEMBL3775140 | Clc1cccc(N2CCN(CCCN3CCC(Cc4ccccc4)CC3)CC2)c1Cl | 2510 |
| CHEMBL4104150 | [O-][S+](Cc1ccc(Cl)s1)C(c1ccccc1)c1ccccc1 | 57200 |
| CHEMBL728 | CN1CCN(CCCN2c3ccccc3Sc3ccc(Cl)cc32)CC1 | 1169 |
| CHEMBL103152 | COC(=O)C1=C(c2ccc3ccccc3c2)CC2CCC1C2 | 1310 |
| CHEMBL1201237 | CC(C)(C)NCC(O)COc1cccc2c1CCCC2=O | 4898 |
| CHEMBL1644609 | N#Cc1ccccc1-c1cccc(OC2CC3CCC(C2)N3)n1 | 8981 |
| CHEMBL54 | O=C(CCCN1CCC(O)(c2ccc(Cl)cc2)CC1)c1ccc(F)cc1 | 3386 |
| CHEMBL1643659 | Clc1ccc(C2CC3CCC(S3)C2c2cc(-c3ccccc3)no2)cc1 | 10000 |
| CHEMBL4083042 | FC(F)(F)c1cccc(-c2nnnn2CCCN2CCC(Cc3ccccc3)CC2)c1 | 8810 |
| CHEMBL4091567 | Fc1ccc(C(OCCN2CCN(c3ccc(Cl)cc3)CC2)c2ccc(F)cc2)cc1 | 6910 |
| CHEMBL1643642 | Cc1cc(C2=C(c3ccccc3)CC3CCC2S3)on1 | 10000 |
| CHEMBL1214114 | CCCn1c(-c2ccccc2)cc(C(=O)NCCCN2CCN(c3cccc(OC)c3OC)CC2)c1C | 1981 |
| CHEMBL221753 | CC(C)(C)CC(C)(C)c1ccc(OCCOCC[N+](C)(C)Cc2ccccc2)cc1.[Cl-] | 2438 |
| CHEMBL153479 | Nc1c(Br)cc(Br)cc1CNC1CCC(O)CC1 | 3391 |
| CHEMBL1812740 | Cc1ccc(C2CC3CCC(C2C(=O)OC2CCC2)N3C)cc1 | 2010 |
| CHEMBL153062 | CCCCCCCCCc1ccc(O)cc1 | 3380 |
| CHEMBL3799131 | CN(C)CCCN1c2ccccc2CCc2ccc(C#CCOCCCN3c4ccccc4CCc4ccccc43)cc21 | 1600 |
| CHEMBL330250 | COC(=O)C1=C(c2ccc(F)cc2)CC2C(O)CC1N2C | 20000 |
| CHEMBL109776 | O=C(CC1CCN(Cc2ccc(F)cc2)CC1)NC(c1ccccc1)c1ccccc1 | 49179 |
| CHEMBL3705035 | O=C(O)COc1cc(CC2CCNCC2)ccc1Br | 10000 |
| CHEMBL89754 | COC(=O)C1=C(c2ccccc2)CC2CC(O)C1N2C | 10000 |
| CHEMBL3216707 | Cc1c(C(=O)NCCCN2CCN(c3ccncc3)CC2)cc(-c2ccccc2)n1C.Cl.Cl | 10000 |
| CHEMBL314912 | COC(=O)C1C(OC(c2ccc(F)cc2)c2ccc(F)cc2)CC2CC(O)C1N2C | 9560 |
| CHEMBL3216497 | CCCc1nc(C(=O)NCC(O)CN2CCN(c3cccc(Cl)c3Cl)CC2)cn1-c1ccc(Cl)cc1.Cl.Cl.Cl.Cl.Cl | 1276 |
| CHEMBL4064805 | Fc1ccc(C(OCCN2CCN(c3ccccc3Cl)CC2)c2ccc(F)cc2)cc1 | 10000 |
| CHEMBL219703 | CCCCCC(c1ccc(Cl)cc1)C1CCCCN1 | 1700 |
| CHEMBL100010 | COC(=O)C1=C(c2ccccc2)CC2CCC1C2 | 10000 |
| CHEMBL1643648 | Fc1ccc(C2=C(c3cc(-c4ccccc4)no3)C3CCC(C2)S3)cc1 | 10000 |
| CHEMBL201666 | CCCC(C(=O)c1ccc(C(=O)OC)cc1)N1CCCC1 | 2350 |
| CHEMBL227233 | COc1ccc(-c2cnc(C3C(c4ccc(Cl)cc4)CC4CCC3N4C)s2)cc1 | 6700 |
| CHEMBL444412 | CCN(CC)CCn1c2ccccc2c(=O)c2ccccc21 | 10000 |
| CHEMBL388759 | CN1C2CCC1C(c1ncc(-c3ccc(Br)cc3)s1)C(c1ccc(Cl)cc1)C2 | 8220 |
| CHEMBL3775703 | Clc1ccc(N2CCN(CCCN3CCC(Cc4ccccc4)CC3)CC2)c(Cl)c1 | 1460 |
| CHEMBL205028 | CCCCNC(CCC)C(=O)c1ccc(Cl)c(Cl)c1 | 10000 |
| CHEMBL317757 | CN1C2CCC1CC(OC(c1ccc(F)cc1)c1ccc(F)cc1)C2 | 5340 |
| CHEMBL2096857 | COC(=O)C1C(c2ccc(F)cc2)CC2CCC1N2C | 10000 |
| CHEMBL89709 | COC(=O)C1C(c2ccc(F)cc2)CC2CC(O)C1N2C | 20000 |
| CHEMBL88563 | COC(=O)C1C(c2ccccc2)CC2CC(O)C1N2C | 10000 |
| CHEMBL388832 | COc1ccc(-c2cnc(C3C(c4ccc(Cl)cc4)CC4CCC3N4C)s2)cc1OC | 6800 |
| CHEMBL2113279 | COc1ccc(-c2cnc(C3C(c4ccc(C)cc4)CC4CCC3N4C)s2)cc1OC | 10000 |
| CHEMBL4105112 | Fc1ccc(C(OCCN2CCN(c3ncccn3)CC2)c2ccc(F)cc2)cc1 | 1414 |
| CHEMBL420333 | COC(=O)C1C(OC(c2ccc(I)cc2)c2ccc(I)cc2)CC2CCC1N2C | 8550 |
| CHEMBL184867 | COC(=O)C1=C(c2ccccc2)CC2CCC1O2 | 10000 |
| CHEMBL4074417 | CC(=O)c1cccc(-c2nnnn2CCN2CCC(Cc3ccccc3)CC2)c1 | 4050 |
| CHEMBL4091738 | COc1ccc(-c2nnnn2CCCN2CCC(Cc3ccccc3)CC2)cc1 | 4050 |
| CHEMBL88246 | COC(=O)C1C(c2ccc(Cl)c(Cl)c2)CC2C(O)CC1N2C | 10700 |
| CHEMBL508678 | NCC1CCC(CCc2cccnc2)O1 | 2721 |
| CHEMBL4103183 | c1ccc(CN2CCN(CCCCn3nnnc3C(c3ccccc3)c3ccccc3)CC2)cc1 | 1500 |
| CHEMBL1644600 | FC(F)(F)c1cc(OC2CC3CCC(C2)N3)cc(-c2ccccc2)c1 | 4624 |
| CHEMBL1812743 | COc1ccc(-c2cc(C3C(c4ccc(Cl)cc4)CC4CCC3N4C)on2)cc1 | 2000 |
| CHEMBL3613152 | O=C(NCCN1CCC(Cc2ccccc2)CC1)C(c1ccc(Cl)cc1)c1ccc(Cl)cc1 | 10000 |
| CHEMBL219335 | CC(C)CC(c1cccc(Cl)c1)C1CCCCN1 | 2100 |
| CHEMBL204254 | CCCC(C(=O)c1cccc(C)c1)N1CCCC1 | 4400 |
| CHEMBL4073602 | Clc1ccc(-c2nnnn2CCCN2CCC(Cc3ccccc3)CC2)cc1 | 3130 |
| CHEMBL364146 | COC(=O)C1=C(c2ccc(I)cc2)CC2CCC1S2 | 8700 |
| CHEMBL4063764 | Fc1ccc(C(OCCN2CCN(c3cccc(Cl)c3Cl)CC2)c2ccc(F)cc2)cc1 | 5000 |
| CHEMBL589250 | COc1cccc(C(=O)C(C)NC2CCCC2)c1 | 6900 |
| CHEMBL333331 | COC(=O)C1C(OC(c2ccc(Cl)cc2)c2ccc(Cl)cc2)CC2CCC1N2C | 3360 |
| CHEMBL2107797 | C#CC1(O)CCC2C3CCC4=CC(=O)CCC4C3CCC21CC | 1119 |
| CHEMBL322923 | COC(=O)C1=C(c2ccc(Cl)c(Cl)c2)CC2CCC1C2 | 5160 |
| CHEMBL3334795 | CN(C)CCC(c1csc2ccccc12)N1CCOCC1 | 1187.8 |
| CHEMBL482294 | CN(C)Cc1cc(S(N)(=O)=O)ccc1[S+]([O-])c1ccc(C(F)(F)F)cc1 | 1900 |
| CHEMBL2115225 | COC(=O)C1C(c2ccc(Cl)c(Cl)c2)CC2CC(=O)C1N2C | 7038 |
| CHEMBL329454 | COC(=O)C1C2CCC(CC1c1ccc(F)cc1)O2 | 2580 |
| CHEMBL63703 | c1ccc(CN2CCC(NCCOC(c3ccccc3)c3ccccc3)CC2)cc1 | 1860 |
| CHEMBL2012095 | COC(=O)C1=C(c2nccs2)CC2CCC1O2 | 3000 |
| CHEMBL1165775 | CSc1ccc(CC(Cc2ccc(SC)cc2)NC=O)cc1 | 15700 |
| CHEMBL378243 | CN1C2CCC1C1COC(=O)CCSCC(=O)Nc3ccc(cc3)C1C2 | 6730 |
| CHEMBL121857 | COC(=O)C1C(OC(c2ccccc2)c2ccc(I)cc2)CC2CCC1N2C | 9280 |
| CHEMBL2391541 | C=CCN(CC=C)CCc1c[nH]c2ccc(OC)cc12 | 8709.64 |
| CHEMBL218999 | CC(C)CC(c1ccc(Cl)c(Cl)c1)C1CCCCN1 | 1100 |
| CHEMBL2096875 | Cc1ccc(C2CC3CCC(C2c2ccccc2)N3C)cc1 | 16000 |
| CHEMBL339401 | COC(=O)C1C(OC(c2ccc(Br)cc2)c2ccc(I)cc2)CC2CCC1N2C | 4390 |
| CHEMBL104 | Clc1ccccc1C(c1ccccc1)(c1ccccc1)n1ccnc1 | 5760 |
| CHEMBL3798819 | CN(C)CCCN1c2ccccc2CCc2ccc(C#CCOCCOc3cccc4ccccc34)cc21 | 4800 |
| CHEMBL633 | CCCCc1oc2ccccc2c1C(=O)c1cc(I)c(OCCN(CC)CC)c(I)c1 | 2305 |
| CHEMBL1201203 | CN1C2CCC1CC(OC(c1ccccc1)c1ccccc1)C2 | 24100 |
| CHEMBL379333 | CN1C2CCC1C1COC(=O)CCC(=O)CCC(=O)Nc3ccc(cc3)C1C2 | 5010 |
| CHEMBL2096858 | COC(=O)C1C(c2ccc(Cl)cc2)CC2CCC1N2C | 1450 |
| CHEMBL1306 | CC(C)N1CCN(c2ccc(OCC3COC(Cn4cncn4)(c4ccc(Cl)cc4Cl)O3)cc2)CC1 | 11456 |
| CHEMBL4083969 | Clc1ccccc1-c1nnnn1CCCN1CCC(Cc2ccccc2)CC1 | 1320 |
| CHEMBL202565 | CCCC(CN1CCCC1)C(=O)c1ccc(C)cc1 | 4800 |
| CHEMBL1173772 | CCC1NC(C)(C)COC1(O)c1cccc(Cl)c1 | 2500 |
| CHEMBL894 | CC(NC(C)(C)C)C(=O)c1cccc(Cl)c1 | 1.00E+05 |
| CHEMBL43064 | C(=Cc1ccccc1)CN1CCN(C(c2ccccc2)c2ccccc2)CC1 | 1090 |
| CHEMBL60889 | CCOc1cc(N)c(Cl)cc1C(=O)NCC1CN(Cc2ccc(F)cc2)CCO1 | 3811.7 |
| CHEMBL1271930 | CSc1nc(C)cc(C(=O)NCCN2CCN(c3cccc(C)c3C)CC2)n1.Cl | 1014 |
| CHEMBL169429 | COC(=O)C1=C(c2ccc(Cl)c(Cl)c2)CC2CCC1O2 | 1900 |
| CHEMBL1643656 | Cc1cc(C2C3CCC(CC2c2ccc(Cl)c(Cl)c2)S3)on1 | 2000 |
| CHEMBL4073939 | c1ccc(CC2CCN(CCCn3nnnc3C(c3ccccc3)c3ccccc3)CC2)cc1 | 10000 |
| CHEMBL2012086 | COC(=O)C1=C(c2cnc(-c3cccc(Br)c3)s2)CC2CCC1O2 | 3000 |
| CHEMBL1644478 | Fc1cccc(OC2CC3CCC(C2)N3)c1 | 1245 |
| CHEMBL3323187 | CN1Cc2cc(NCCO)ccc2C(c2ccc3sccc3c2)C1 | 2500 |
| CHEMBL4101959 | Clc1cccc(-c2nnnn2CCCN2CCC(Cc3ccccc3)CC2)c1 | 1350 |
| CHEMBL363510 | COC(=O)C1=C(c2ccc(F)cc2)CC2CCC1S2 | 28000 |
| CHEMBL1672015 | NC(Cc1ccc2c(c1)OCO2)Cc1ccc2c(c1)OCO2 | 1600 |
| CHEMBL318592 | COC(=O)C1C2CCC(C2)CC1c1ccccc1 | 20000 |
| CHEMBL2012085 | COC(=O)C1=C(c2cccc(-c3nccs3)c2)CC2CCC1O2 | 3000 |
| CHEMBL379755 | CN1C2CCC1C1COC(=O)CCCCCCCCCCC(=O)Nc3ccc(cc3)C1C2 | 1200 |
| CHEMBL3681351 | CCCc1nc(C(=O)NCCN2CCN(c3cccc(Cl)c3Cl)CC2)cn1-c1ccc(Cl)cc1 | 1093 |
| CHEMBL4073400 | Fc1ccc(C(OCCN2CCC(Cc3ccccc3)CC2)c2ccc(F)cc2)cc1 | 1670 |
| CHEMBL1643649 | Clc1ccc(C2=C(c3cc(-c4ccccc4)no3)C3CCC(C2)S3)cc1 | 10000 |
| CHEMBL3774556 | Clc1ccc(N2CCN(CCCN3CCC(Cc4ccccc4)CC3)CC2)cc1 | 2030 |
| CHEMBL1643650 | Brc1ccc(C2=C(c3cc(-c4ccccc4)no3)C3CCC(C2)S3)cc1 | 10000 |
| CHEMBL4092998 | c1ccc(CC2CCN(CCn3nnnc3C(c3ccccc3)c3ccccc3)CC2)cc1 | 7700 |
| CHEMBL2112595 | O=C(CCc1ccc(Br)cc1)NC=CNCCOC(c1ccc(F)cc1)c1ccc(F)cc1 | 4500 |
| CHEMBL1812741 | CCc1cc(C2C(c3ccc(C)cc3)CC3CCC2N3C)on1 | 2600 |
| CHEMBL1173274 | CC1NC(C)(C)COC1(O)c1ccccc1 | 1.00E+05 |
| CHEMBL86710 | COC(=O)C1=C(c2ccc(Cl)c(Cl)c2)CC2CC(O)C1N2C | 6000 |
| CHEMBL55174 | COC(=O)C1C(OC(c2ccc(F)cc2)c2ccc(F)cc2)CC2CCC1N2Cc1ccccc1 | 4970 |
| CHEMBL227179 | CN1C2CCC1C(c1ncc(-c3ccccc3)s1)C(c1ccc(Cl)cc1)C2 | 2710 |
| CHEMBL1812739 | Cc1ccc(C2CC3CCC(C2C(=O)OC(C)C)N3C)cc1 | 2830 |
| CHEMBL1173354 | CC1NC(C)(C)COC1(O)c1ccc(Cl)cc1 | 4600 |
| CHEMBL6437 | CN1CCN2c3ccccc3Cc3ccccc3C2C1 | 2068 |
| CHEMBL2012091 | COC(=O)C1=C(c2cc3ccccc3[nH]2)CC2CCC1O2 | 3000 |
| CHEMBL66112 | Fc1ccc(CCN2CCCC(CNCCOC(c3ccccc3)c3ccccc3)C2)cc1 | 1130 |
| CHEMBL219224 | CC(c1ccc(Cl)cc1)C1CCCCN1 | 1900 |
| CHEMBL1173598 | CC1N(C)C(C)(C)COC1(O)c1ccc(Cl)cc1 | 6480 |
| CHEMBL369997 | CCCC(C(=O)c1ccc(-c2cccs2)cc1)N1CCCC1 | 1960 |
| CHEMBL2112594 | O=C(CCc1ccccc1)NCCNCCOC(c1ccc(F)cc1)c1ccc(F)cc1 | 1140 |
| CHEMBL4068516 | O=[N+]([O-])c1ccc(-c2nnnn2CCN2CCC(Cc3ccccc3)CC2)cc1 | 1240 |
| CHEMBL491753 | CCN(CC)CCCn1c2ccccc2c(=O)c2ccccc21 | 10000 |
| CHEMBL417049 | CN(CCOC(c1ccccc1)c1ccccc1)C1CCN(CCCc2ccccc2)CC1 | 1910 |
| CHEMBL327963 | COC(=O)C1=C(c2ccc(F)cc2)CC2CCC1C2 | 35400 |
| CHEMBL379446 | CCCC(C(=O)c1ccc([N+](=O)[O-])cc1)N1CCCC1 | 1110 |
| CHEMBL436132 | Cc1ccc(C2CC3CCC(C2c2ncc(-c4cccc(Br)c4)s2)N3C)cc1 | 3860 |
| CHEMBL3216744 | Cc1cccc(N2CCN(CC(O)CNC(=O)c3nc(C)n(CC(C)C)c3C)CC2)c1C.Cl.Cl | 1262 |
| CHEMBL566001 | CC(NC(C)(C)C)C(=O)c1ccc(Br)cc1 | 4508 |
| CHEMBL279331 | O=C(CCC(=O)OCC1C2CCC(CC1c1ccc(Cl)c(Cl)c1)S2)OCC1C2CCC(CC1c1ccc(Cl)c(Cl)c1)S2 | 2200 |
| CHEMBL229699 | Cc1ccc(C2CC3CCC(C2c2ncc(-c4ccc(F)cc4)s2)N3C)cc1 | 3200 |
| CHEMBL361493 | COC(=O)C1=C(c2ccc(F)cc2)CC2CCC1O2 | 50000 |
| CHEMBL1173770 | CC1NC(C)(C)COC1(O)c1cccc(-c2cccc3ccccc23)c1 | 1565 |
| CHEMBL1173427 | CC1NC(C)(C)COC1(O)c1ccc(F)c(F)c1 | 1.00E+05 |
| CHEMBL1765600 | CCCN1C(C)C(O)(c2cccc(Cl)c2)OCC1(C)C | 2900 |
| CHEMBL135 | CC12CCC3c4ccc(O)cc4CCC3C1CCC2O | 10755 |
| CHEMBL375118 | Clc1ccc(C(C2CCCC2)C2CCCCN2)cc1 | 4600 |
| CHEMBL4083782 | Fc1ccc(C(OCCN2CCN(c3ccc(Cl)cc3Cl)CC2)c2ccc(F)cc2)cc1 | 3910 |
| CHEMBL1643660 | Brc1ccc(C2CC3CCC(S3)C2c2cc(-c3ccccc3)no2)cc1 | 10000 |
| CHEMBL1643643 | Cc1cc(C2=C(c3ccc(F)cc3)CC3CCC2S3)on1 | 10000 |
| CHEMBL4063684 | c1ccc(CCCN2CCN(CCCn3nnnc3C(c3ccccc3)c3ccccc3)CC2)cc1 | 8000 |
| CHEMBL426304 | CCCC(C(=O)c1cccc(I)c1)N1CCCC1 | 1070 |
| CHEMBL941 | Cc1ccc(NC(=O)c2ccc(CN3CCN(C)CC3)cc2)cc1Nc1nccc(-c2cccnc2)n1 | 1402 |
| CHEMBL273575 | CN1Cc2c(N)cccc2C(c2ccccc2)C1 | 8709.64 |
| CHEMBL1644470 | c1ccc(-c2cccc(OC3CC4CCC(C3)N4)n2)cc1 | 7644 |
| CHEMBL796 | COC(=O)C(c1ccccc1)C1CCCCN1 | 5100 |
| CHEMBL491754 | CCN(CC)CCCCCn1c2ccccc2c(=O)c2ccccc21 | 7800 |
| CHEMBL4091502 | c1ccc(CCCN2CCN(CCn3nnnc3C(c3ccccc3)c3ccccc3)CC2)cc1 | 3860 |
| CHEMBL4094228 | Brc1ccccc1-c1nnnn1CCCN1CCC(Cc2ccccc2)CC1 | 5130 |
| CHEMBL990 | CN(Cc1ccc(C(C)(C)C)cc1)Cc1cccc2ccccc12 | 1411.8 |
| CHEMBL375685 | Cc1ccc(C2CC3CCC(C2c2ncc(-c4ccc(Br)cc4)s2)N3C)cc1 | 1840 |
| CHEMBL2012087 | COC(=O)C1=C(c2cc3ccccc3o2)CC2CCC1O2 | 3000 |
| CHEMBL296419 | COc1ccc(CCN2CCC(Nc3nc4ccccc4n3Cc3ccc(F)cc3)CC2)cc1 | 1318 |
| CHEMBL333667 | O=C(Cc1ccccc1)NCCNCCOC(c1ccc(F)cc1)c1ccc(F)cc1 | 1300 |
| CHEMBL185792 | COC(=O)C1=C(c2ccc(Cl)cc2)CC2CCC1O2 | 60000 |
| CHEMBL1162 | C#CC1(O)CCC2C3CCC4=CC(=O)CCC4C3CCC21C | 4537 |
| CHEMBL334255 | CCCCCCCCCCCCCCCC[n+]1ccccc1.[Br-] | 1680 |
| CHEMBL4103476 | [O-][S+](Cc1nc(C(F)(F)F)no1)C(c1ccccc1)c1ccccc1 | 209500 |
| CHEMBL426659 | CN1C2CCC1CC(O)(c1cccc(-c3ccccc3)c1)C2 | 3770 |
| CHEMBL186242 | COC(=O)C1=C(c2ccc(Cl)c(Cl)c2)CC2CCC1S2 | 3600 |
| CHEMBL227231 | CN1C2CCC1C(c1ncc(-c3cccc(Br)c3)s1)C(c1ccc(Cl)cc1)C2 | 4000 |
| CHEMBL2012092 | COC(=O)C1=C(c2cccs2)CC2CCC1O2 | 3000 |
| CHEMBL4072566 | O=[N+]([O-])c1ccc(-c2nnnn2CCCN2CCC(Cc3ccccc3)CC2)cc1 | 5250 |
| CHEMBL1644601 | Brc1cc(OC2CC3CCC(C2)N3)cc(-c2ccccc2)c1 | 2610 |
| CHEMBL3775008 | Clc1ccc(N2CCN(CCN3CCC(Cc4ccccc4)CC3)CC2)c(Cl)c1 | 4320 |
| CHEMBL4094047 | Fc1ccc(C(OCCN2CCN(c3ccccc3)CC2)c2ccc(F)cc2)cc1 | 1906 |
| CHEMBL566000 | CC(NC(C)(C)C)C(=O)c1ccc(Cl)cc1 | 9800 |
| CHEMBL534479 | CCOC(=O)C(c1ccccc1)C1CCCCN1.Cl | 10000 |
| CHEMBL338313 | COC(=O)C1C(OC(c2ccccc2)c2ccc(Cl)cc2)CC2CCC1N2C | 5960 |
| CHEMBL199598 | CN1C2CCC1CC(O)(C1c3ccccc3-c3ccc(Br)cc31)C2 | 2570 |
| CHEMBL370158 | C[N+]1(CCl)C2C=C(c3ccccc3-c3ccccc3)CC1CC2 | 16100 |
| CHEMBL456017 | COc1ccc(F)cc1C1CCC(CCN)O1 | 2576 |
| CHEMBL253376 | CN(Cc1cc(Br)cc(Br)c1N)C1CCCCC1 | 2091 |
| CHEMBL1812744 | Cc1cc2nc(C3C(c4ccc(Cl)cc4)CC4CCC3N4C)[nH]c2cc1C | 1.00E+05 |
| CHEMBL381127 | CCCC(C(=O)c1ccc(OC)cc1)N1CCCC1 | 2430 |
| CHEMBL200473 | CN1C2CCC1CC(O)(c1ccccc1-c1ccccc1)C2 | 2520 |
| CHEMBL1173277 | Cc1cccc(C2(O)OCC(C)(C)NC2C)c1 | 1.00E+05 |
| CHEMBL89208 | COC(=O)C1C(c2ccccc2)CC2C(O)CC1N2C | 20000 |
| CHEMBL69380 | CN(CCc1ccc(F)cc1)CC1CCCN(CCOC(c2ccccc2)c2ccccc2)C1 | 2130 |
| CHEMBL89469 | COC(=O)C1=C(c2ccc(F)cc2)CC2CCC1N2C | 7990 |
| CHEMBL417035 | Fc1ccc(CCNCC2CCCN(CCOC(c3ccccc3)c3ccccc3)C2)cc1 | 1830 |
| CHEMBL110530 | Fc1ccc(CN2CCC(CCNC(c3ccccc3)c3ccccc3)CC2)cc1 | 5848 |
| CHEMBL1173599 | CC1NC(C)(C)COC1(O)c1cccnc1 | 1.00E+05 |
| CHEMBL1644607 | N#Cc1cccc(-c2cccc(OC3CC4CCC(C3)N4)c2)c1 | 10000 |
| CHEMBL373710 | CC(C)CCC(c1ccc(Cl)cc1)C1CCCCN1 | 8300 |
| CHEMBL118 | Cc1ccc(-c2cc(C(F)(F)F)nn2-c2ccc(S(N)(=O)=O)cc2)cc1 | 6276 |
| CHEMBL2012090 | COC(=O)C1=C(c2cc3ccccc3s2)CC2CCC1O2 | 2000 |
| CHEMBL3775012 | Brc1ccc(N2CCN(CCCN3CCC(Cc4ccccc4)CC3)CC2)cc1 | 2250 |
| CHEMBL441864 | Clc1ccc(C(CCCc2ccccc2)C2CCCCN2)cc1 | 1600 |

**Supplementary Table 2 The test1 data set**

| CHEMBL ID | smiles | IC50/nM |
|---|---|---|
| CHEMBL808 | Clc1ccc(COC(Cn2ccnc2)c2ccc(Cl)cc2Cl)cc1 | 423 |
| CHEMBL1081302 | Cc1c(C(=O)NCCCN2CCN(c3cccc(Cl)c3)CC2)cc(-c2ccccc2)n1Cc1ccccc1 | 235 |
| CHEMBL513360 | O=C(C1CCC1)N(Cc1ccc(Cl)cc1Cl)C1CCNC1 | 41 |
| CHEMBL1214178 | CCCn1c(-c2ccccc2)cc(C(=O)N(C)CCCN2CCN(c3cccc(Cl)c3Cl)CC2)c1C.Cl | 102 |
| CHEMBL3963829 | Cc1ccc(OC(c2ccc(Cl)c(Cl)c2)C2CNC2)cc1 | 5.67 |
| CHEMBL4083258 | CCCCCN(C)CCCC1(c2ccc(F)cc2)OCc2cc(C#N)ccc21 | 75 |
| CHEMBL1852373 | COc1ccc(-n2cc(C(=O)NCCCN3CCN(c4cccc(C)c4C)CC3)nc2-c2ccccc2)cc1 | 42.6 |
| CHEMBL1214046 | CCCn1c(-c2ccccc2)cc(C(=O)NCCCN2CCN(c3cccc(Br)c3)CC2)c1C | 225 |
| CHEMBL173344 | C=C(C)c1ccc(C2CC3CCC(N3)C2C(=O)CC)cc1 | 0.16 |
| CHEMBL2338039 | Cc1cc(Cl)ccc1OC(c1ccccc1)C1CCNC1 | 1 |
| CHEMBL460204 | COc1ccc(F)cc1CCCCC1CCC(CCN)O1 | 31 |
| CHEMBL1237044 | COc1cccc(C2(O)CCCCC2CN(C)C)c1 | 372 |
| CHEMBL4092411 | CNc1nc(C#Cc2ccc(Cl)s2)nc2c1ncn2C1OC(C(=O)OC)C(O)C1O | 10 |
| CHEMBL395408 | Cc1cccc(CN(C2CCOCC2)C2CCNC2)c1C | 43 |
| CHEMBL2338055 | Clc1cc(Cl)c(OC(c2ccccc2)C2CCNC2)c(-c2ccccn2)c1 | 31.62 |
| CHEMBL326466 | CN(C)CC1C2CCC(C2)C1c1ccc2ccccc2c1 | 9 |
| CHEMBL214209 | CN1C2CCC1C1COC(=O)CCCCCC(=O)Nc3ccc(cc3)C1C2 | 200 |
| CHEMBL479783 | CSc1ccc(Sc2ccc(N(C)S(C)(=O)=O)cc2CN(C)C)cc1 | 5 |
| CHEMBL4064213 | CN(C)CCCC1(c2ccc(F)cc2)OCc2cc3ccccc3cc21 | 8 |
| CHEMBL3216702 | CCCn1c(-c2ccccn2)cc(C(=O)NCCCN2CCN(c3cccc(Cl)c3Cl)CC2)c1C.Cl.Cl.Cl.Cl | 397 |
| CHEMBL1088163 | CCCn1c(-c2ccccc2)cc(C(=O)NCCCN2CCN(c3cccc(C)c3C)CC2)c1C | 76 |
| CHEMBL3216962 | COc1ccccc1-n1c(C)nc(C(=O)NCCCN2CCN(c3cccc(Cl)c3C)CC2)c1C.Cl.Cl.Cl | 16.8 |
| CHEMBL471211 | NCC1CC1(C(=O)N1Cc2ccccc2C1)c1cccs1 | 13 |
| CHEMBL1852362 | CCCc1nc(C(=O)NCC(O)CN2CCN(c3cccc(C)c3C)CC2)c(C)n1-c1ccc(OC)cc1 | 6 |
| CHEMBL2338034 | Fc1cc(Cl)ccc1OC(c1ccccc1)C1CCNC1 | 12.59 |
| CHEMBL121611 | Fc1ccc(C(OCCNCCCNCCc2ccccc2)c2ccc(F)cc2)cc1 | 249 |
| CHEMBL493476 | CN(C)Cc1ccccc1C(O)c1ccc(Cl)c(Cl)c1 | 250 |
| CHEMBL370805 | COC(=O)C1C(OC(=O)c2ccccc2)CC2CCC1N2C | 325 |
| CHEMBL316428 | COC(=O)C1=C(c2ccc3ccccc3c2)CC2C(O)CC1N2C | 260 |
| CHEMBL258062 | CNCc1cc(C(=O)N(C)C)ccc1Oc1ccc(Cl)cc1Cl | 11 |
| CHEMBL225235 | Cc1ccc(C2CC3CCC(C2c2ccc(F)cc2)N3C)cc1 | 100 |
| CHEMBL204759 | CCCC(C(=O)c1ccc2ccccc2c1)N1CCCC1 | 46 |
| CHEMBL419728 | Cc1ccc(C2CC3CCC4C2C(O)(c2ccc(Cl)cc2)CN34)cc1 | 13 |
| CHEMBL3216259 | Cc1c(C(=O)NCCCN2CCN(c3cccc(Cl)c3Cl)CC2)cc(-c2ccccn2)n1C.Cl.Cl.Cl.Cl | 110 |
| CHEMBL1644477 | Ic1cccc(OC2CC3CCC(C2)N3)c1 | 106 |
| CHEMBL2338043 | Clc1ccc(Cl)c(OC(c2ccccc2)C2CCNC2)c1Cl | 3.981 |
| CHEMBL2338033 | Fc1cc(F)cc(OC(c2ccccc2)C2CCNC2)c1 | 3.981 |
| CHEMBL404317 | CNCc1cc(C(=O)NC)ccc1Oc1ccc(Cl)cc1C | 14 |
| CHEMBL213030 | Clc1ccccc1CC(c1nccs1)N1CCNCC1 | 79 |
| CHEMBL512967 | CCC(=O)N(Cc1ccc(Cl)cc1Cl)C1CCNC1 | 25 |
| CHEMBL247340 | Cc1cccc(CN(C2CCOCC2)C2CCNC2)c1Cl | 5 |
| CHEMBL2219883 | Clc1ccc(C(Oc2ccccc2)C2CNC2)cc1 | 9.2 |
| CHEMBL401554 | CN1C2CCC1C(COC(=O)CCC(=O)OCC1C3CCC(CC1c1ccc(Cl)c(Cl)c1)O3)C(c1ccc(Cl)c(Cl)c1)C2 | 405 |
| CHEMBL256818 | CNCc1ccccc1Oc1ccc(Cl)cc1OC | 13 |
| CHEMBL2012104 | COC(=O)C1C2CCC(CC1c1ccc(-c3ccco3)cc1)O2 | 30 |
| CHEMBL3216303 | Cc1nc(C(=O)NCCCN2CCN(c3cccc(Cl)c3Cl)CC2)c(C)n1C1CCCC1.Cl.Cl.Cl.Cl | 35 |
| CHEMBL1644480 | O=[N+]([O-])c1cccc(OC2CC3CCC(C2)N3)c1 | 86 |
| CHEMBL2338045 | Fc1cc(F)c(Cl)c(OC(c2ccccc2)C2CCNC2)c1Cl | 7.943 |
| CHEMBL3216095 | Cc1c(C(=O)NCC(O)CN2CCN(c3cccc(Cl)c3Cl)CC2)nc(-c2ccccc2)n1-c1ccccc1.Cl.Cl.Cl.Cl | 71.1 |
| CHEMBL54628 | COC(=O)C1C(c2ccc(F)cc2)CC2CCC1N2CCCc1ccccc1 | 133 |
| CHEMBL3216741 | Cc1cccc(N2CCN(CCCNC(=O)c3nc(C)n(-c4ccccc4F)c3C)CC2)c1C.Cl.Cl | 6.9 |
| CHEMBL3216960 | Cc1c(Cl)cccc1N1CCN(CC(O)CNC(=O)c2nc(-c3ccccc3)n(-c3ccc(F)cc3)c2C)CC1.Cl.Cl.Cl | 232 |
| CHEMBL1086613 | CCn1c(-c2ccccc2)cc(C(=O)NCCCN2CCN(c3cccc(C)c3C)CC2)c1C | 36 |
| CHEMBL67387 | COC(=O)C1C(c2cccc(I)c2)CC2CCC1N2C | 9.88 |
| CHEMBL3216293 | CCCc1nc(C(=O)NCC(O)CN2CCN(c3cccc(Cl)c3Cl)CC2)c(C)n1-c1ccc(OC)cc1.Cl.Cl.Cl.Cl | 51.8 |
| CHEMBL3215626 | Cc1cccc(N2CCN(CC(O)CNC(=O)c3nc(C)n(-c4ccccc4Cl)c3C)CC2)c1C.Cl.Cl.Cl | 5 |
| CHEMBL4062666 | CN(C)CCCC1(c2ccc(Cl)c(Cl)c2)OCc2ccccc21 | 13 |
| CHEMBL362716 | O=C1OC2(CCC(N3CCC(Cc4cc(F)ccc4Br)CC3)CC2)c2ccc3c(c21)OCCO3 | 77 |
| CHEMBL3216979 | CCCn1c(C)nc(C(=O)NCCCN2CCN(c3cccc(Cl)c3Cl)CC2)c1C.Cl.Cl.Cl.Cl | 18 |
| CHEMBL2338035 | Fc1ccc(OC(c2ccccc2)C2CCNC2)c(Cl)c1 | 15.85 |
| CHEMBL257424 | CN(C)Cc1ccccc1Oc1cccc(Cl)c1 | 190 |
| CHEMBL1214425 | CCn1c(-c2ccc(Cl)cc2)cc(C(=O)NCCCN2CCN(c3cccc(C)c3C)CC2)c1C.Cl | 93 |
| CHEMBL488326 | CSc1ccc(Oc2ccncc2CN(C)C)cc1 | 2 |
| CHEMBL2337596 | Clc1ccc(OC(c2ccccc2)C2CCNC2)cc1 | 7.943 |
| CHEMBL3216964 | COc1ccc(-n2c(-c3ccccc3)nc(C(=O)NCCCN3CCN(c4cccc(Cl)c4C)CC3)c2C)cc1.Cl.Cl.Cl | 25 |
| CHEMBL3216746 | Cc1cccc(N2CCN(CCCNC(=O)c3nc(C)n(CC(C)C)c3C)CC2)c1C.Cl.Cl | 436.2 |
| CHEMBL562115 | COC(=O)C1C(c2ccc(C)cc2)CC2CCC1N2CC#CCF | 239.77 |
| CHEMBL2337590 | Cc1ccccc1OC(c1ccccc1)C1CCNC1 | 63.1 |
| CHEMBL3216966 | CCCc1nc(C(=O)NCC(O)CN2CCN(c3cccc(Cl)c3C)CC2)c(C)n1-c1ccc(OC)cc1.Cl.Cl.Cl | 17 |
| CHEMBL180899 | Oc1nc2cc3c(cc2c2c1C(N1CCC(Cc4cc(F)ccc4F)CC1)CCC2)OCO3 | 14 |
| CHEMBL3799503 | CN(C)CCCN1c2ccccc2CCc2ccc(C#CCOSCc3cccc4ccccc34)cc21 | 350 |
| CHEMBL1852346 | COc1ccccc1-n1c(C)nc(C(=O)NCCCN2CCN(c3cccc(C)c3C)CC2)c1C | 9 |
| CHEMBL469068 | CC(C)(C)CC(=O)N(Cc1ccc(Cl)cc1Cl)C1CCNC1 | 27 |
| CHEMBL3799356 | CN(C)CCCN1c2ccccc2CCc2ccc(CCCOCCCN3c4ccccc4CCc4ccccc43)cc21 | 270 |
| CHEMBL2380983 | CN(C)CC1CC1c1ccc2[nH]cc(C#N)c2c1 | 6 |
| CHEMBL331799 | CNCC1C2CCC(C2)C1c1ccc2ccccc2c1 | 50 |
| CHEMBL1852516 | Cc1nc(C(=O)NCCN2CCN(c3cccc(Cl)c3Cl)CC2)c(C)n1-c1ccccc1 | 66 |
| CHEMBL595063 | NCCC(Oc1cccc2ccccc12)c1cccs1 | 25.6 |
| CHEMBL3612833 | O=C(NCCCN1CCC(Cc2ccccc2)CC1)c1ccc2ccccc2c1 | 108 |
| CHEMBL492842 | CSc1ccc(Cc2ccccc2CN(C)C)cc1 | 5 |
| CHEMBL449145 | CC(C)(C)C(=O)N(Cc1ccc(Cl)cc1Cl)C1CCNC1 | 41 |
| CHEMBL404754 | CNCc1cc(C(=O)NC)ccc1Oc1ccc(Cl)cc1Cl | 13 |
| CHEMBL3323185 | CN1Cc2cc(N3CCCCC3)ccc2C(c2cc3ccccc3s2)C1 | 38 |
| CHEMBL375131 | FC(F)(F)Oc1ccccc1CC(c1ccccc1)N1CCNCC1 | 13 |
| CHEMBL421867 | c1ccc(CCCNC2CCN(CCOC(c3ccccc3)c3ccccc3)CC2)cc1 | 390 |
| CHEMBL603620 | COc1cccc(C(=O)C(C)NC(C)(C)C)c1 | 100 |
| CHEMBL27227 | CN1CC=C(c2c[nH]c3ccc(Br)cc23)CC1 | 210 |
| CHEMBL459367 | COc1ccc(F)cc1CCCC1CCC(CN)O1 | 2.6 |
| CHEMBL3323096 | CN1Cc2ccccc2C(c2ccc3ccsc3c2)C1 | 40 |
| CHEMBL477374 | CC(C)C(=O)N(Cc1ccc(Cl)c(Cl)c1Cl)C1CCNC1 | 5 |
| CHEMBL457968 | Cc1cc(CCC2CCC(CCN)O2)ccc1F | 30 |
| CHEMBL2380980 | CN(C)CC1CCCC1c1ccc2[nH]cc(C#N)c2c1 | 3.8 |
| CHEMBL3216075 | CCCc1nc(C(=O)NCC(O)CN2CCN(c3cccc(Cl)c3C)CC2)cn1-c1ccc(Cl)cc1.Cl.Cl.Cl.Cl | 80 |
| CHEMBL3334796 | CN(C)CCC(c1csc2ccccc12)N1CCCC1 | 425.3 |
| CHEMBL214121 | Fc1cccc(C(Cc2ccccc2Cl)N2CCNCC2)c1 | 15 |
| CHEMBL1214045 | CCCn1c(-c2ccccc2)cc(C(=O)NCCCN2CCN(c3ccc(Cl)c(Cl)c3)CC2)c1C.Cl | 284 |
| CHEMBL491601 | CN(C)Cc1cc(NS(C)(=O)=O)ccc1Oc1ccc2c(c1)OCCS2 | 15 |
| CHEMBL3216506 | COc1ccc(-n2c(C)nc(C(=O)NCC(O)CN3CCN(c4cccc(C)c4C)CC3)c2C)cc1.Cl.Cl | 21.6 |
| CHEMBL605501 | CC(NC1CCCC1)C(=O)c1cccc(F)c1 | 100 |
| CHEMBL3799596 | CN(C)CCCN1c2ccccc2CCc2ccc(CCCOCCOCCc3cccc4ccccc34)cc21 | 280 |
| CHEMBL562854 | Cc1cccc(OC(c2cccnc2)C2CCNCC2)c1C | 3 |
| CHEMBL578612 | CC(C(=O)c1cccc(Cl)c1)N(C)C(C)(C)C | 100 |
| CHEMBL504028 | NCC1CCC(c2ccc3ccccc3c2)O1 | 4 |
| CHEMBL447825 | NCC1CCC(c2ccc(Cl)cc2)O1 | 32 |
| CHEMBL507806 | C=CCN(Cc1nccs1)C(=O)C1(c2cccs2)CC1CN | 16 |
| CHEMBL1087895 | Cc1cccc(N2CCN(CCCNC(=O)c3cc(-c4ccccc4)n(Cc4ccccc4)c3C)CC2)c1C | 14.3 |
| CHEMBL209707 | c1ccc(CCC(c2ccccc2)N2CCNCC2)cc1 | 23 |
| CHEMBL56564 | CN1C2CCC1CC(OC(=O)c1c[nH]c3ccccc13)C2 | 488 |
| CHEMBL2447969 | Cc1c(C(=O)NCCN2CCN(c3cccc(Cl)c3Cl)CC2)cc(-c2ccccc2)n1C.Cl | 57 |
| CHEMBL4072807 | CN(C)CCCC(=O)c1ccc2c(c1)COC2(CCCN(C)C)c1ccc(F)cc1 | 1.6 |
| CHEMBL3681355 | CCCn1c(C)nc(C(=O)NCC(O)CN2CCN(c3cccc(Cl)c3C)CC2)c1C | 12.1 |
| CHEMBL2012111 | COC(=O)C1C2CCC(CC1c1cc3ccccc3s1)O2 | 10 |
| CHEMBL404258 | COc1cc(Cl)ccc1Oc1ccccc1CN(C)C | 6 |
| CHEMBL212877 | Clc1ccccc1CC(c1ccccc1)N1CCCNCC1 | 42 |
| CHEMBL469233 | O=C(c1ccccc1)N(Cc1ccc(Cl)cc1Cl)C1CCNC1 | 195 |
| CHEMBL2338038 | COc1cc([N+](=O)[O-])ccc1OC(c1ccccc1)C1CCNC1 | 1.585 |
| CHEMBL1642903 | NCC1Cc2ccccc2C(c2ccc(Cl)c(Cl)c2)C1 | 19 |
| CHEMBL479410 | CN(C)Cc1cc(NS(C)(=O)=O)ccc1Oc1ccc(C(F)(F)F)cc1 | 5 |
| CHEMBL3323093 | CN1Cc2ccccc2C(c2cccc3occc23)C1 | 29 |
| CHEMBL1644604 | N#Cc1cc(OC2CC3CCC(C2)N3)cc(-c2ccccc2)c1 | 100 |
| CHEMBL1221 | Clc1ccc(CSC(Cn2ccnc2)c2ccc(Cl)cc2Cl)cc1 | 141 |
| CHEMBL202348 | C=CCC(C(=O)c1ccc(Cl)c(Cl)c1)N1CCCC1 | 440 |
| CHEMBL3216084 | COc1ccc(-n2c(-c3ccccc3)nc(C(=O)NCC(O)CN3CCN(c4cccc(C)c4C)CC3)c2C)cc1.Cl.Cl | 201 |
| CHEMBL3216737 | Cc1cccc(N2CCN(CCCNC(=O)c3nc(-c4ccccc4)n(-c4ccc(F)cc4)c3C)CC2)c1C.Cl.Cl | 153.2 |
| CHEMBL3217170 | Cc1c(Cl)cccc1N1CCN(CCCNC(=O)c2nc(-c3ccccc3)n(-c3ccccc3)c2C)CC1.Cl.Cl.Cl | 17.8 |
| CHEMBL490205 | COc1ccc(Oc2ccc(SC)cc2)c(CN(C)C)c1 | 6 |
| CHEMBL71 | CN(C)CCCN1c2ccccc2Sc2ccc(Cl)cc21 | 39 |
| CHEMBL1271985 | Cc1cc(C(=O)NCCCN2CCN(c3cccc(C)c3C)CC2)nc(C2CC2)n1.Cl | 347 |
| CHEMBL3681367 | COc1ccc(-n2cc(C(=O)NCC(O)CN3CCN(c4cccc(Cl)c4Cl)CC3)nc2-c2ccccc2)cc1 | 323 |
| CHEMBL379205 | FC(F)(F)COc1ccccc1CC(c1ccccc1)N1CCNCC1 | 38 |
| CHEMBL3215623 | Cc1nc(C(=O)NCCCN2CCN(c3cccc(Cl)c3Cl)CC2)c(C)n1-c1ccccc1F.Cl.Cl.Cl.Cl | 13 |
| CHEMBL3217181 | CCCn1c(C)nc(C(=O)NCC(O)CN2CCN(c3cccc(C)c3C)CC2)c1C.Cl.Cl | 10.4 |
| CHEMBL3323173 | Cc1ccc2c(c1)CN(C)CC2c1ccc2sccc2c1 | 4.9 |
| CHEMBL360790 | O=C1OC2(CCC(N3CCC(Cc4cc(F)ccc4Br)CC3)CC2)c2ccc3c(c21)OCO3 | 82 |
| CHEMBL3334801 | COc1ccc2cc(C(CCN(C)C)N3CCCC3)ccc2c1Cl | 337.6 |
| CHEMBL1644482 | COc1cccc(OC2CC3CCC(C2)N3)c1 | 369 |
| CHEMBL3216756 | CCCc1nc(C(=O)NCC(O)CN2CCN(c3cccc(Cl)c3Cl)CC2)c(C)n1-c1ccccc1.Cl.Cl.Cl | 6.2 |
| CHEMBL3216504 | CCCc1nc(C(=O)NCC(O)CN2CCN(c3cccc(Cl)c3Cl)CC2)c(C)n1-c1ccc2c(c1)OCCO2.Cl.Cl.Cl | 20 |
| CHEMBL1729 | COc1cc(N)c(Cl)cc1C(=O)NC1CCN(CCCOc2ccc(F)cc2)CC1OC | 384 |
| CHEMBL3216298 | Cc1cccc(N2CCN(CCCNC(=O)c3nc(C)n(-c4ccccc4Cl)c3C)CC2)c1C.Cl.Cl.Cl | 17 |
| CHEMBL2096860 | COC(=O)C1C(c2ccc(I)cc2)CC2CCC1N2C | 66.3 |
| CHEMBL1080746 | Cc1c(C(=O)NCCCN2CCN(c3cccc(Cl)c3)CC2)cc(-c2ccccc2)n1N1CCCCC1 | 110 |
| CHEMBL450907 | COc1ccc(F)cc1CCCC1CCC(CCN)O1 | 2.3 |
| CHEMBL183550 | COc1ccc(Br)c(CC2CCN(C3CCC4(CC3)OC(=O)c3c4ccc4c3OCO4)CC2)c1 | 91 |
| CHEMBL3770651 | CN(C)CC1(c2ccc(Cl)c(Cl)c2)CCCC(O)C1 | 44 |
| CHEMBL551556 | FC(F)(F)c1ccccc1OC(c1cccnc1)C1CCNCC1 | 77 |
| CHEMBL283899 | CN1CC=C(c2c[nH]c3ccc(C#N)cc23)CC1 | 19 |
| CHEMBL4066973 | CNc1nc(C#Cc2ccc(Cl)s2)nc2c1ncn2C1C(O)C(O)C2(C(=O)OC)CC12 | 10 |
| CHEMBL1852420 | CCCn1c(C)nc(C(=O)NCC(O)CN2CCN(c3cccc(C)c3C)CC2)c1C | 10.4 |
| CHEMBL2219958 | Clc1ccc(C(OCc2ccccc2)C2CNC2)cc1Cl | 7.81 |
| CHEMBL1213934 | Cc1c(C(=O)NCCCN2CCN(c3cccc4cccnc34)CC2)cc(-c2ccccc2)n1C | 22 |
| CHEMBL471002 | C=CCN(CC)C(=O)C1(c2cccs2)CC1CN | 140 |
| CHEMBL2338031 | Fc1ccc(OC(c2ccccc2)C2CCNC2)c(F)c1 | 7.943 |
| CHEMBL490204 | CSc1ccc(Oc2ccc(C)cc2CN(C)C)cc1 | 5 |
| CHEMBL1080712 | Cc1cccc(N2CCN(CCCCNC(=O)c3cc(-c4ccccc4)[nH]c3C)CC2)c1C | 21 |
| CHEMBL3917453 | Cc1cccc(OC(c2ccc(Cl)c(Cl)c2)C2CNC2)c1 | 19.8 |
| CHEMBL3216505 | Cc1c(Cl)cccc1N1CCN(CC(O)CNC(=O)c2nc(C)n(C3CCCC3)c2C)CC1.Cl.Cl.Cl | 39.1 |
| CHEMBL1852422 | COc1ccccc1-n1c(C)nc(C(=O)NCCCN2CCN(c3cccc(Cl)c3Cl)CC2)c1C | 15.8 |
| CHEMBL1852505 | COc1ccc(-n2c(C)nc(C(=O)NCCCN3CCN(c4cccc(Cl)c4C)CC3)c2C)cc1 | 12.4 |
| CHEMBL489166 | CNS(=O)(=O)c1ccc(Oc2ccc(SC)cc2)c(CN(C)C)c1 | 9 |
| CHEMBL564100 | Clc1ccccc1OC(c1cccnc1)C1CCNCC1 | 9 |
| CHEMBL255140 | CCCCCCC(=O)OCC1C(c2ccc(Cl)c(Cl)c2)CC2CCC1N2C | 53 |
| CHEMBL376113 | Cc1ccc(C2CC3CCC(C2c2ncc(-c4ccc(Cl)c(Cl)c4)s2)N3C)cc1 | 4400 |
| CHEMBL1173355 | COc1ccc(C2(O)OCC(C)(C)NC2C)cc1 | 100000 |
| CHEMBL1852514 | Cc1cccc(N2CCN(CC(O)CNC(=O)c3nc(C)n(CC(C)C)c3C)CC2)c1C | 1262 |
| CHEMBL3774732 | COc1ccccc1N1CCN(CCCN2CCC(Cc3ccccc3)CC2)CC1 | 1760 |
| CHEMBL1405 | CC12CCC3c4ccc(O)cc4CCC3C1CCC2=O | 5301 |
| CHEMBL125017 | COC(=O)C1C(OC(c2ccccc2)c2ccccc2)CC2CCC1N2C | 10100 |
| CHEMBL314402 | COC(=O)C1C2CCC(CC1c1ccccc1)O2 | 10000 |
| CHEMBL122452 | CN(CCCN(C)CCc1ccccc1)CCOC(c1ccccc1)c1ccccc1 | 1700 |
| CHEMBL382768 | CCCC(C(=O)c1ccc(OC)c(OC)c1)N1CCCC1 | 1540 |
| CHEMBL2096876 | Cc1ccc(C2C(c3ccccc3)CC3CCC2N3C)cc1 | 11000 |
| CHEMBL4064788 | Cc1ccc(C(c2ccccc2)[S+]([O-])Cc2cccs2)cc1 | 118400 |
| CHEMBL612 | CC(N)Cc1ccccc1 | 16595.87 |
| CHEMBL4081927 | Brc1cccc(-c2nnnn2CCCN2CCC(Cc3ccccc3)CC2)c1 | 10000 |
| CHEMBL1271874 | Cc1cc(C(=O)NCCN2CCN(c3cccc(C)c3C)CC2)nc(C(C)(C)C)n1.Cl | 6374 |
| CHEMBL1271815 | Cc1nc(C(=O)NCCCN2CCN(c3cccc(C)c3C)CC2)cc(C(C)C)n1.Cl | 1213 |
| CHEMBL2430686 | O=C1NCC(c2cccc(Cl)c2)C1c1ccc(Cl)c(Cl)c1 | 3757 |
| CHEMBL218848 | CC(C)C(c1ccc(Cl)cc1)C1CCCCN1 | 3300 |
| CHEMBL4086279 | Fc1ccc(C(OCCN2CCN(c3cccc(Cl)c3)CC2)c2ccc(F)cc2)cc1 | 4960 |
| CHEMBL4095847 | c1ccc(CC2CCN(CCn3nnnc3-c3cccc4ccccc34)CC2)cc1 | 10000 |
| CHEMBL2012080 | COC(=O)C1=C(c2ccc(-c3cccn3C)cc2)CC2CCC1O2 | 3000 |
| CHEMBL323348 | COC(=O)c1ccccc1C(=O)OC | 100000 |
| CHEMBL224062 | COc1ccc(C2C(c3ccc(C)cc3)CC3CCC2N3C)cc1 | 1669 |
| CHEMBL3774957 | Clc1cccc(N2CCN(CCCN3CCC(Cc4ccccc4)CC3)CC2)c1 | 1180 |
| CHEMBL4101780 | Fc1ccc(C(OCCN2CCN(c3ccc(Cl)c(Cl)c3)CC2)c2ccc(F)cc2)cc1 | 5220 |
| CHEMBL726 | OCCN1CCN(CCCN2c3ccccc3Sc3ccc(C(F)(F)F)cc32)CC1 | 1492 |
| CHEMBL186447 | COC(=O)C1=C(c2ccc(Cl)cc2)CC2CCC1S2 | 10000 |
| CHEMBL7002 | CC1(COc2ccc(CC3SC(=O)NC3=O)cc2)CCCCC1 | 14948 |
| CHEMBL643 | CC(CN1c2ccccc2Sc2ccccc21)N(C)C | 4010 |
| CHEMBL1173275 | CC1NC(C)(C)COC1(O)c1cccc(F)c1 | 100000 |
| CHEMBL382955 | CN1C2CCC1CC(O)(C1c3ccccc3-c3ccccc31)C2 | 35400 |
| CHEMBL1164191 | CSc1ccc(CC(N)Cc2ccc(SC)cc2)cc1 | 2500 |
| CHEMBL339481 | CCC(OC1CC2CCC(C1C(=O)OC)N2C)c1ccccc1 | 21000 |
| CHEMBL432571 | COC(=O)C1=C(c2ccc(Cl)c(Cl)c2)CC2C(O)CC1N2C | 3320 |
| CHEMBL420900 | O=C(Cc1ccccc1)NCCCNCCOC(c1ccc(F)cc1)c1ccc(F)cc1 | 1230 |
| CHEMBL372312 | CCCC(C(=O)c1ccc(C)cc1)N1CCCC1 | 1070 |
| CHEMBL1165303 | CCNC(Cc1ccc(SC)cc1)Cc1ccc(SC)cc1 | 2200 |
| CHEMBL365092 | COC(=O)C1=C(c2ccc3ccccc3c2)CC2CCC1S2 | 1300 |
| CHEMBL444592 | Fc1ccc(C(NCCC2CCN(Cc3ccccc3)CC2)c2ccc(F)cc2)cc1 | 10657 |
| CHEMBL340521 | COC(=O)C1C(OC(c2ccccc2)c2ccc(Br)cc2)CC2CCC1N2C | 5770 |
| CHEMBL227180 | CN1C2CCC1C(c1ncc(-c3ccc(Cl)cc3)s1)C(c1ccc(Cl)cc1)C2 | 4400 |
| CHEMBL1173279 | COc1cccc(C2(O)OCC(C)(C)NC2C)c1 | 100000 |
| CHEMBL1172928 | CC1NC(C)(C)COC1(O)c1cccc(Cl)c1 | 100000 |
| CHEMBL1643667 | c1ccc(-c2cc(C3C4CCC(CC3c3ccccc3)S4)on2)cc1 | 10000 |
| CHEMBL282575 | CC12CCC3c4ccc(OC(=O)c5ccccc5)cc4CCC3C1CCC2O | 3672 |
| CHEMBL1173429 | CC1NC(C)(C)COC1(O)c1cc(F)cc(F)c1 | 100000 |
| CHEMBL26320 | c1ccc(CCCN2CCN(CCOC(c3ccccc3)c3ccccc3)CC2)cc1 | 3710 |
| CHEMBL201976 | CCCC(C(=O)c1ccc(-c2ccco2)cc1)N1CCCC1 | 2180 |
| CHEMBL1765599 | CCN1C(C)C(O)(c2cccc(Cl)c2)OCC1(C)C | 1500 |
| CHEMBL89662 | COC(=O)C1=C(c2ccccc2)CC2CCC1N2C | 28600 |
| CHEMBL1171260 | CC1N(C)C(C)(C)COC1(O)c1cccc(Cl)c1 | 100000 |
| CHEMBL157455 | CC(NC(C)(C)C)C(O)c1ccccc1 | 1240 |
| CHEMBL378483 | CN1C2CCC1C1COC(=O)CCC3(CCC(=O)Nc4ccc(cc4)C1C2)c1ccccc1-c1ccccc13 | 2000 |
| CHEMBL1173276 | CC1NC(C)(C)COC1(O)c1cccc(Br)c1 | 100000 |
| CHEMBL218193 | CC(C)CC(c1ccc(C(C)C)cc1)C1CCCCN1 | 4700 |
| CHEMBL1214047 | CCCn1c(-c2ccccc2)cc(C(=O)NCCCN2CCN(c3ccccc3OC)CC2)c1C | 1084 |
| CHEMBL1642917 | CNC1CCc2ccccc2C1c1ccc(Cl)c(Cl)c1 | 2827 |
| CHEMBL1644471 | c1ccc(OC2CC3CCC(C2)N3)cc1 | 5655 |
| CHEMBL211494 | CN1C2CCC1C1COC(=O)CCCCC(=O)Nc3ccc(cc3)C1C2 | 2000 |
| CHEMBL376852 | CCC(c1ccc(Cl)cc1)C1CCCCN1 | 2400 |
| CHEMBL841 | CN(C)C(=O)C(CCN1CCC(O)(c2ccc(Cl)cc2)CC1)(c1ccccc1)c1ccccc1 | 2202 |
| CHEMBL219916 | O=c1[nH]c2ccccc2n1CCCN1CCC(n2c(=O)[nH]c3cc(Cl)ccc32)CC1 | 3055 |
| CHEMBL458639 | NCC1CCC(c2ccccc2)O1 | 1714 |
| CHEMBL1670 | Clc1ccc(C(c2ccccc2Cl)C(Cl)Cl)cc1 | 3015 |
| CHEMBL83 | CCC(=C(c1ccccc1)c1ccc(OCCN(C)C)cc1)c1ccccc1 | 2334 |
| CHEMBL187695 | COC(=O)C1=C(c2ccc(Br)cc2)CC2CCC1S2 | 25000 |
| CHEMBL1644472 | CN1C2CCC1CC(Oc1ccccc1)C2 | 10000 |
| CHEMBL1173600 | CC1NC(C)(C)COC1(O)c1ccccn1 | 100000 |
| CHEMBL2012096 | COC(=O)C1=C(c2ccc(-c3ccccc3)o2)CC2CCC1O2 | 3000 |
| CHEMBL4068156 | [O-][S+](Cc1cccs1)C(c1ccc(F)cc1)c1ccc(F)cc1 | 120500 |
| CHEMBL3775177 | c1ccc(CC2CCN(CCCN3CCN(c4ccccc4)CC3)CC2)cc1 | 1360 |
| CHEMBL1173701 | CC1NC(C)(C)COC1(O)c1ccc(-c2ccccc2)cc1 | 100000 |
| CHEMBL186488 | COC(=O)C1=C(c2ccc(Br)cc2)CC2CCC1O2 | 30000 |
| CHEMBL2012097 | COC(=O)C1=C(c2ccc(-c3ccccc3)s2)CC2CCC1O2 | 3000 |
| CHEMBL411 | CCC(=C(CC)c1ccc(O)cc1)c1ccc(O)cc1 | 12492 |
| CHEMBL1643644 | Cc1cc(C2=C(c3ccc(Cl)cc3)CC3CCC2S3)on1 | 10000 |
| CHEMBL1643647 | c1ccc(C2=C(c3cc(-c4ccccc4)no3)C3CCC(C2)S3)cc1 | 10000 |
| CHEMBL3774516 | FC(F)(F)c1ccc(N2CCN(CCCN3CCC(Cc4ccccc4)CC3)CC2)cc1 | 2610 |
| CHEMBL204538 | CC#Cc1ccc(C(=O)C(CCC)N2CCCC2)cc1 | 3300 |
| CHEMBL219653 | Clc1ccc(C(CC2CCCC2)C2CCCCN2)cc1 | 2100 |
| CHEMBL435990 | COC(=O)C1C(OC(c2ccc(C)cc2)c2ccc(C)cc2)CC2CCC1N2C | 9800 |

**Supplementary Table 3 The test2 data set**

| smiles | LABEL |
|---|---|
| Clc1ccc(COC(Cn2ccnc2)c2ccc(Cl)cc2Cl)c(Cl)c1 | non-inhibitor |
| NC(Cc1ccc(N(CCCl)CCCl)cc1)C(=O)O | non-inhibitor |
| C[N+](C)(C)CC(O)CC(=O)[O-] | non-inhibitor |
| CC12CCC3c4ccc(O)cc4CCC3C1CCC2=O | non-inhibitor |
| O=S(O)CNc1ccc(S(=O)(=O)c2ccc(NCS(=O)O)cc2)cc1 | non-inhibitor |
| Nc1ccc(N=Nc2ccccc2)c(N)n1 | non-inhibitor |
| OCCN1CCN(CCCN2c3ccccc3C=Cc3ccccc32)CC1 | non-inhibitor |
| C=C1C(=CC=C2CCCC3(C)C2CCC3C(C)CCCC(O)(C(F)(F)F)C(F)(F)F)CC(O)CC1O | non-inhibitor |
| O=C1C2CCCCC2C(=O)N1CCCCN1CCN(c2nsc3ccccc23)CC1 | non-inhibitor |
| CC1(C)OC2COC3(COS(N)(=O)=O)OC(C)(C)OC3C2O1 | non-inhibitor |
| Cc1onc(-c2ccccc2)c1-c1ccc(S(N)(=O)=O)cc1 | non-inhibitor |
| CC(Cn1cnc2c(N)ncnc21)OCP(=O)(O)O | non-inhibitor |
| Cc1cccc(N(C)C(=S)Oc2ccc3ccccc3c2)c1 | non-inhibitor |
| COC(=O)C1C(c2ccc(I)cc2)CC2CCC1N2CCCF | non-inhibitor |
| CC(=O)Oc1ccc(C(c2ccc(OC(C)=O)cc2)c2ccccn2)cc1 | non-inhibitor |
| COC1OC(CO)C(OC2OC(CO)C(OC3OC(CO)C(OC4OC(CO)C(OC5OC(CO)C | non-inhibitor |
| CC(=O)NCCCS(=O)(=O)O | non-inhibitor |
| CCN(CC)CCOCCOC(=O)C1(c2ccccc2)CCCC1 | non-inhibitor |
| CNC(=C[N+](=O)[O-])NCCSCc1ccc(CN(C)C)o1 | non-inhibitor |
| CC12CCC3=C4CCC(=O)C=C4CCC3C1CCC2(O)CC#N | non-inhibitor |
| CC(=N)N1CCC(SC2=C(C(=O)O)N3C(=O)C(C(C)O)C3C2)C1 | non-inhibitor |
| COC1=C(C(=O)O)N2C(=O)C(NC(=O)C(N)C3C=CCC=C3)C2SC1 | non-inhibitor |
| Nc1nc(F)nc2c1ncn2C1OC(CO)C(O)C1O | non-inhibitor |
| O=C(Nc1c(Cl)cncc1Cl)c1ccc(OC(F)F)c(OCC2CC2)c1 | non-inhibitor |
| COC1C(OC(N)=O)C(O)C(Oc2ccc3c(=O)c(NC(=O)c4ccc(O)c(CC=C(C)C)c4)c(O)oc3c2C)OC1(C)C | non-inhibitor |
| CON=C(C(=O)NC1C(=O)N2C(C(=O)O)=C(COC(C)=O)CSC12)c1csc(N)n1 | non-inhibitor |
| CCC(NC(C)C)C(O)c1ccc(O)c2[nH]c(=O)ccc12 | non-inhibitor |
| Nc1nc(N)c2nc(-c3ccccc3)c(N)nc2n1 | non-inhibitor |
| NC(=O)N1c2ccccc2C=Cc2ccccc21 | non-inhibitor |
| CC(N)(Cc1ccc(O)c(O)c1)C(=O)O | non-inhibitor |
| CC(CN(C)C)CN1c2ccccc2CCc2ccccc21 | inhibitor |
| COCCCCC(=NOCCN)c1ccc(C(F)(F)F)cc1 | inhibitor |
| CNCCCC1c2ccccc2C=Cc2ccccc21 | inhibitor |
| COC(=O)C1C(OC(=O)c2ccccc2)CC2CCC1N2C | inhibitor |
| CCc1nn(CCCN2CCN(c3cccc(Cl)c3)CC2)c(=O)n1CCOc1ccccc1 | inhibitor |
| CN(C)CC=C(c1ccc(Br)cc1)c1cccnc1 | inhibitor |
| N#Cc1ccc2[nH]cc(CCCCN3CCN(c4ccc5oc(C(N)=O)cc5c4)CC3)c2c1 | inhibitor |